\documentclass[showpacs,showkeys,nofootinbib,aps]{revtex4}
\usepackage{graphicx}% Include figure files
\usepackage{dcolumn}% Align table columns on decimalpoint
\usepackage{bm}% bold math
\usepackage{graphicx}
\usepackage{longtable}
\usepackage{color}
\usepackage{latexsym}
\usepackage{amssymb}
\begin{document}
\newcommand{\ti}[1]{\mbox{\tiny{#1}}}
\newcommand{\im}{\mathop{\mathrm{Im}}}
\def\be{\begin{equation}}
\def\ee{\end{equation}}
\def\bea{\begin{eqnarray}}
\def\eea{\end{eqnarray}}
\newcommand{\il}{~}
\newcommand{\uQ}{\sqrt{8\pi}M_{\ti{Pl}}/m}
\newcommand{\uR}{1/m}
\newcommand{\uN}{M_{\ti{Pl}}^2/m^2}
\newcommand{\uM}{M_{\ti{Pl}}^2/m}
\newcommand{\uq}{\sqrt{8\pi}m/M_{\ti{Pl}}}
\newcommand{\uQM}{\sqrt{8\pi}/M_{\ti{Pl}}}
\newcommand{\uQMR}{m\sqrt{8\pi}/M_{\ti{Pl}}}
\newcommand{\uQR}{\sqrt{8\pi}M_{\ti{Pl}}}
\newcommand{\tb}[1]{\textbf{\texttt{#1}}}
\newcommand{\rtb}[1]{\textcolor[rgb]{1.00,0.00,0.00}{\tb{#1}}}
\newcommand{\btb}[1]{\textcolor[rgb]{0.00,0.00,1.00}{\tb{#1}}}
\newcommand{\ytb}[1]{\textcolor[rgb]{0.85,0.80,0.48}{\tb{#1}}}
%\textcolor[rgb]{0.85,0.80,0.48}{}

\title{On charged boson stars}

\author{Daniela Pugliese$^{1}$, Hernando Quevedo$^{2,3,4,5}$, Jorge A. Rueda H.$^{2,3}$, and Remo Ruffini$^{2,3}$}
\email{d.pugliese.physics@gmail.com, jorge.rueda@icra.it,quevedo@nucleares.unam.mx, ruffini@icra.it}
\affiliation{$^1$ School of Mathematical Sciences, Queen Mary University of London,
Mile End Road, London E1 4NS, United Kingdom\\
$^2$Dipartimento di Fisica and ICRA, Sapienza Universit\`a di Roma, Piazzale Aldo Moro 5, I--00185 Roma, Italy\\
$^3$ ICRANet, Piazza della Repubblica 10, I--65122 Pescara, Italy \\
$^4$Instituto de Ciencias Nucleares, Universidad Nacional Aut\'onoma de M\'exico,
AP 70543, M\'exico, DF 04510, Mexico\\
$^5$ Instituto de Cosmologia, Relatividade e Astrofisica and ICRANet - CBPF,
Rua Dr. Xavier Sigaud, 150, CEP 22290-180, Rio de Janeiro, Brazil
}
\date{\today}
\begin{abstract}
We study static, spherically symmetric, self-gravitating
systems minimally coupled to a scalar field with $U(1)$ gauge symmetry: charged
boson stars. We find numerical solutions to the Einstein-Maxwell
equations coupled to the relativistic Klein-Gordon equation. It is shown
that bound stable configurations exist only for values of the coupling constant
less than or equal to a certain critical value. The metric coefficients and the
relevant physical quantities such as the total mass and charge, turn out to
be in general bound functions of the radial coordinate, reaching their maximum values
at a critical value of the scalar field at the origin. {We discuss the stability problem both from the quantitative and qualitative point of view.
Taking properly  into account the electromagnetic contribution to the total mass,
the stability issue is faced also following an  indication of the binding energy
per particle, verifying then
 the existence of  configurations having  positive binding energy:  objects apparently bound can be unstable
against small perturbations, in full analogy with what observed in the mass-radius relation of neutron stars.}
\end{abstract}
%self-gravitating systems, 04.40.-b
%Boson systems, 05.30.Jp
%BH classical, 04.70.Bw
%%% ----------------------------------------------------------------------
\maketitle
%%% ----------------------------------------------------------------------
%%%%%%%%%%%%%%%%%%%%%%%%%%%%%%%%%%%%%%%%%%%%%%%%%%%%%%%%%%%%%%%
\section{Introduction}
Spherically symmetric charged boson stars
are solutions of the Einstein-Maxwell system  of equations coupled to the  general relativistic Klein-Gordon equations of a
complex scalar field with a local $U(1)$ symmetry.
The study of the phenomena related
to the formation and stability of self-gravitating systems is of major interest in astrophysics. It has been conjectured, for instance, that a
boson star could model Bose-Einstein
condensates on astrophysical scales \cite{SM99,SM98,MS96,JS94,JB01,BLV01,JoBern2001,ChaHar2012}.
The collapse of charged compact objects composed by bosons could
lead in principle to charged black holes (see e.g.~\cite{Jetzer:1993nk,Kleihaus:2009kr} and also \cite{TCL00}). Compact
boson objects play an important role in astrophysics since these configurations may represent also an
initial condition for the process of gravitational collapse \cite{RR2007};
see also \cite{Lieb-Pale12} for a recent review. Moreover, boson stars have been shown to be able to mimic
 the power spectrum of accretion disks around black holes (see, for example, \cite{Guzman-RUBE2009}).
Scalar fields are also implemented in many cosmological models either to regulate the inflationary scenarios
\cite{Gorini:2003wa,LlinMota13,Bertolami:2004nh,Fay04} or to describe dark matter and dark energy
(see e.g.~\cite{BerCarPar2012,GaoKinkLiPar10,MainColoBono05,Arbey09}).
 On the other hand, in the Glashow-Weinberg-Salam Standard Model of elementary particles, a real scalar particle, the Higgs boson,
 is introduced in order to provide leptons and vector bosons with
 mass after symmetry breaking; in this respect, the latest results of the Large Hadron Collider experiments \cite{Chatrchyan:2012ufa} reflect the importance of the scalar fields in particle physics. Scalar fields are also found within superstring theories as dilaton fields, and, in the low energy limit of string theory, give rise to various scalar-tensor theories for the gravitational interaction \cite{polchinski}.

Ruffini and Bonazzola \cite{RR} quantized a real scalar field and found a spherically symmetric solution of the Einstein-Gordon system of equations.
The general relativistic treatment eliminates completely some difficulties of the Newtonian approximation, where an
increase of the number of particles corresponds to an increase of the
total energy of the system until the energy reaches a maximum value
and then decreases to assume negative values. It was also shown in \cite{RR} that for these many boson systems
 the assumption of perfect fluid does not apply any longer since the pressure of the
system is anisotropic.
On the other hand, this treatment introduces for the
first time the concept of a critical mass for these objects. Indeed, in full analogy with
white dwarfs and neutron stars,
there is a critical mass and a critical number of particles and, for
charged objects, a critical value of the total charge, over
which this system is unstable against gravitational collapse
to a black hole.

In \cite{Jetzer:1989av,Jetzer:1989us,Jetzer:1990wr} the study of
the charged boson stars was introduced  solving	numerically the
Einstein-Maxwell-Klein-Gordon equations.
In \cite{Jetzer:1989us}  charged
boson configurations were studied for non singular
asymptotically-flat solutions. In particular, it was shown the
existence of a critical value for the central density, mass and number of particles.
{The gravitational attraction of spherically symmetric self-gravitating systems of bosons (charged and neutral)
balances the  kinetic  and Coulomb  repulsion. On the other hand, the Heisenberg
uncertainty principle prevents neutral boson stars from a
gravitational collapse. Furthermore, in order to avoid
gravitational collapse the radius $R$ must
satisfy the condition $ R\geq3R_{\mbox{\tiny{S}}}$ where
$R_{\mbox{\tiny{S}}}$ is Schwarzschild radius  \cite{Schunck:2003kk,Jetzer:1989fx}. On the other hand, stable charged boson stars can
exist  if the gravitational attraction is larger than the
Coulomb repulsion: if the repulsive Coulomb force overcomes
 the attractive gravitational force, the system becomes
unstable \cite{Schunck:2003kk, Jetzer:1991jr,Kusmartsev:2008py,Schunck:1999pm,Mielke:1997re}.
Moreover, as for other  charged objects, if the
radius of these systems	is less than the electron Compton
wavelength and if they are super--critically charged, then pair
production of electrons and positrons occurs.}

These previous works restricted the boson charge to the so-called ``critical''
value (in {Lorentz-Heaviside}  units)
 $q_{\ti{crit}}^{2}=4\pi\left(m/M_{\mbox{\tiny{Pl}}}\right)^{2}$ for a particle of mass $m$ where $M_{\mbox{\tiny{Pl}}}$ is the Planck mass. This value comes out from equating the Coulomb and gravitational forces, so it is expected that for a boson charge $q>q_{\ti{crit}}$, the repulsive Coulomb force be larger than the attractive gravitational one. However, such a critical particle charge does not take into account the gravitational binding energy per particle and so there may be the possibility of having {stable} configurations for bosons with $q>q_{\ti{crit}}$.

Thus, in this work we numerically integrate the coupled system of
Einstein-Maxwell-Klein-Gordon equations, focusing our attention on
configurations characterized by a value of the boson charge close to or larger than $q_{\ti{crit}}$. We will not consider the excited state for the boson fields, consequently we study  only the zero-nodes solutions.

We here show that stable charged configurations of self-gravitating charged
bosons are possible with particle charge $q=q_{\ti{crit}}$. In addition,  it can be shown by means of numerical calculations that for values $q>q_{\ti{crit}}$ {localized} solutions
are possible
only for values of the central density smaller than some critical value over which the
boundary conditions at the origin are not satisfied. We also study the behavior of the radius as well as of
the total charge and mass of the system for $q\simeq q_{\ti{crit}}$.

The plan of the paper is the following: In Sec.\il\ref{Sec:EQUACBS}, we set up the problem by introducing the general formalism  and  writing the system of Einstein-Maxwell-Klein-Gordon equations for charged boson stars.
In Sec.\il\ref{Sec:CBSCRM}, we discuss the concepts of charge, radius, mass and particle number.
In Sec.\il\ref{Sec:CBSNI}, we show the results of the numerical integration.
Finally, in Sec.\il\ref{Sec:CBSCONCL}, we  summarize  and discuss the results.
To compare our results with those of uncharged configurations, we include in the Appendix
the numerical analysis of the limiting case of neutral boson stars.

%%%%%%%%%%%%%%%%%%%%%%%%%%%%%%%%%%%%%%%%%%%%%%%%%%%%%%%%%%%%%%%%%%%%%%%%%
%%%%%%%%%%%%%%%%%%%%%%%%%%%%%%%%%%%%%%%%%%%%%%%%%%%%%%%%%%%%%%%%%%%%%%%%%
\section{The Einstein-Maxwell-Klein-Gordon equations}
\label{Sec:EQUACBS}
We consider static, spherically symmetric self-gravitating
systems of a scalar field minimally coupled to a
$U(1)$ gauge field: charged boson stars.
The Lagrangian density of the massive electromagnetically coupled scalar
field $\Phi$, in units with $\hbar=c=1$, is
\begin{equation}\label{Man-n}
\mathcal{L}_{\mbox{\tiny{M}}}=\sqrt{-g}\left[g^{\mu\nu}\left(D_{\mu}
\Phi\right)\left(D_{\nu}\Phi\right)^{*}-m^{2}
\Phi\Phi^{*}-\frac{1}{4}F_{\mu\nu}F^{\mu\nu}\right],
\end{equation}
where  $g\equiv \det g_{\mu\nu}$,
$m$ is the scalar field mass and $D_{\mu}\equiv\nabla_{\mu}+\imath q
A_{\mu}$, where the constant $q$  is the boson charge,
$\nabla$ stands for the covariant derivative, the
asterisk denotes the complex conjugation, $A_{\mu}$ is the
electromagnetic vector potential, while $F_{\mu\nu}=
\partial_{\mu} A_{\nu}-\partial_{\nu}A_{\mu}$ is the electromagnetic
field tensor \cite{Jetzer:1990wr,Fulling,BD}.
We use a metric $g_{\mu\nu}$ with signature $(+,-,-,-)$;
Greek indices run from 0 to 3, while Latin indices run from 1 to 3.

Therefore the total Lagrangian density $\mathcal{L}$
for the field $\Phi$ minimally coupled to gravity and  to a $U(1)$
gauge field is
\begin{equation}\label{total}
\mathcal{L}=\sqrt{-g}\frac{R}{16 \pi
G_{\mbox{\tiny{N}}}}+\mathcal{L}_{\mbox{\tiny{M}}},
\end{equation}
where $R$ is the scalar curvature,
$M_{\mbox{\tiny{Pl}}}=G_{\mbox{\tiny{N}}}^{-1/2}$ is the Planck
mass, and $G_{\mbox{\tiny{N}}}$ is the gravitational constant.

The Lagrangian density is invariant under a local $U(1)$ gauge transformation
(of the field $\Phi$). The corresponding conserved Noether density
current $J^{\mu}$ is given by
\begin{equation}\label{R-d-o}
J^{\mu}=\sqrt{-g}g^{\mu\nu}\left[\imath q\left(\Phi^*\partial_{\nu}
\Phi-\Phi \partial_{\nu}\Phi^{*}\right)-2q^{2}A_{\nu}
\Phi\Phi^*\right],
\end{equation}
while the energy-momentum tensor $T_{\mu\nu}$ is
\bea\nonumber
T_{\mu\nu}=\left(D_{\mu}
\Phi\right)^{*}\left(D_{\nu}\Phi\right)+\left(D_{\mu}
\Phi\right)\left(D_{\nu}\Phi\right)^{*}-g_{\mu\nu}g^{\alpha\beta}\left(D_{\alpha}
\Phi\right)^{*}\left(D_{\beta}\Phi\right)\\\label{colon-so-ra}-g_{\mu\nu}m^{2}
\Phi\Phi^*+\frac{1}{4}g_{\mu\nu}F_{\alpha\beta}F^{\alpha\beta}-g^{\alpha\beta}F_{\mu\alpha}F_{\nu\beta}.
\eea
In the case of spherical symmetry, the general line element
can be written  in standard Schwarzschild-like coordinates $(t,r,\vartheta,\varphi)$
as
\begin{equation}\label{Les-aut}
ds^2=e^\nu dt^2-e^\lambda
dr^2-r^2\left(d\vartheta^2+\sin\vartheta^2d\varphi^2\right),
\end{equation}
where $\nu$ and
$\lambda$ are  functions of the radial coordinate $r$ only.
Since we want to study only static solutions, the metric
and energy-momentum tensor must be time-independent even if the
matter field $\Phi$ may depend on time.

Then, we set the following stationarity
ansatz \cite{Ad02,De64,Schunck:2003kk,Jetzer:1997zx,Wy81}
\begin{equation}\label{Pre-v-t}
\Phi(r,t)=\phi(r)e^{\imath \omega t},
\end{equation}
where $\phi$ is in general a complex field. Equation (\ref{Pre-v-t}) describes a
spherically symmetric bound state of scalar fields
with positive (or negative) frequency $\omega$.
Accordingly, the electromagnetic four-potential is
$A_\mu(r)=\left(A_t(r)=A(r),A_r=0,A_\theta=0, A_\varphi=0\right)$.

From Eq.\il(\ref{colon-so-ra}) we obtain the following non-zero components of
the energy-momentum tensor $T^{\mu}_{\phantom\ \nu}$:
\begin{eqnarray}
% \nonumber to remove numbering (before each equation)
\label{20-20208Set} T^{0}_{\phantom\ 0}&=& \left[m^{2}+e^{-\nu}
\left(\omega+qA\right)^2\right] \phi^2+\frac{ e^{-\lambda-\nu}
(A')^2}{2}+ \phi'^2e^{-\lambda}, \\
\label{20-20208Setbis} T^{1}_{\phantom\ 1} &=&
\left[m^{2}-e^{-\nu} \left(\omega+qA\right)^2\right]
\phi^2+\frac{e^{-\lambda-\nu}
(A')^2}{2}- \phi'^2e^{-\lambda}, \\
\label{20-20208Settris} T^{2}_{\phantom\ 2} &=& T^{3}_{\phantom\
3}=\left[m^{2}-e^{-\nu} \left(\omega+qA\right)^2\right]
\phi^2-\frac{e^{-\lambda-\nu} (A')^2}{2}+ \phi'^2e^{-\lambda},
\end{eqnarray}
where the prime denotes the differentiation with  respect to $r$.
Let us note from Eqs.~(\ref{20-20208Set}--\ref{20-20208Settris}) that the
energy-momentum tensor is not isotropic.

Finally, the set of Euler-Lagrange equations for the system
described by Eq.\il(\ref{total}) gives the two following independent equations for the
metric components:
\begin{eqnarray}
% \nonumber to remove numbering (before each equation)
\nonumber \lambda'&=&\frac{1-e^{\lambda}}{r}+ 8\pi
G_{\mbox{\tiny{N}}}r e^{\lambda}\left\{\left[m^{2}+e^{-\nu}
\left(\omega+qA\right)^2\right] \phi^2+\frac{ e^{-\lambda-\nu}
(A')^2}{2}+ \phi'^2e^{-\lambda}\right\},
\\\label{Pu-c-ni1}\\
\nonumber \nu'&=&\frac{-1+e^{\lambda}}{r}+ 8\pi G_{\mbox{\tiny{N}}}r
e^{\lambda}\left\{\left[-m^{2}+e^{-\nu}
\left(\omega+qA\right)^2\right] \phi^2-\frac{e^{-\lambda-\nu}
(A')^2}{2}+ \phi'^2e^{-\lambda}\right\},
\\\label{Pu-c-ni2}
\end{eqnarray}
which are equivalent to the Einstein equations $ G^{\mu}_{\phantom\ \nu}=8\pi
G_{\mbox{\tiny{N}}}T^{\mu}_{\phantom\ \nu}$, where
$G^{\mu}_{\phantom\ \nu}= R^{\mu}_{\phantom\
\nu}-\frac{1}{2}\delta^{\mu}_{\phantom\ \nu}R$ is the Einstein
tensor. Then the Maxwell equations are simply
\begin{equation}\label{Met-tan-80}
A''+\left( \frac{2}{r}-\frac{\nu'+\lambda'}{2}\right)A'-2 q
e^{\lambda} \phi^2 \left(\omega+qA\right)=0,
\end{equation}
and the Klein-Gordon equation is
\begin{equation}\label{JaJLev80}
\phi''+\left( \frac{2}{r}+\frac{\nu'-\lambda'}{2}\right)\phi'+
e^{\lambda} \left[
\left(\omega+qA\right)^{2}e^{-\nu}-m^2\right]\phi=0.
\end{equation}
In order to have a localized particle distribution,  we impose the
following boundary conditions:
\be
\phi(\infty)=0, \quad\phi'(\infty)=0,\quad
\mbox{and}\quad \phi(0)=\mbox{constant},\quad\phi'(0)=0 .\label{som.la}
\ee
We also impose  the electric field to be vanishing  at the origin so that
\begin{equation}\label{cos_ui}
 A'(0)=0\ ,
\end{equation}
and we demand that
\begin{equation}\label{Don-ze-ti}
A(\infty)=0\ , \quad  A'(\infty)=0.
\end{equation}

Furthermore, we impose the following two conditions on the metric
components:
\begin{eqnarray}
% \nonumber to remove numbering (before each equation)
\label{Nem-r-o} g^{tt}(\infty)&=&1, \\
\label{dul-c-ra}g^{rr}(0)&=&1\ .
\end{eqnarray}
Equation (\ref{Nem-r-o}) implies that the spacetime is asymptotically the ordinary Minkowski manifold,
while Eq.\il(\ref{dul-c-ra}) is a regularity
condition \cite{Jetzer:1989av}.

We can read Eqs.~(\ref{Pu-c-ni1}--\ref{JaJLev80}),
with these boundary conditions, as eigenvalue equations for the
frequency $\omega$. They form a system of four coupled ordinary
differential equations to be solved numerically.

It is also possible to make the following rescaling of variables:
\begin{eqnarray}\label{dot-re}
% \nonumber to remove numbering (before each equation)
\omega\rightarrow m\omega, \quad &\quad&q\rightarrow q \ m\sqrt{8\pi
G_{\mbox{\tiny{N}}}},\quad \phi(r)\rightarrow\phi(r)\left/ \sqrt{8\pi
G_{\mbox{\tiny{N}}}}\right.   \\
\label{dot-re0} r \rightarrow r/m, &\quad&
A(r) q+\omega
\rightarrow C(r),
\end{eqnarray}
in order to simplify the integration of the system \cite{Jetzer:1989av,Jetzer:1993nk}.

Using Eqs.~(\ref{dot-re}) and (\ref{dot-re0}),  Eqs.\il(\ref{Pu-c-ni1}--\ref{JaJLev80}) become:
\begin{eqnarray}
% \nonumber to remove numbering (before each equation)
\label{tre} \phi''&=&-e^{\lambda} \left(-1+ e^{-\nu} C^2\right)
\phi+e^{\lambda} r \phi^2 \phi'-\left(\frac{e^{\lambda}+1}{r}-
\frac{e^{-\nu}r C'^2}{2q^{2}}\right) \phi'
\\\label{sir-ga}
C''&=&2 e^{\lambda}  q^2 C \phi^2+ e^{\lambda -\nu} r C^2 \phi^2
C'-C' \left(\frac{2}{r}-r \phi'^2\right),
\\\label{mar-e-la}
\lambda '&=&-\frac{(e^{\lambda}-1)}{r}+r e^{\lambda}
(C^{2}e^{-\nu}+1)\phi^{2}+r \phi'^{2}+ \frac{C'^{2}}{2q^{2}}r
e^{-\nu},\\\label{li} \nu'&=&\frac{e^{\lambda}-1}{r}+r
e^{\lambda}(C^{2} e^{-\nu}-1)\phi^{2}+r
\phi'^{2}-\frac{C'^{2}}{2q^{2}}r e^{-\nu}.
\end{eqnarray}

It is worth to  note  that these equations are invariant under the
following rescaling:
%\footnote{Where:
%\begin{multicol}{2}
\begin{equation}\label{le-ott-si}
C\rightarrow \gamma C, \quad e^{\nu(r)}\rightarrow \gamma^{2}
e^{\nu(r)},
\end{equation}
%\end{multicol}
%
where $\gamma$ is a constant. Therefore, since we impose the
conditions at infinity:
\begin{equation}\label{c0}
g^{tt}(\infty)=1,\quad A(\infty)=0,
\end{equation}
we can use this
remaining invariance to make  $C(0)=1$. Thus the equations become  eigenvalue
equations for $e^{\nu(0)}$ and not for $\omega$. For each field value $\phi(0)>0$, one can solve the equations and study the
behavior of the solutions for different values of the charge $q$
imposing
\bea\label{cond-net}
\lambda(0)&=&0,\quad\phi'(0)=0,\\
\label{cond-neta}
C(0)&=&1, \quad C'(0)=0,
\eea
and looking for $\nu(0)$ in such a way that $\phi$ be a smoothly decreasing function and approaches  zero at infinity. (See
also \cite{Jetzer:1989us}).
\section{Charge, radius, mass and particle number}\label{Sec:CBSCRM}
The  locally conserved Noether density current (\ref{R-d-o})
provides  a definition for the total charge $Q$ of the system with
%, leading
%
\begin{equation}\label{G-ia-ta}
Q=\int d^3x J^{0}=8 \pi q\int^{\infty}_{0}dr r^2
\left(\omega+qA\right)\phi^2e^{\frac{\lambda-\nu}{2}}.
\end{equation}
Assuming   the
bosons to have  the identical charge $q$, the total number $N$ can be
related to  $Q$ by  $Q=qN$, so $N$ is given by
as \cite{Schunck:2003kk}
\begin{equation}\label{tri-fo}
N\equiv 8 \pi \int^{\infty}_{0}dr r^2
\left(\omega+qA\right)\phi^2e^{\frac{\lambda-\nu}{2}}.
\end{equation}
%
%\{
For the mass of the system it has been widely used in the literature the expression (see e.g.~\cite{Jetzer:1989av,Jetzer:1989us, Jetzer:1990wr})
\begin{equation}\label{F-ro}
M=4 \pi \int^{\infty}_{0} dr
r^2\left\{\left[\left(\omega+qA\right)^{2}e^{-\nu}+m^2\right]\phi^2+\phi'^{2}
e^{-\lambda}+\frac{1}{2}A'^{2}e^{-\left(\lambda+\nu\right)}\right\},
\end{equation}
which follows from the definition
$M=4\pi \int  dr r^ 2 T^0_0$, using Eq.~(\ref{20-20208Set}). Notice that, however,
such a mass does not satisfy the matching condition with an exterior Reissner-Nordstr\"om spacetime, which
relates the actual mass $M^*$, and charge $Q$ of the system with the metric function $\lambda$ through the relation
\begin{equation}\label{Ros-na}
e^{-\lambda}=1-\frac{2M^*G_{\mbox{\tiny{N}}}}{r}+\frac{Q^2G_{\mbox{\tiny{N}}}}{4\pi
r^2},
%\quad r\rightarrow\infty
\end{equation}
where $Q$ is given by the integral (\ref{G-ia-ta}). Thus, the contribution of the scalar field to the exterior gravitational field is encoded
in the mass and charge only, see e.g.~\cite{Jetzer:1993nk,Jetzer:1989us,Jetzer:1989av,Jetzer:1990wr,Schunck:2003kk,Jetzer:1989fx,Prikas:2002ij}.

The masses $M$ and $M^*$ given by Eqs.~(\ref{F-ro}) and (\ref{Ros-na}), respectively, are related each other as
\be\label{Mstar}
M^*\equiv M+\frac{Q^2}{8 \pi r},
\ee
so the difference $\Delta M\equiv M^*-M$ gives the electromagnetic contribution to the total mass. Using the variables (\ref{dot-re}) and (\ref{dot-re0}), Eq.~(\ref{Mstar}) reads
\be
M^*\equiv M+\frac{Q^2}{ r}.
\ee

We will discuss below the difference both from the quantitative and qualitative point of view of using the mass definitions $(M, M^*)$ for different boson star configurations.

Finally, we define the radius of the charged boson star
as
\begin{equation}\label{do-s-lio}
R\equiv\frac{1}{qN}\int d^3xJ^{0}r=\frac{8 \pi}{N} \int^{\infty}_{0}
drr^3\left(\omega+qA\right)\phi^2e^{\frac{\lambda-\nu}{2}},
\end{equation}
where $N$ is given by the integral (\ref{tri-fo}).
This formula relates the radius $R$ to the particle number $N$ and
to the charge $q$ (and also to the total charge $Q$) \cite{Jetzer:1990wr}.
Using the variables (\ref{dot-re}) and (\ref{dot-re0}), the
expressions (\ref{G-ia-ta}--\ref{do-s-lio}) become
\begin{eqnarray}
% \nonumber to remove numbering (before each equation)
M &=&\frac{1}{2}\int^{\infty}_{0} r^{2} \left(e^{-\nu}  C^{2}
\phi^{2}+ e^{-\lambda}\phi'^{2} + \phi^{2}
+\frac{1}{2}\frac{e^{-(\lambda +\nu )} C'^{2}}{q^{2}}\right)dr,
\\
N&=&\int^{\infty}_{0}dr r^{2}C e^{\frac{(\lambda -\nu
)}{2}}\phi^{2},\\
Q&=&q\int^{\infty}_{0}dr r^{2}C e^{\frac{(\lambda -\nu )}{2}}\phi^{2},\\
R&=& \frac{1}{N} \int^{\infty}_{0} drr^{3} C e^{\frac{(\lambda -\nu
)}{2}}\phi^{2}.
\end{eqnarray}
Note that $M$ is measured in units of $M_{\mbox{\tiny{Pl}}}^{2}/m$, the
particle number $N$ in units of $M_{\mbox{\tiny{Pl}}}^{2}/m^2$, the
charge $q$ in units of $\sqrt{8\pi}m/M_{\mbox{\tiny{Pl}}}$ and $R$
and $Q=q N$ in units of $1/m$  and $\sqrt{8\pi}M_{\mbox{\tiny{Pl}}}/
m$, respectively\cite{Jetzer:1989av}. 

We notice that in these units the critical boson charge defined above becomes $q^2_{crit}=1/2$ or $|q_{crit}|=1/\sqrt{2}\approx 0.707$. Thus, the construction of configurations with boson charge $q^2\simeq 1/2$
 will be particularly interesting.
%%%%%%%%%%%%%%%%%%%%%%%%%%%%%%%%%%%%%%%%%%%%%%%%%%%%%%%%%%%%%%%%%%%%
%%%%%%%%%%%%%%%%%%%%%%%%%%%%%%%%%%%%%%%%%%%%%%%%%%%%%%%%%%%%%%%%%%%%
\section{Numerical integration}\label{Sec:CBSNI}
We carried out a numerical integration of
Eqs.\il(\ref{tre}--\ref{li}) for different values of the radial function $\phi(r)$
at the origin and for different values of the boson charge. The results
are summarized in Figs.~\ref{Vi-ce-te}--
\ref{Ag-08-ven} and in Tables~\ref{tab:nove-t}--
\ref{tab:circ-max}. We pay special attention to the study of zero-nodes solutions.

We found, in particular,  that
bounded  configurations of self-gravitating charged bosons
exist with particle charge $q\leq q_{\ti{crit}}$, and for  values $q>q_{\ti{crit}}$
{localized} solutions are possible only for low values of the central density, {that is  for $\phi(0)<0.3$}.
{For instance, for
$q=0.8 $ we found  {localized} zero-nodes solutions only at $\phi(0)=0.1$.}
On the other hand, for $q>q_{\ti{crit}}$ and higher central densities, the
boundary conditions at the origin are not satisfied any more and only
bounded configurations with one or more nodes, i.e. excited states, could be possible (see also \cite{Jetzer:1990wr}).

In Sec.\il\ref{som-like}, we analyze the features of the metric functions $(e^{\nu}, e^{\lambda})$ and the Klein--Gordon field $\phi$. In Sec.\il\ref{s-on-ua} we focus on the charge and mass, total particle number and radius of the bounded configuration.
Since we have integrated the system
(\ref{Pu-c-ni1}--\ref{JaJLev80}) using the Eqs.\il
(\ref{dot-re}--\ref{dot-re0}), i.e.,
\begin{equation}\label{Cor-le}
\bar{\omega}=m\omega \, \quad C(r)=q A + \omega \ ,
\end{equation}
to obtain Eqs.\il(\ref{tre}--\ref{li}), we may use the asymptotic
assumption  $A(\infty)=0$ for the potential so that
\begin{equation}\label{Pas-ra-le}
C(\infty)=\omega\ .
\end{equation}
Different values of $\omega$ are listed in
Table\il\ref{tab:Mont-di}.
\begin{table}[h!]
\begin{tabular}{lcccr}
%\begin{tabularx}{\textheight}{|WWWWW|}
\hline
$\phi(0)$&$\omega$&$\omega$&$\omega$&$\omega$\\
&q=0.5&q=0.65&q=0.7&$q=1/\sqrt{2}$
\\
\hline
0.1&1.03433&1.05912&1.07885&1.07464\\
0.2&1.10489&1.24457&1.43809 &1.38764 \\
0.3&1.17031&1.44232&1.96159&2.05052\\
0.4&1.22759&1.61925& 2.15270&2.30157\\
0.5&1.27042&1.72349& 2.25687&2.38881\\
0.6&1.29767&1.78963& 2.26809&2.39819\\
0.7&1.30994&1.79980&2.26305&2.37248\\
0.8&1.30852&1.76409&2.16889&2.25671\\
0.9&1.29720&1.72279&2.08637 &2.12406\\
1&1.28046&1.67749&1.99464&2.08020\\
\hline
\end{tabular}
\caption[font={footnotesize,it}]{\footnotesize{\label{tab:Mont-di}Table
provides the eigenvalues $\omega$ for different values of the
central density $\phi(0)$ and for different values of the charge
$q$. The charge $q$ in measured in  units of
$\sqrt{8\pi}m/M_{\mbox{\tiny{Pl}}}$}}
\end{table}

We recall that the mass  $M$ is measured in units of $M_{\mbox{\tiny{Pl}}}^{2}/m$, the
particle number $N$ in units of $M_{\mbox{\tiny{Pl}}}^{2}/m^2$, the
charge $q$ in units of $\sqrt{8\pi}m/M_{\mbox{\tiny{Pl}}}$ and $R$
and $Q$ in units of $1/m$  and $\sqrt{8\pi}M_{\mbox{\tiny{Pl}}}/
m$, respectively.

%%%%%%%%%%%%%%%%%%%%%%%%%%%%%%%%%%%%%%%%%%%%%%%%%%%%%%%%%%%%%%%%%%%%%%%%%%%%%%%%%%%%%
%%%%%%%%%%%%%%%%%%%%%%%%%%%%%%%%%%%%%%%%%%%%%%%%%%%%%%%%%%%%%%%%%%%%%%%%%%%%%%%%%%%%
\subsection{Klein-Gordon field and metric functions}\label{som-like}
\begin{figure}[h!]
\centering
% \begin{minipage}[b]{1cm}
\includegraphics[width=0.5\hsize,clip]{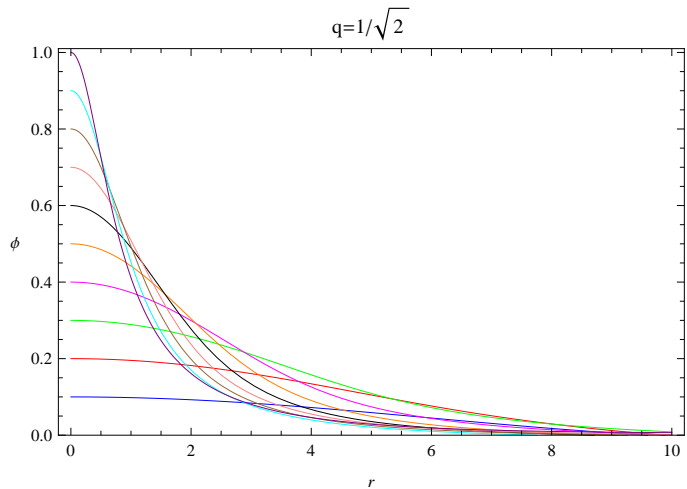}
%\end{minipage}
\caption[font={footnotesize,it}]{\footnotesize{(Color online) The radial function of the scalar field, for
a fixed value of the charge $q=q_{\ti{crit}}=1/\sqrt{2}$ in units of $\sqrt{8\pi}m/M_{\mbox{\tiny{Pl}}}$, is plotted as a function
of the radial coordinate $r$  for different values at the origin. The
radial function decreases monotonically as the dimensionless radius increases.
}} \label{Vi-ce-te}
\end{figure}

In Fig.\il\ref{Vi-ce-te}, the scalar field $\phi$, at
fixed value of the charge $q=q_{\mbox{\tiny{crit}}}$, is plotted as
a function of the radial coordinate $r$ and for different values at the
origin $\phi(0)$. The shape of the function does not change significantly for different values of the boson
charge, i.e,  the electromagnetic repulsion between particles has a weak
influence on the behavior of  $\phi$.
%\begin{center}
%\end{center}

\begin{figure}[h!]
\centering
%\begin{tabular}{cc}
% \begin{minipage}[b]{1cm}
\includegraphics[width=0.33\hsize,clip]{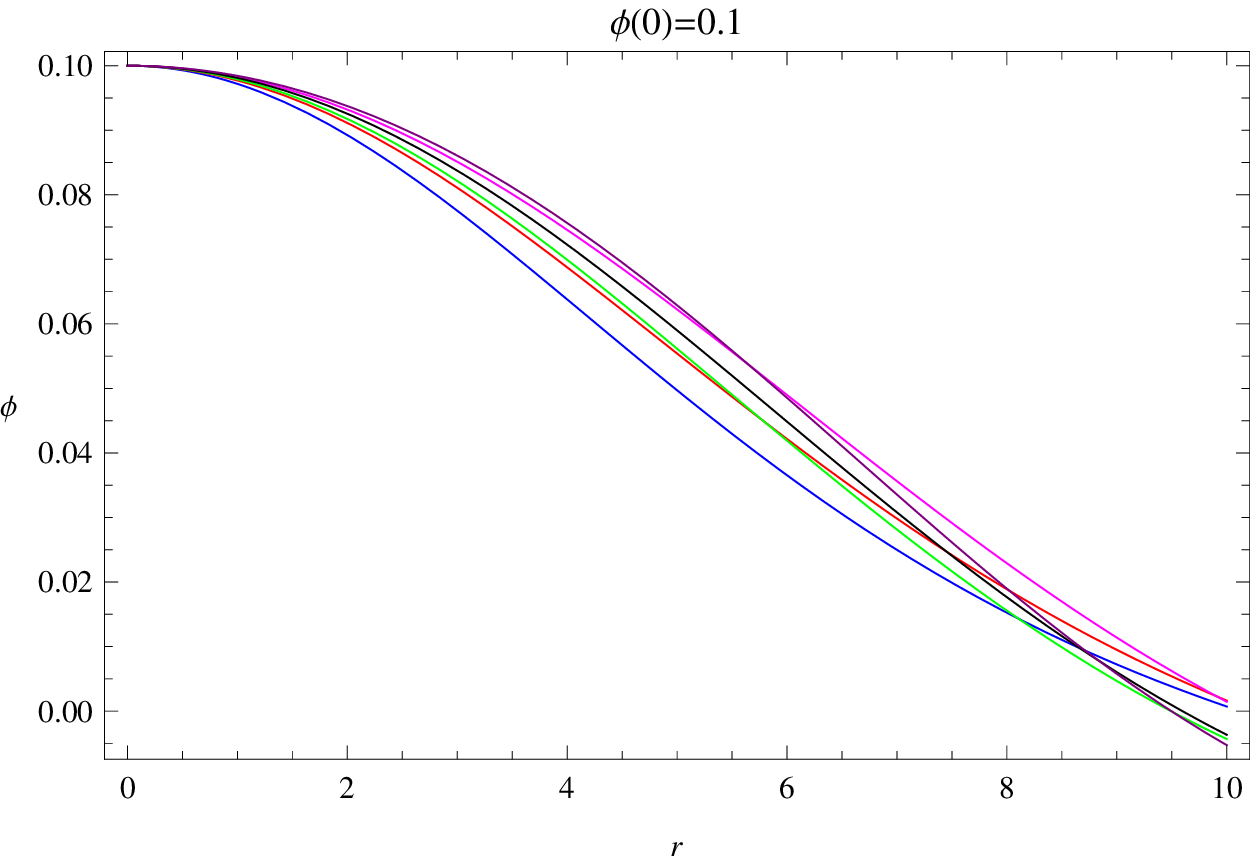}% "%" necessario
\includegraphics[width=0.33\hsize,clip]{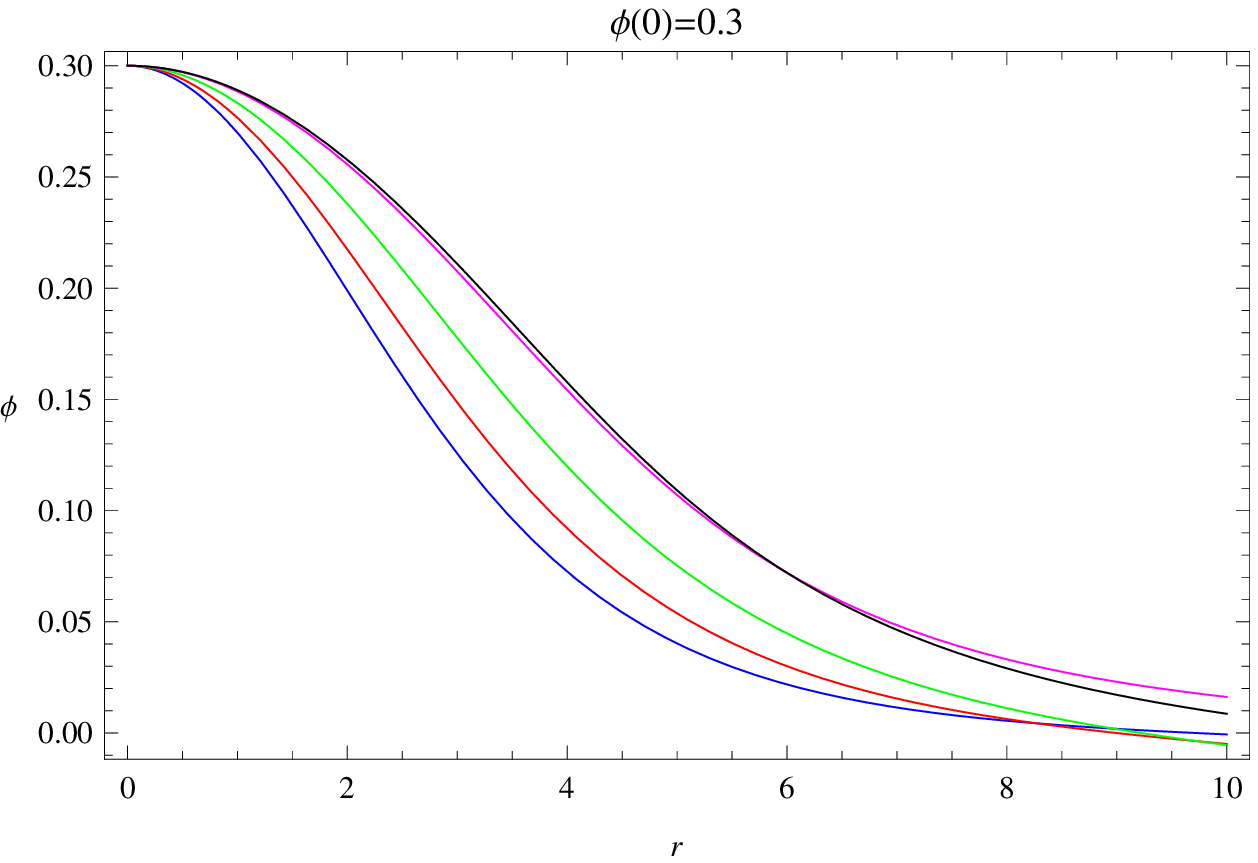}
\includegraphics[width=0.33\hsize,clip]{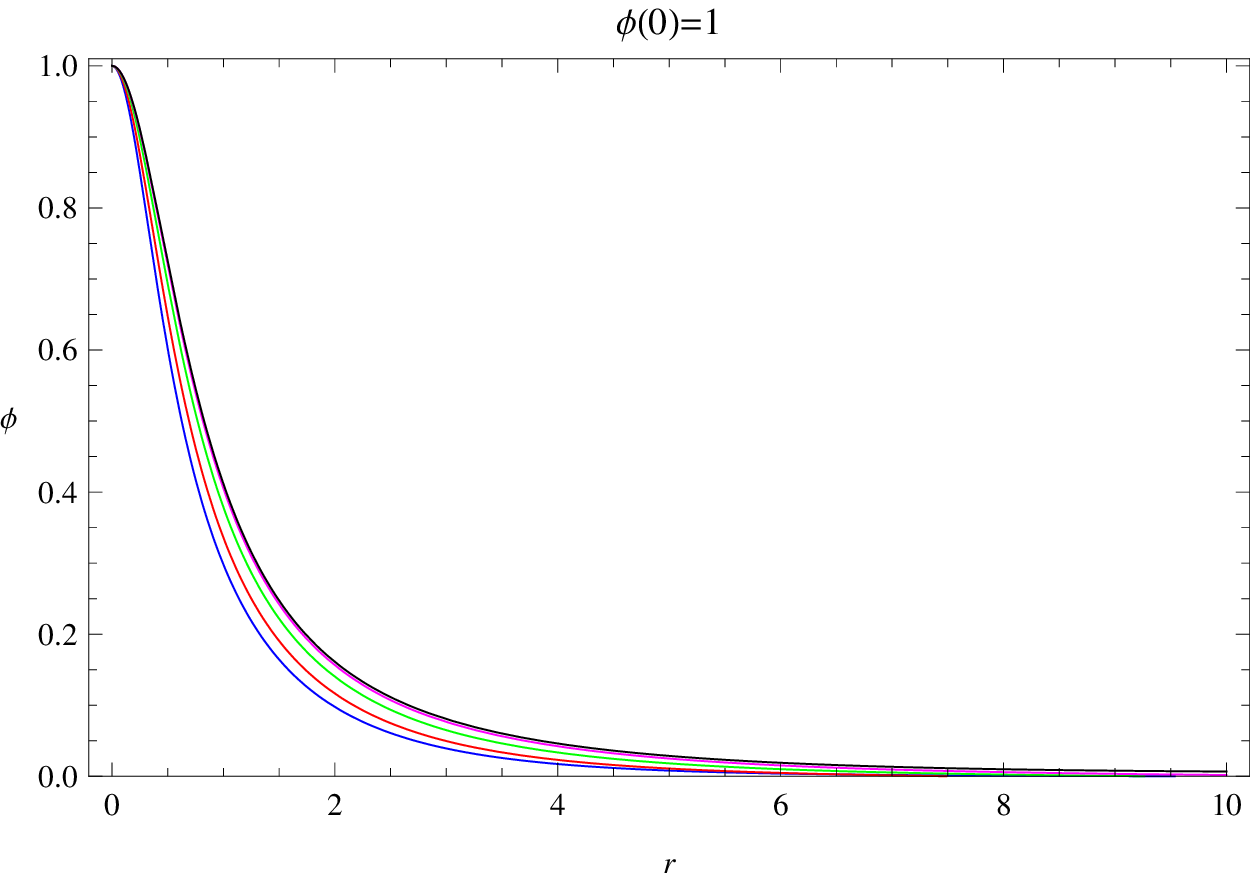}
%\end{minipage}
%\end{tabular}
\caption[font={footnotesize,it}]{\footnotesize{(Color online) The radial function
$\phi$ is plotted as a function of $r$ (dimensionless) for selected
values of the radial function  at the origin and different values of
the charge: $q=0$ (blue line),  $q=0.5$ (red line),  $q=0.65$ (green line),  $q=0.7$ (magenta line),  $q=1/\sqrt{2}$ (black line), in units of $\sqrt{8\pi}m/M_{\mbox{\tiny{Pl}}}$.
}} \label{ermi-1}
\end{figure}

In  Fig.\il\ref{ermi-1},
the radial function $\phi$ is plotted for different initial values at the origin and for
different values of the charge $q$.
As  expected $\phi$ decreases monotonically as the radius $r$ increases.
Moreover, we see that for a fixed value of $r$ and of the central density, an increase of the boson charge corresponds to larger values of $\phi$.

\begin{center}
\begin{figure}[h!]
\centering
%\begin{tabular}{cc}
% \begin{minipage}[b]{1cm}
\includegraphics[width=0.32\hsize,clip]{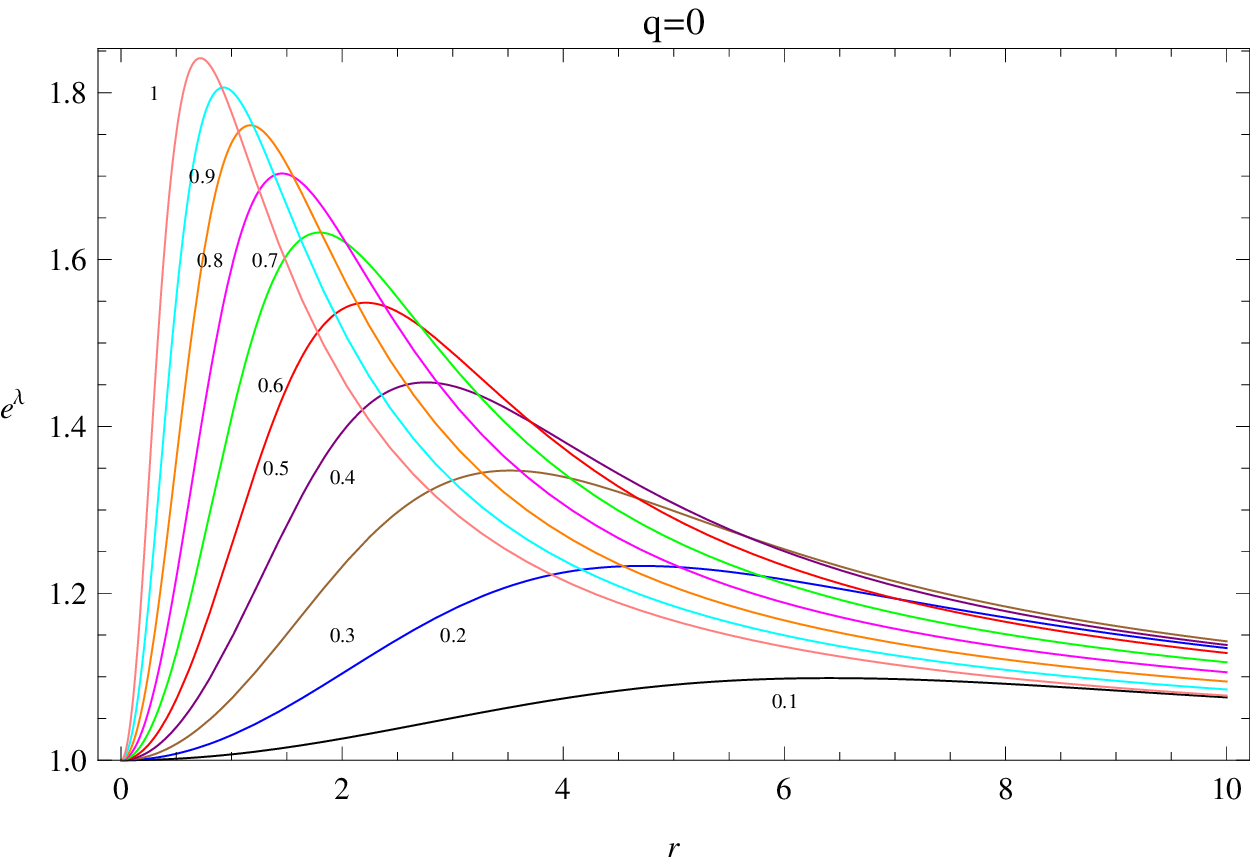}
\includegraphics[width=0.32\hsize,clip]{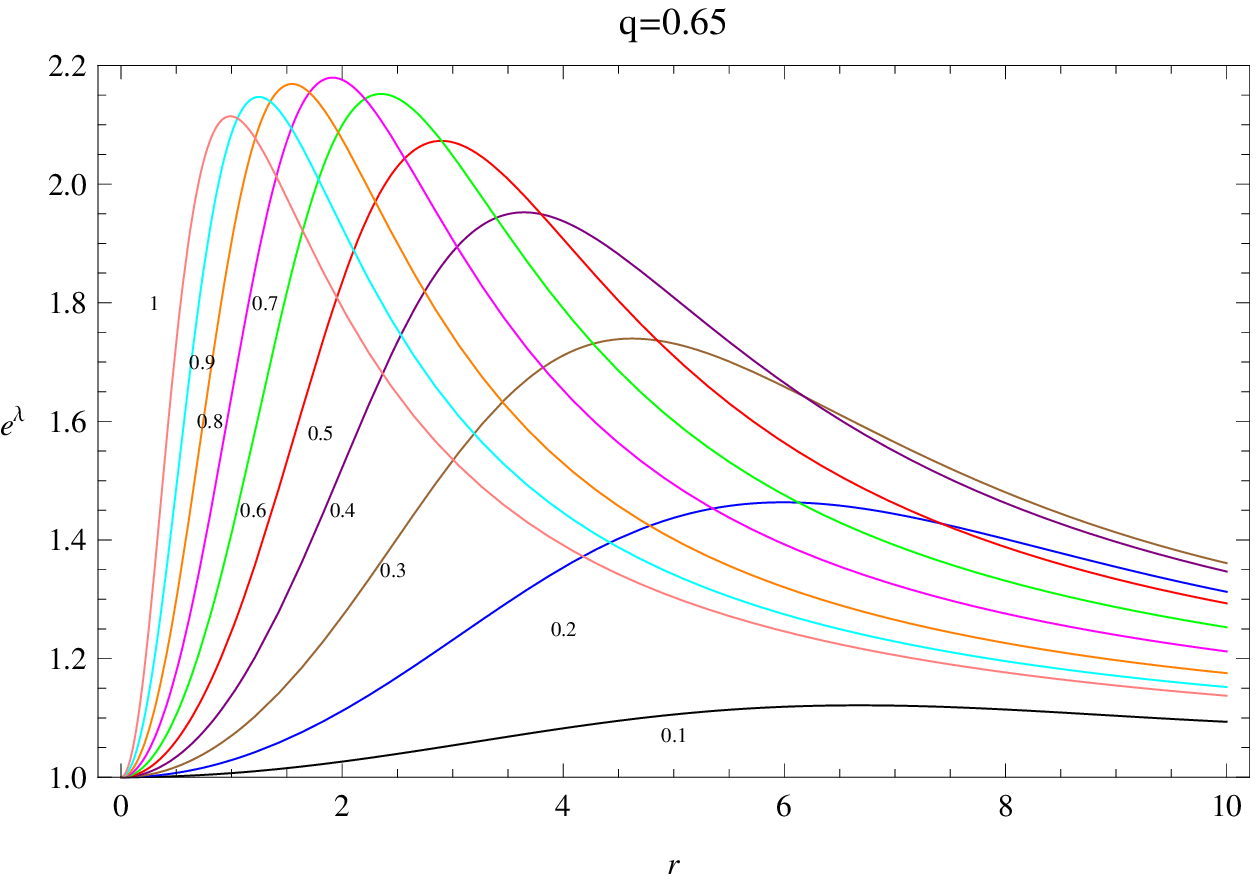}
\includegraphics[width=0.32\hsize,clip]{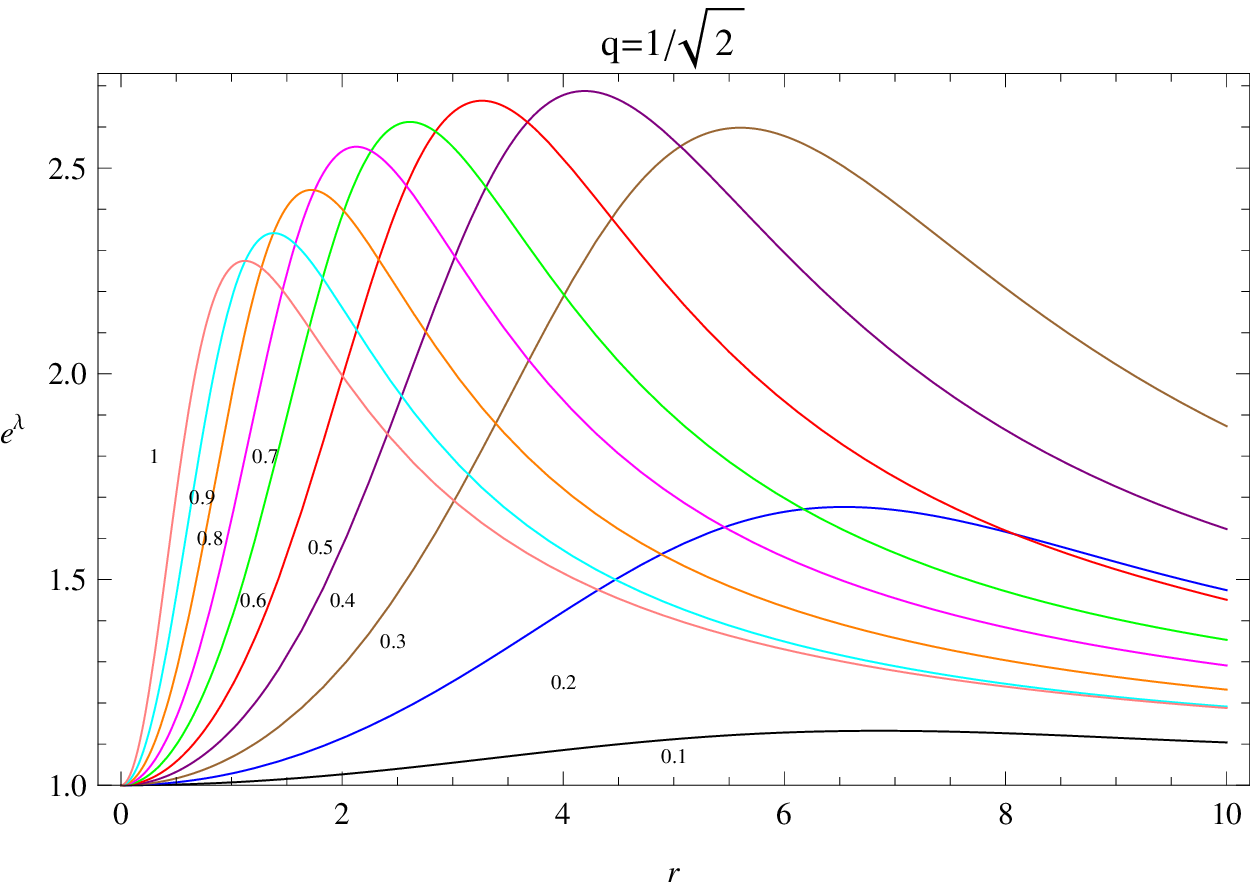}
%\end{minipage}
%\end{tabular}
\caption[font={footnotesize,it}]{\footnotesize{(Color online) The coefficient $e^{\lambda}$ of the metric is plotted as a function of $r$
(dimensionless) for different values of the radial function  at the
origin and for selected values of  the charge $q$(in units of
$\sqrt{8\pi}m/M_{\mbox{\tiny{Pl}}}$).
}}\label{tra-vi-ru}
\end{figure}
\end{center}

\begin{center}
\begin{figure}[h!]
\centering
%\begin{tabular}{cc}
%\begin{minipage}[b]{1cm}
\includegraphics[width=0.33\hsize,clip]{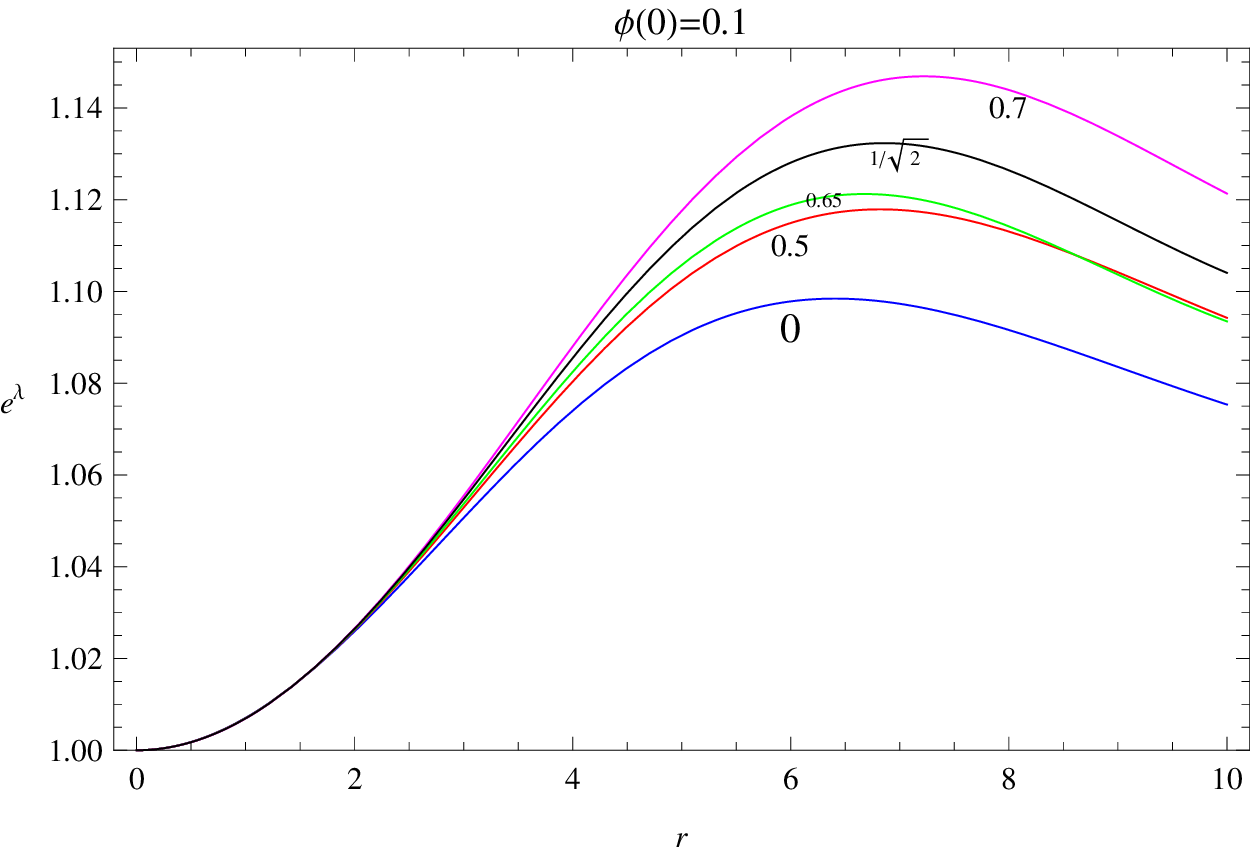}% "%" necessario
\includegraphics[width=0.33\hsize,clip]{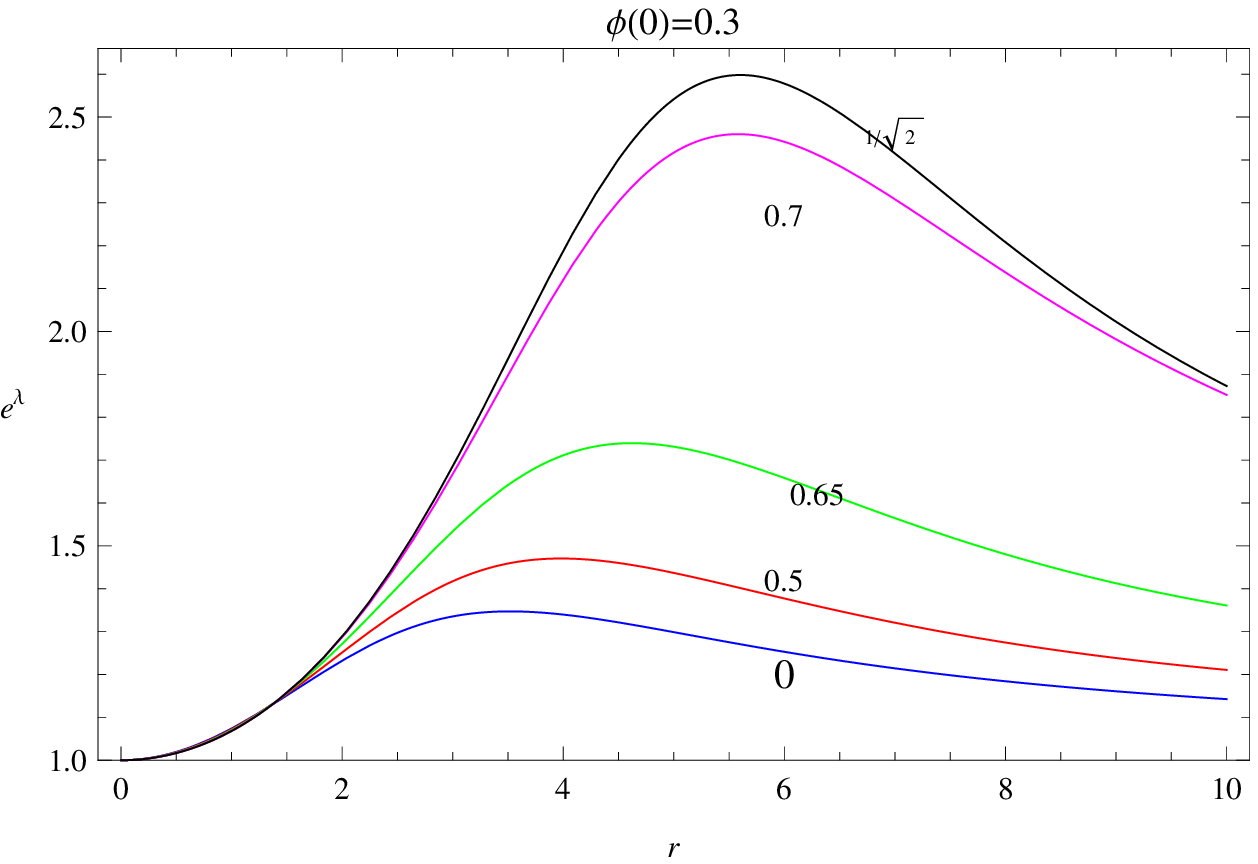}
\includegraphics[width=0.33\hsize,clip]{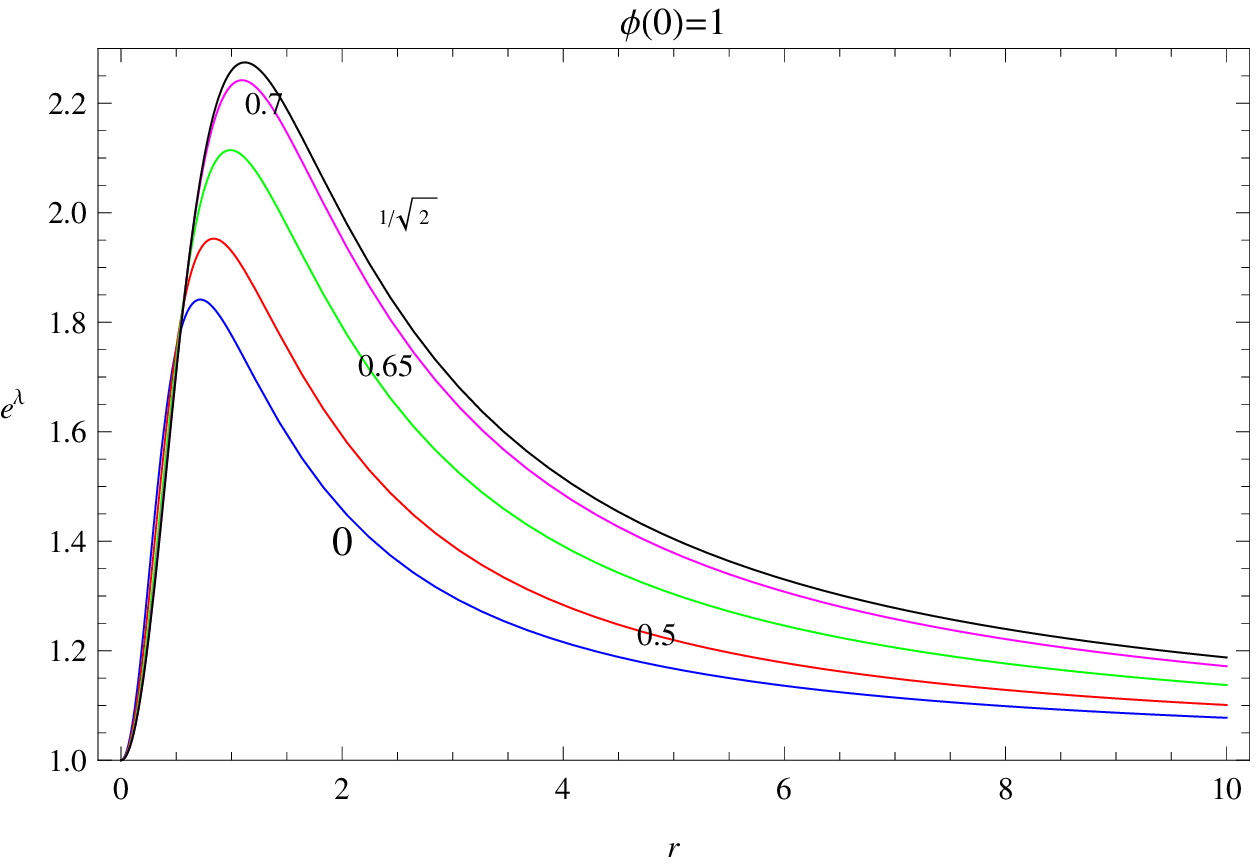}
%\end{minipage}
%\end{tabular}
\caption[font={footnotesize,it}]{\footnotesize{(Color online) The coefficient $e^{\lambda}=-g_{11}$
($e^{\lambda(r)}$) of the metric is plotted as a function of $r$
(dimensionless) for fixed values of the radial function  at the
origin and for different values of the charge $q$(in units of
$\sqrt{8\pi}m/M_{\mbox{\tiny{Pl}}}$).
}}. \label{al-par-m}
\end{figure}
\end{center}

In Figs.\il\ref{tra-vi-ru} and \ref{al-par-m} the metric function $e^{\lambda}=-g_{11}$
 is plotted as a function of the dimensionless radius $r$
 for different values of the radial function $\phi(r)$ at the
origin and for a selected values of  the charge $q$.

%in units of
%$M_{\mbox{\tiny{Pl}}}/\sqrt{8 \pi }m$.
In general, we observe that  $e^{\lambda}$ reaches its maximum value
at the value, say,  $r_{\mbox\tiny{Max}}$ of the radial
coordinate. Once the maximum is reached, the function decreases monotonically as $r$
increases and tends asymptotically to 1, in accordance with the imposed
asymptotic behavior. For a fixed value of the charge, the value of
$r_{\mbox\tiny{Max}}$ decreases as the central density increases.

Table\il\ref{tab:nove-t} provides  the maximum values of
$e^{\lambda(r)}$ and the corresponding radial coordinate
$r_{\mbox\tiny{Max}}$, for different values of the central density. In the  case of a neutral configuration \cite{RR}, $q=0$,
     the  boson star radius is defined as the value
$r_{\mbox\tiny{Max}}$  corresponding to the  maximum value of $e^{\lambda(r)}=-g_{11}$. Then,  the values of $r_{\mbox\tiny{Max}}$, listed in Table\il\ref{tab:nove-t}  can be assumed as good  estimates of the  radius of the corresponding
charged configurations.
\begin{table}[h!]
\begin{center}
%\resizebox{1.2\textwidth}{!}{%
\resizebox{.71\textwidth}{!}{%
\begin{tabular}{l|llllllllll}
%\cline{2-11}
\hline\hline
q&\multicolumn{10}{c}{$\phi(0)$}
\\
\cline{2-11}
&\multicolumn{2}{c}{0.1}&\multicolumn{2}{c}{0.2}&\multicolumn{2}{c}{0.3}
&\multicolumn{2}{c}{0.4}&\multicolumn{2}{c}{0.5}
\\
\hline\hline
& $e^{\lambda_{\ti{Max}}}$ &$r_{\ti{Max}}$ &
$e^{\lambda_{\ti{Max}}}$ & $r_{\ti{Max}}$ & $e^{\lambda_{\ti{Max}}}$ & $r_{\ti{Max}}$&$e^{\lambda_{\ti{Max}}}$
&$r_{\ti{Max}}$&$e^{\lambda_{\ti{Max}}}$&$r_{\ti{Max}}$\\
0   &1.0984&
6.4060&1.2328&4.7090&1.3471&3.5184&1.4528&2.7582&1.5482&
2.2144\\
0.5  &1.1179&
6.8172&1.3132&5.2940&1.4705&3.9800&1.6094&3.1282&1.7234&2.512\\
0.65  &  1.1212&6.6753 &1.4635&5.9912&1.7395&
4.6225&1.9524&3.6451&2.0729&2.8975\\
0.7&1.1469&7.2189 &1.8154&
7.0841&2.4602&5.5813&2.5159& 4.1125&2.5409& 3.2099\\
$1/\sqrt{2}$&1.1323& 6.8585& 1.6764& 6.5422& 2.5981& 5.6016&
2.6873& 4.1943& 2.6637& 3.2650\\
\hline\hline
&\multicolumn{10}{c}{$\phi(0)$}
\\
\cline{2-11}
&\multicolumn{2}{c}{0.6}&\multicolumn{2}{c}{0.7}&\multicolumn{2}{c}{0.8}
&\multicolumn{2}{c}{0.9}&\multicolumn{2}{c}{1}
\\
0   &1.6324&1.7964&1.7031&1.4561&1.7609&1.1707&
1.8064&0.92659&1.8414&0.7175\\
0.5  &1.8124&2.0394&1.8774&1.6600&1.9197&1.3430
&1.9432&1.0722&1.9526&0.8383\\
0.65  &2.1520&2.3519&2.1795&1.9132&2.1688
&1.5471&2.1470&1.2464&2.1144&0.9895\\
0.7&2.5116&
2.5649&2.4798&2.0919&2.3972&1.6925&2.3222
& 1.3706&2.2420&1.0939\\
$1/\sqrt{2}$& 2.6123& 2.6133& 2.5519&
2.1275& 2.4469&1.7199& 2.3420 &1.3806& 2.2745 &1.1198\\
\hline \hline
%\end{center}
%\end{center}
\end{tabular}}
\caption[font={footnotesize,it}]{\footnotesize{The maximum values of
$e^{\lambda(r)}=-g_{11}$ and the corresponding radial
coordinate $r_{\mbox\tiny{Max}}$ for different values of the
central density. For a fixed $\phi(0)$, an increase of the boson
charge $q$ generates an increase of
the maximum value of $e^{\lambda}$ and of the value of
the radius $r_{\mbox\tiny{Max}}$.
}}\label{tab:nove-t}
\end{center}
\end{table}
Moreover, in Figs.\il\ref{al-par-m}, the coefficient
$e^{\lambda}$ of the metric is plotted as a function of the radial coordinate $r$ for
fixed values of the radial function  at the origin and different
values of the charge $q$. For a fixed $\phi(0)$, an
increase in the maximum value of $g_{11}$ and of the value of
$r_{\mbox\tiny{Max}}$, corresponds to an increase of the boson
charge $q$. At a fixed value of $r$, the value of the coefficient $-g_{11}$ increases with an
increase of the central density, reaching the maximum value at $\phi_{\mbox{\tiny{Max}}}(0)\simeq 0.3$.

%%%%%%%%%%%%%%%%%%%%%%%%%%%%%%%%%%%%%%%%%%%%%%%%%%%%%%%%%%%%%%%%%%%%%%%%%%
%%%%%%%%%%%%%%%%%%%%%%%%%%%%%%%%%%%%%%%%%%%%%%%%%%%%%%%%%%%%%%%%%%%%%%%%%

\subsection{Mass, charge, radius and particle number}\label{s-on-ua}
\begin{center}
\begin{figure}[h!]
\centering
%\begin{tabular}{ccc}
\includegraphics[width=0.32\hsize,clip]{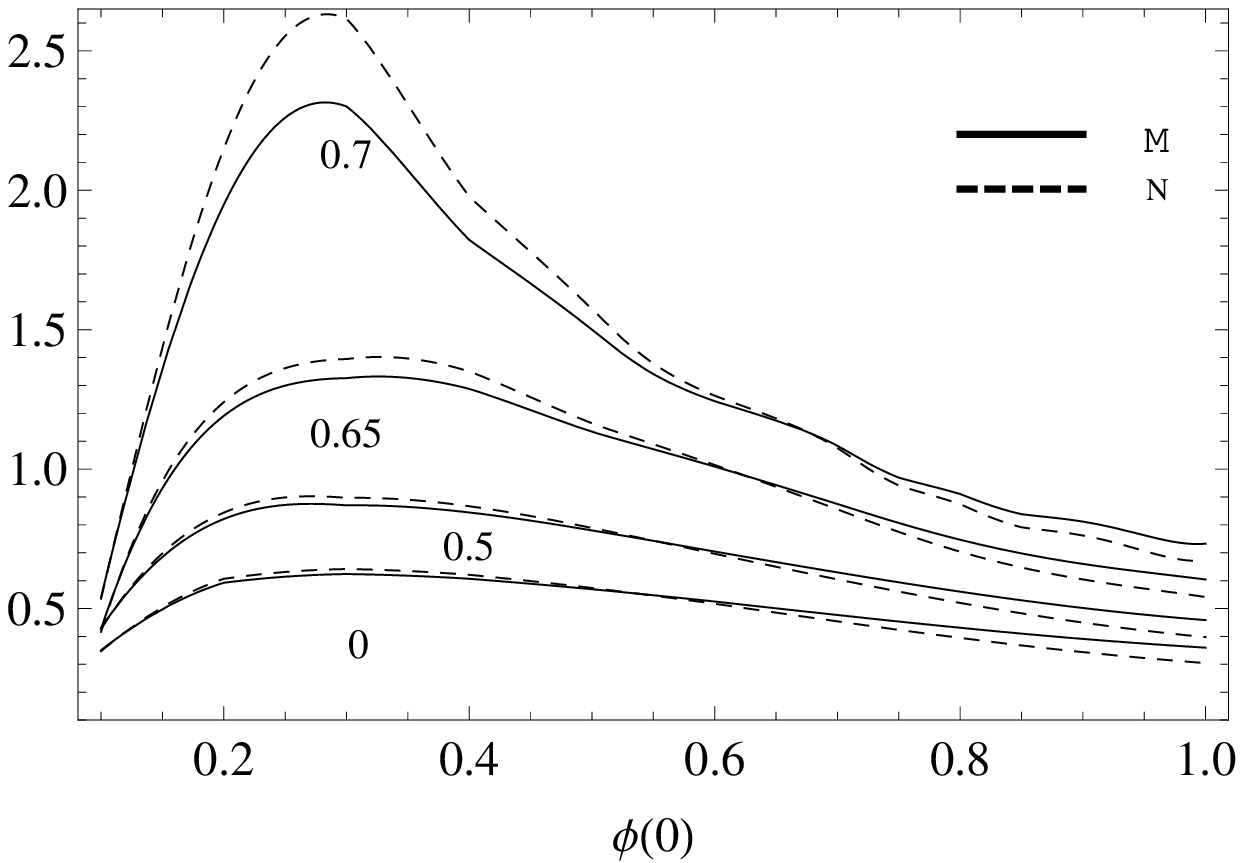}
\includegraphics[width=0.32\hsize,clip]{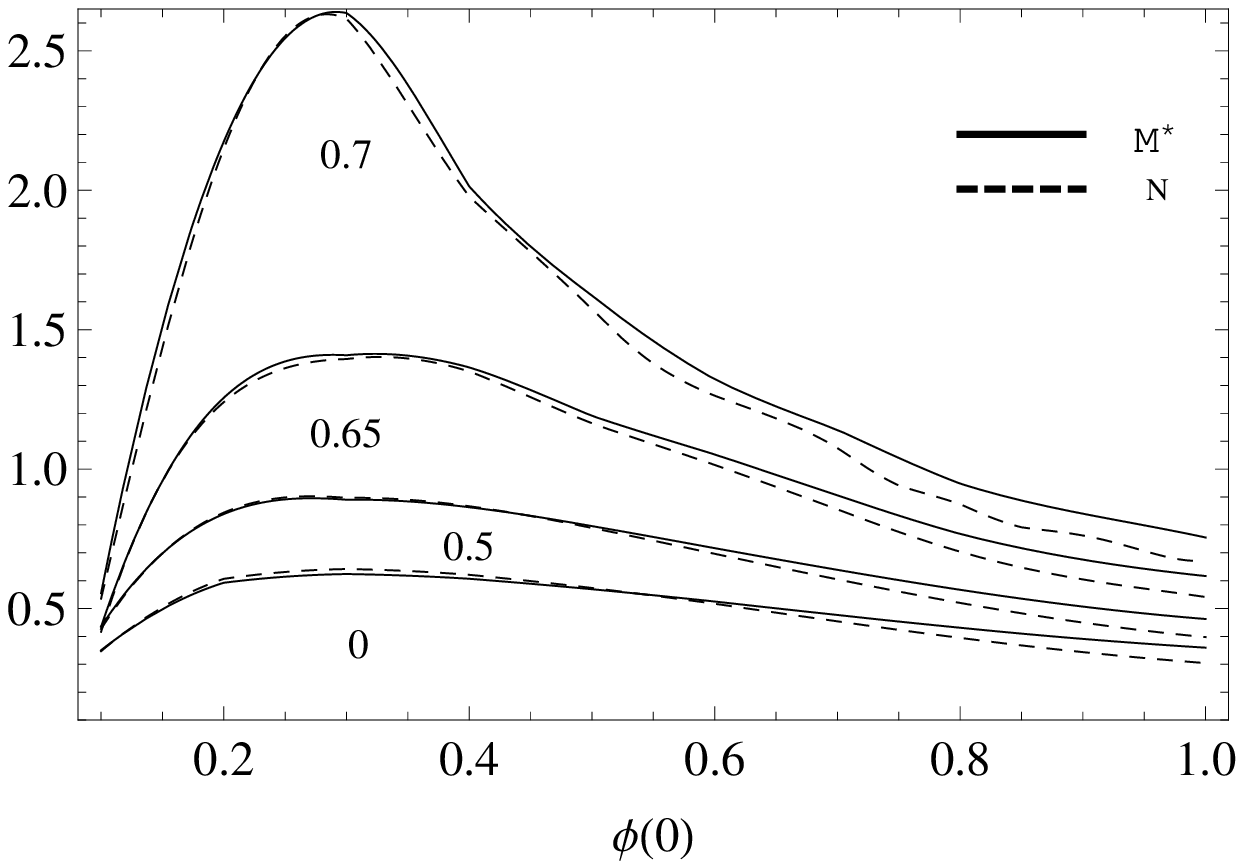}
\includegraphics[width=0.32\hsize,clip]{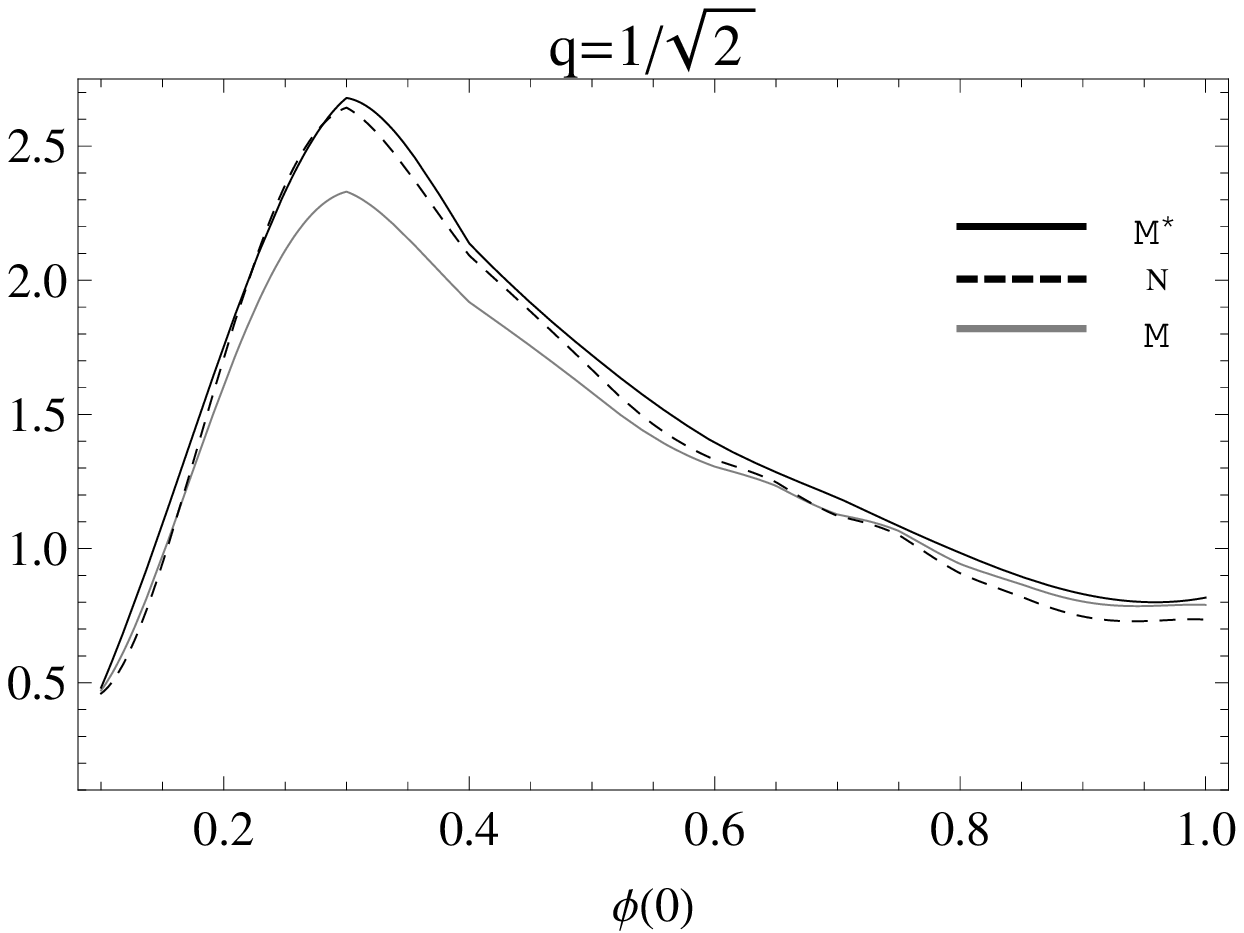}
%\end{tabular}
\caption[font={footnotesize,it}]{\footnotesize{The charged boson star
mass $M$ and  mass $M^*$ in units of $M_{\mbox{\tiny{Pl}}}^{2}/m$ (solid line), and
particle number $N$ in units of $M_{\mbox{\tiny{Pl}}}^{2}/m^2$
(dashed line) are plotted as functions of the central density
$\phi(0)$ for different values of the charge $q$ (in units of
$M_{\mbox{\tiny{Pl}}}/\sqrt{8 \pi }m$).
{We note that in particular at the critical density $\phi(0)\approx0.3$  for $q=1/\sqrt{2}$ it is  $M^*>N$.} }} \label{er-ge}
\end{figure}
\end{center}

The masses $M$ and $M^*$ of the system, in units of $M_{\mbox{\tiny{Pl}}}^{2}/m$, and the
particle number $N$, in units of $M_{\mbox{\tiny{Pl}}}^{2}/m^2$, are
plotted in the Fig.~\ref{er-ge} as functions of the central
density $\phi(0)$, for different values of the boson charge $q$.
%%%
%

\begin{figure}[h!]
\centering
\begin{tabular}{cc}
% \begin{minipage}[b]{1cm}Plotcaricamassaraggio205
\includegraphics[width=0.33\hsize,clip]{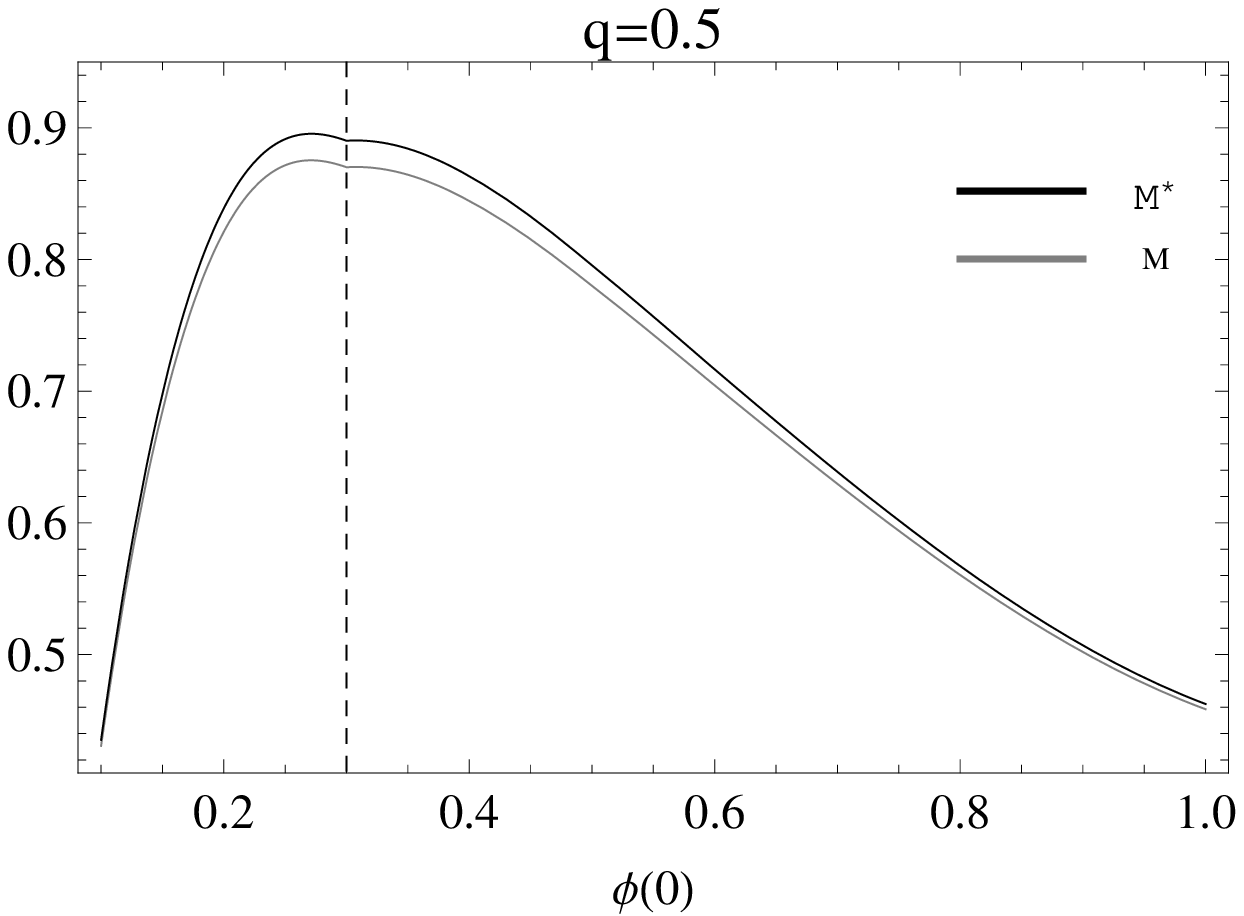}
\includegraphics[width=0.33\hsize,clip]{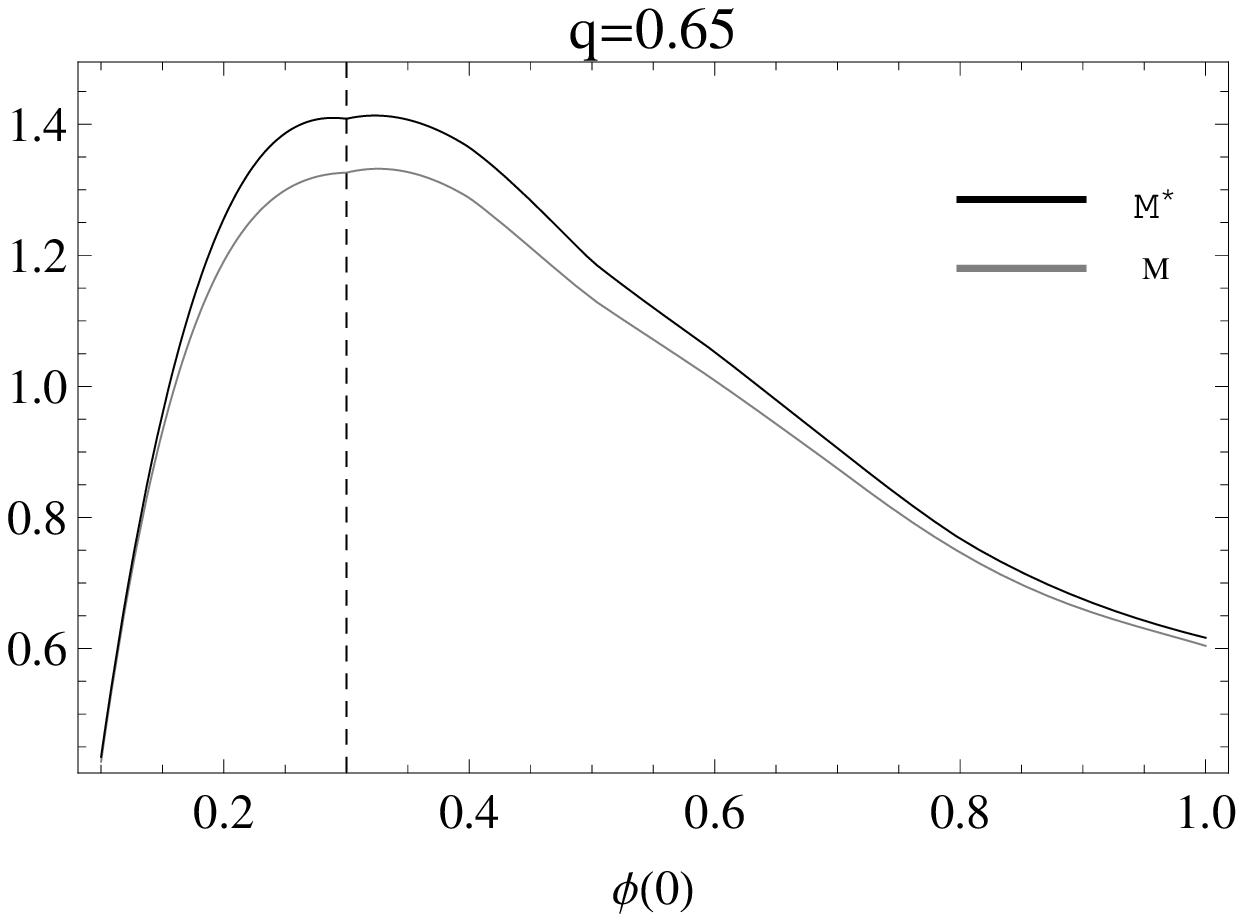}
\includegraphics[width=0.33\hsize,clip]{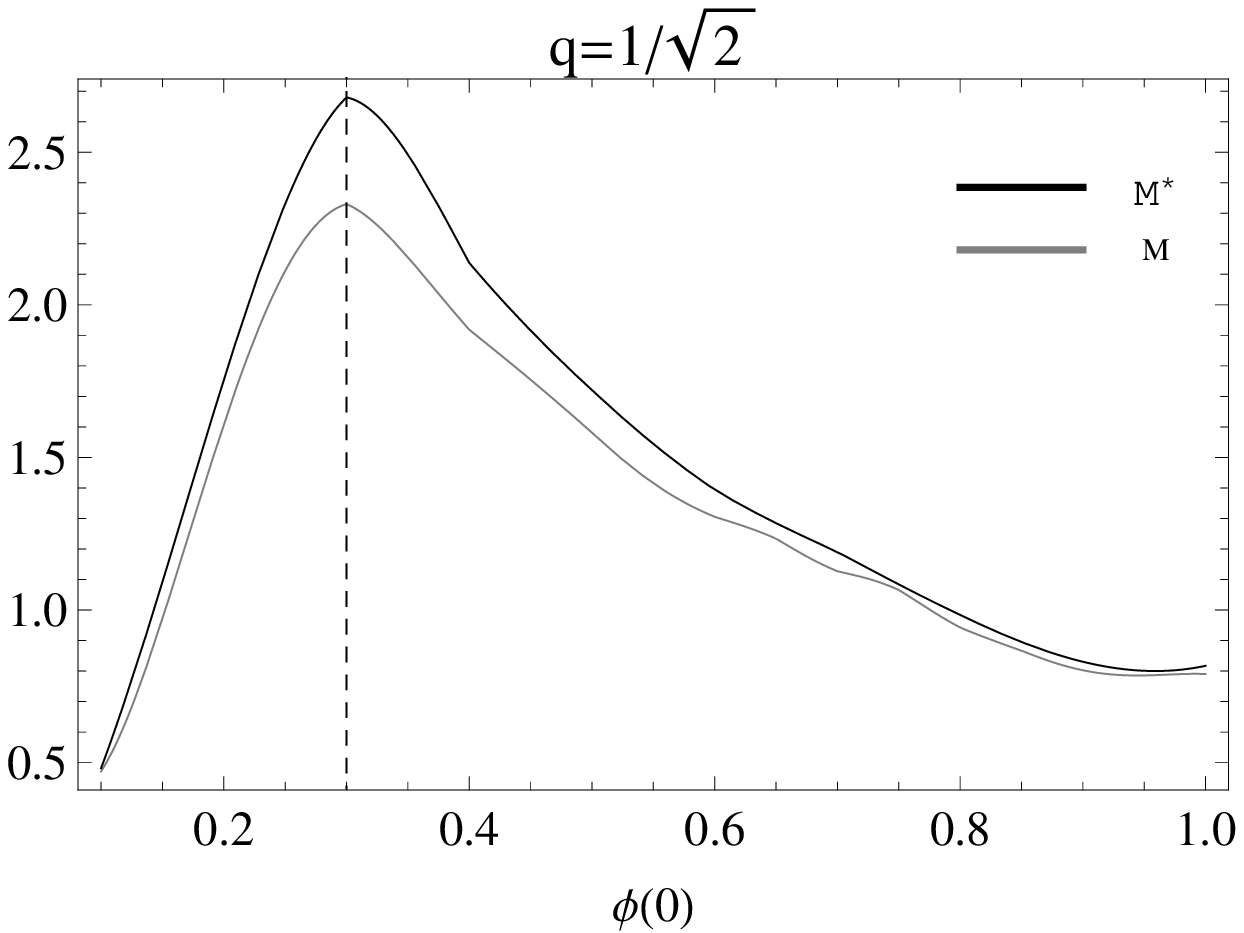}
%\end{minipage}
\end{tabular}
\caption[font={footnotesize,it}]{\footnotesize{The mass $M$ (in units of $\uM$) and the mass $M^*$ are
plotted as functions  of the central density $\phi(0)$ and for
different values of the boson charge $q$ (in units of
$\sqrt{8\pi}m/M_{\mbox{\tiny{Pl}}}$).
}}\label{Ang-9}
\end{figure}

In Fig.~\ref{Ang-9}, the masses $M$ and $M^*$ are plotted as functions of the scalar central density for selected values of the boson charge; we have indicated the difference $\Delta =M^*-M$ at a certain $\phi(0)$. This quantity clearly increases with the boson change $q$; as expected.
Analogously to the case of white dwarf and neutron stars, a critical mass $M_{\mbox{\tiny{Max}}}$
and correspondingly  a critical number $N_{\mbox{\tiny{Max}}}$  exist for a central density
   $\phi(0)_{\mbox{\tiny{Max}}}\simeq 0.3$,  independently of the value of $q$.
   Configurations with $\phi(0)>\phi(0)_{\mbox{\tiny{Max}}}$
are gravitationally unstable, see e.g. \cite{Jetzer:1989av,Jetzer:1989us,Jetzer:1990wr, Jetzer:1989fx,Jetzer:1989qp,Jetzer:1989vs,Jetzer:1988af,Jetzer:1988vr}.
In Table\il\ref{tab:max-and-min}, the maximum values of the charged boson star
mass $M_{\mbox{\tiny{Max}}}$,  and  the number of particles
$N_{\mbox{\tiny{Max}}}$, and $\phi_{\mbox{\tiny{Max}}}$  are listed for
selected values of $q$.
\begin{table}[h!]
\begin{center}
\resizebox{0.78\textwidth}{!}{%
\begin{tabular}{lcccccccccr}
\hline\hline
$q$ &$M_{\mbox{\tiny{Max}}}$&$\phi(0)_{\mbox{\tiny{(Max,M)}}}$ &$M^*_{\mbox{\tiny{Max}}}$&$\phi(0)_{\mbox{\tiny{(Max,$M^* $)}}}$ &
$N_{\mbox{\tiny{Max}}}$&$\phi(0)_{\mbox{\tiny{(Max,N)}}}$&
$Q_{\mbox{\tiny{Max}}}$&$\phi(0)_{\mbox{\tiny{(Max,Q)}}}$ &
$R_{\mbox{\tiny{Max}}} $&$\phi(0)_{\mbox{\tiny{(Max,R)}}}$
\\ \hline\hline
0&0.62374&0.3&0.62374&0.3&0.641665&0.3&$\checkmark$&$\checkmark$&4.56589*&0.1*
\\
0.50&0.87536&0.271041&0.895504&0.271444&0.902576& 0.271361&0.448485&0.3&4.79634*& 0.1*
\\
0.65&1.33207&0.325797&1.4133&0.32305&1.402170&0.326816&0.90818&0.318766
&4.63946&0.091957
\\
0.70&2.31504&0.282575&2.63956&0.291404&2.63120&0.284047&1.84184& 0.284047&4.74968*&0.1*
\\
$1/\sqrt{2}$&2.33016&0.3&2.67951&0.3& 2.64329& 0.3& 1.86909& 0.3& $5.14269$&
$0.158228$
\\
\hline\hline
%\end{tabularx}
\end{tabular}
}
\end{center}
\caption[font={footnotesize,it}]{\footnotesize{The maximum values of the
boson star mass $M_{\mbox{\tiny{Max}}}$ and $M^*_{\mbox{\tiny{Max}}}$ (in units of $M_{\mbox{\tiny{Pl}}}^{2}/m$) and of the number of particles
$N_{\mbox{\tiny{Max}}}$ (in units of $M_{\mbox{\tiny{Pl}}}^{2}/m^2$), the  charge $Q_{\mbox{\tiny{Max}}}$ (in units of
$\sqrt{8\pi} M_{\mbox{\tiny{Pl}}}/m$) as functions of
$\phi(0)$, and the corresponding central density values $\phi(0)_{\mbox{\tiny{(Max,M)}}}$  for the mass $M_{\mbox{\tiny{Max}}}$, $\phi(0)_{\mbox{\tiny{(Max,N)}}}$ for  particle number $N_{\mbox{\tiny{Max}}}$, $\phi(0)_{\mbox{\tiny{(Max,Q)}}}$ for  the total charge $Q_{\mbox{\tiny{Max}}}$ are listed   for different $q$.
The entries with a star $(*)$ do not correspond to maximum values but to the initial points of the numerical integration.
(See Fig.\il\ref{far-ros-o})
}}
\label{tab:max-and-min}
\end{table}
Comparing  the plots at different charge values we can see that
the presence  of  charge does not change the behavior qualitatively.
However, to an increase of the boson charge
values corresponds an increase of
$M_{\mbox{\tiny{Max}}}$, $M^*_{\mbox{\tiny{Max}}}$  and $N_{\mbox{\tiny{Max}}}$, and an increase
of  the difference $(N_{\mbox{\tiny{Max}}}-M_{\mbox{\tiny{Max}}})$
between the maximum number of particles and the
mass at a fixed central density.
The critical central density value is
$\phi(0)_{\mbox{\tiny{Max}}}\simeq 0.3$. This value seems to be independent of the charge
values $q$ (see also \cite{Jetzer:1990xa}).

\begin{figure}[h!]
\centering
\begin{tabular}{cc}
% \begin{minipage}[b]{1cm}
\includegraphics[width=0.5\hsize,clip]{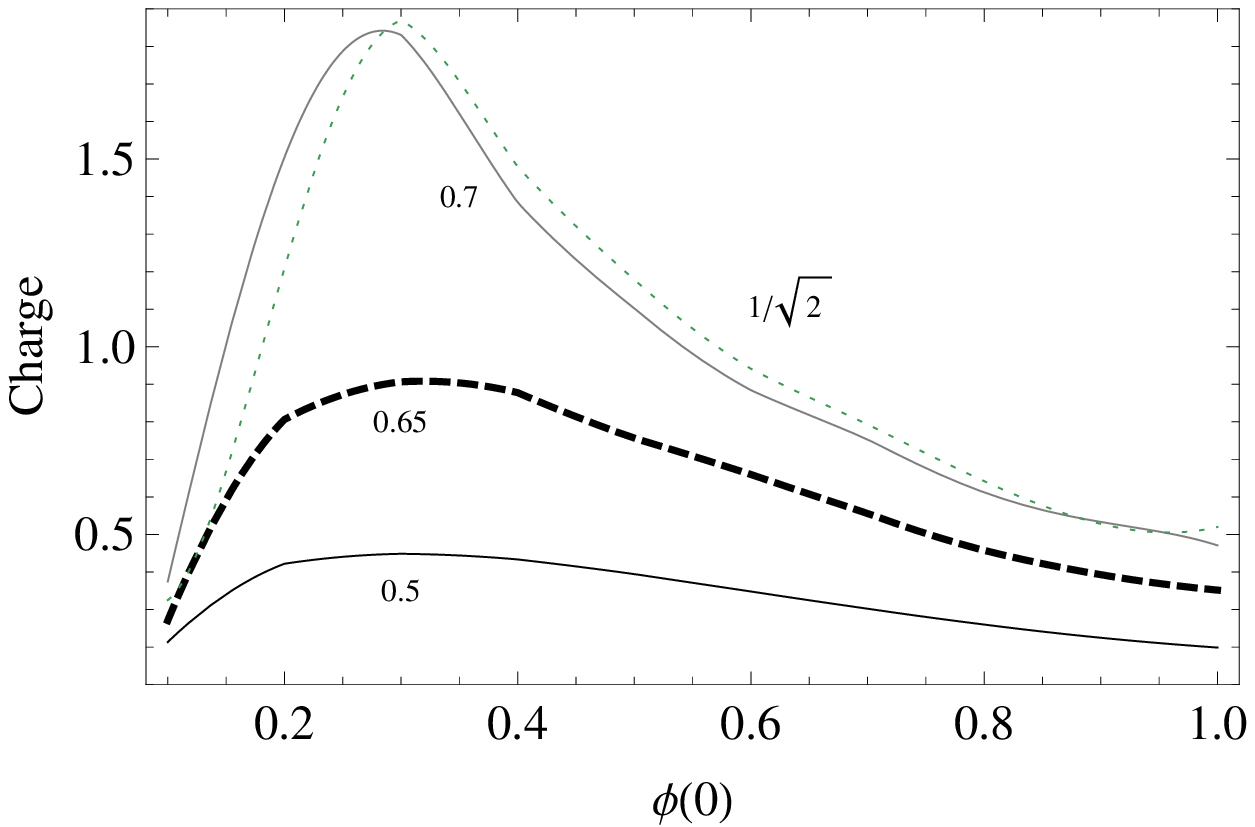}
\includegraphics[width=0.5\hsize,clip]{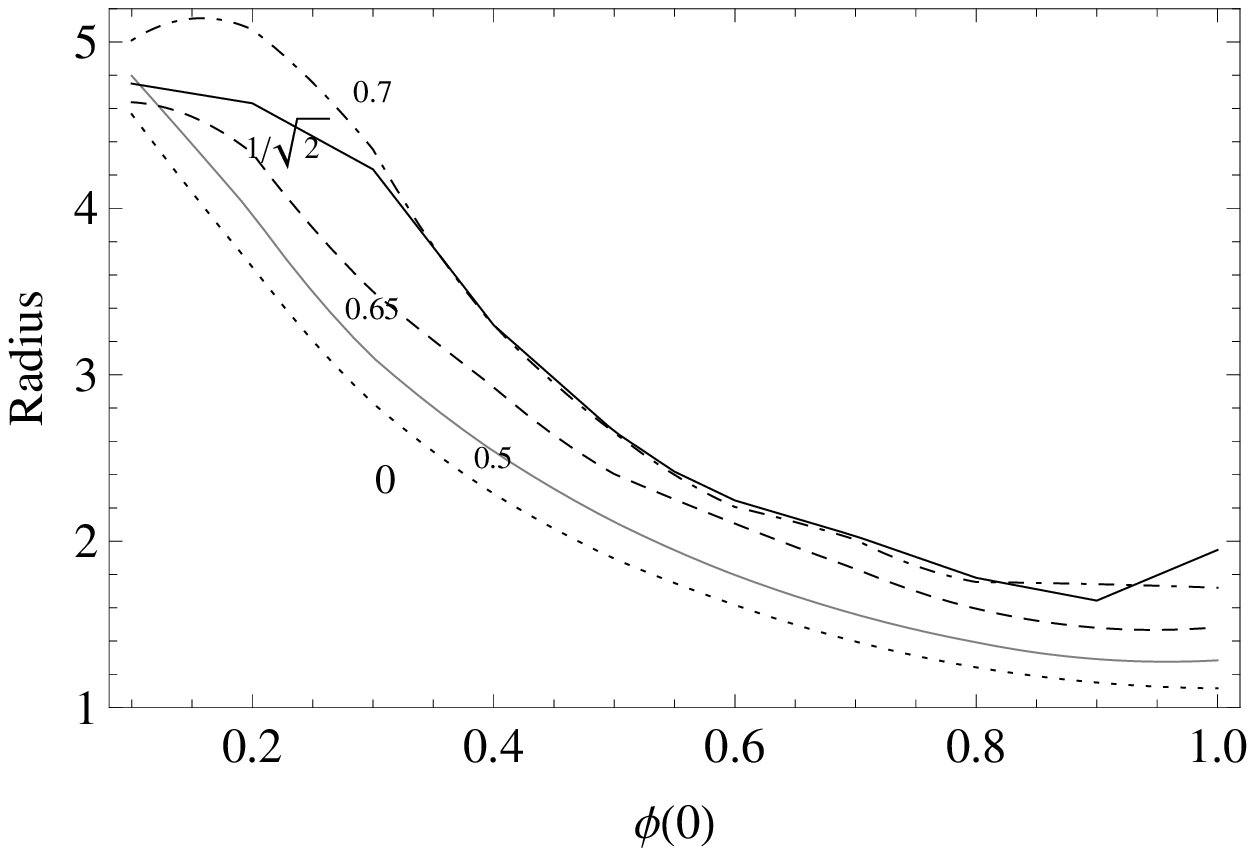}
%\\(a)&(b)\\%\end{minipage}
\end{tabular}
\caption[font={footnotesize,it}]{\footnotesize{This graphic
shows the total charge $Q$ (in units of
$\sqrt{8\pi}M_{\mbox{\tiny{Pl}}}/ m$) and the radius $R$ (in
units of $1/m$) as  functions of the
central density $\phi(0)$, for different values of the boson charge
$q$ (in units $\sqrt{8\pi}m/M_{\mbox{\tiny{Pl}}}$).
}
}\label{far-ros-o}
\end{figure}

\begin{figure}[h!]
\centering
\begin{tabular}{cc}
% \begin{minipage}[b]{1cm}Plotcaricamassaraggio205
\includegraphics[width=0.5\hsize,clip]{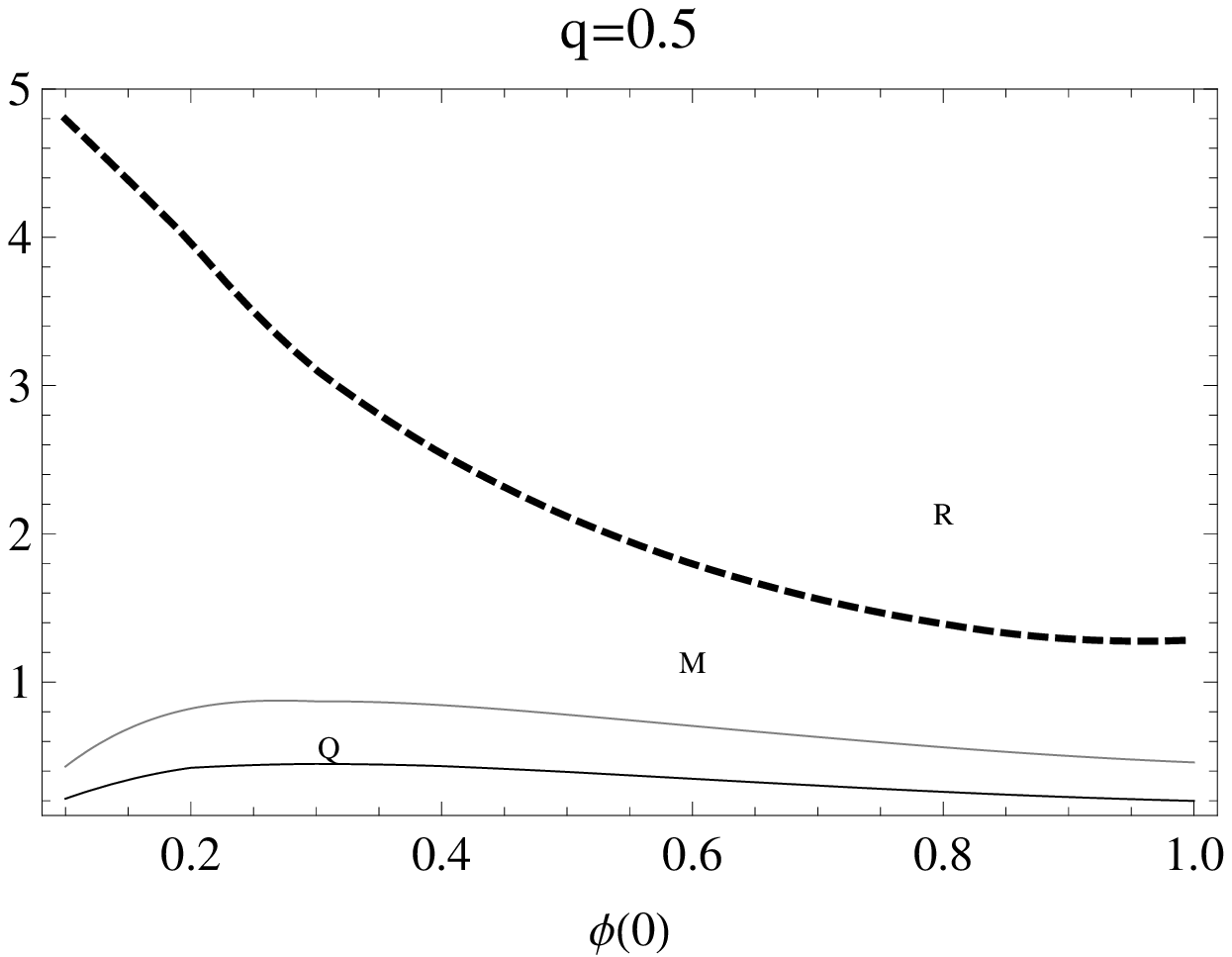}
\includegraphics[width=0.5\hsize,clip]{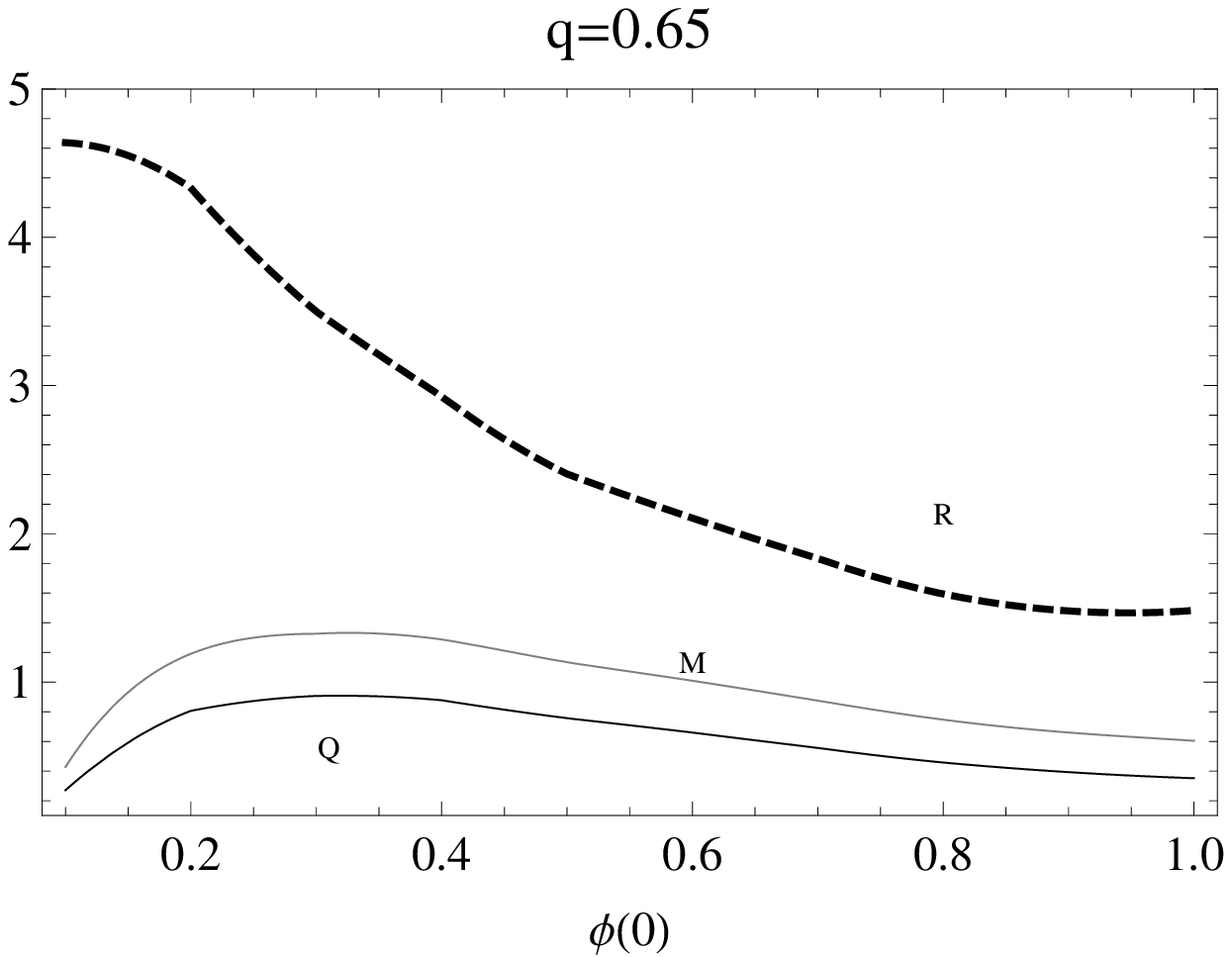}\\
\includegraphics[width=0.5\hsize,clip]{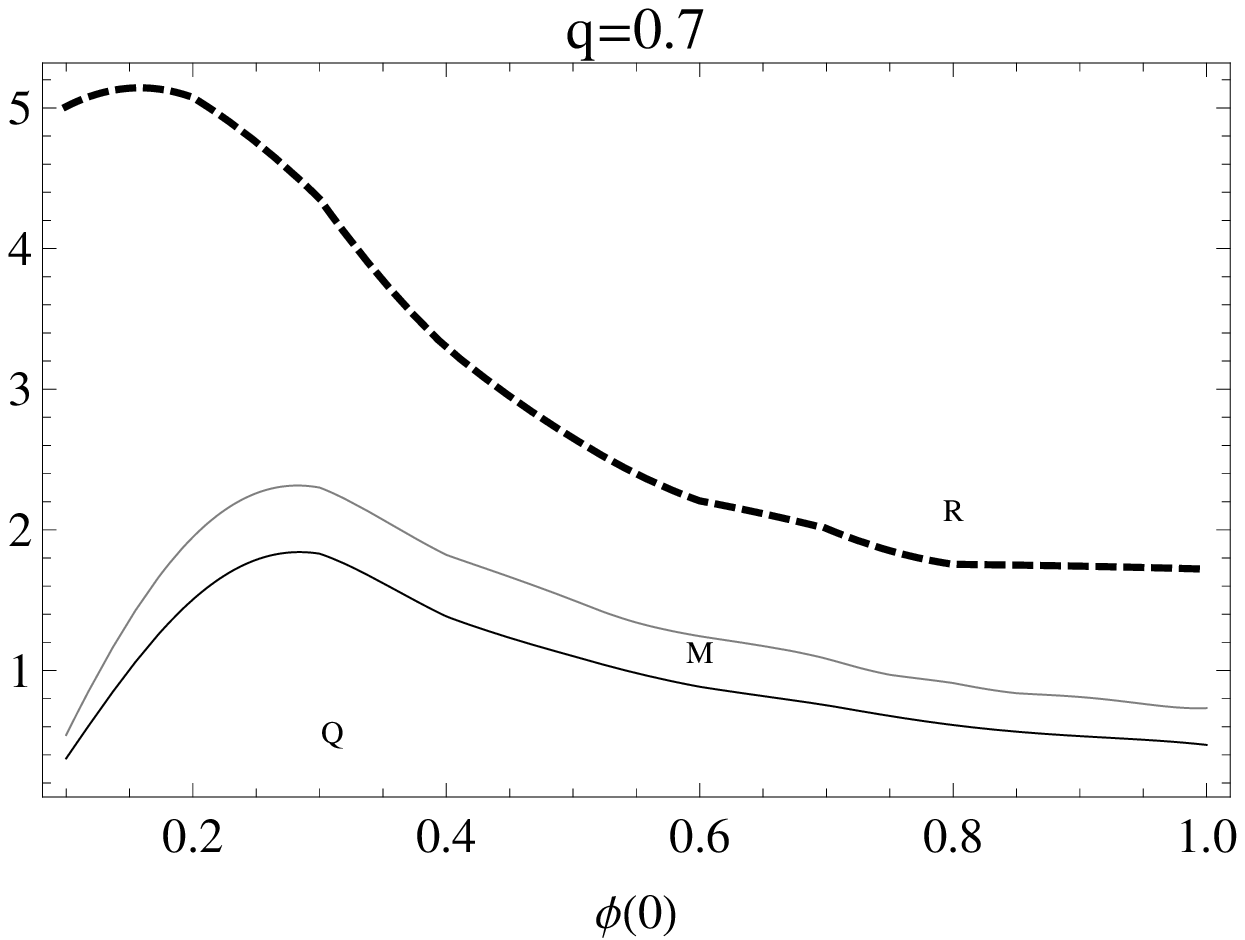}
\includegraphics[width=0.5\hsize,clip]{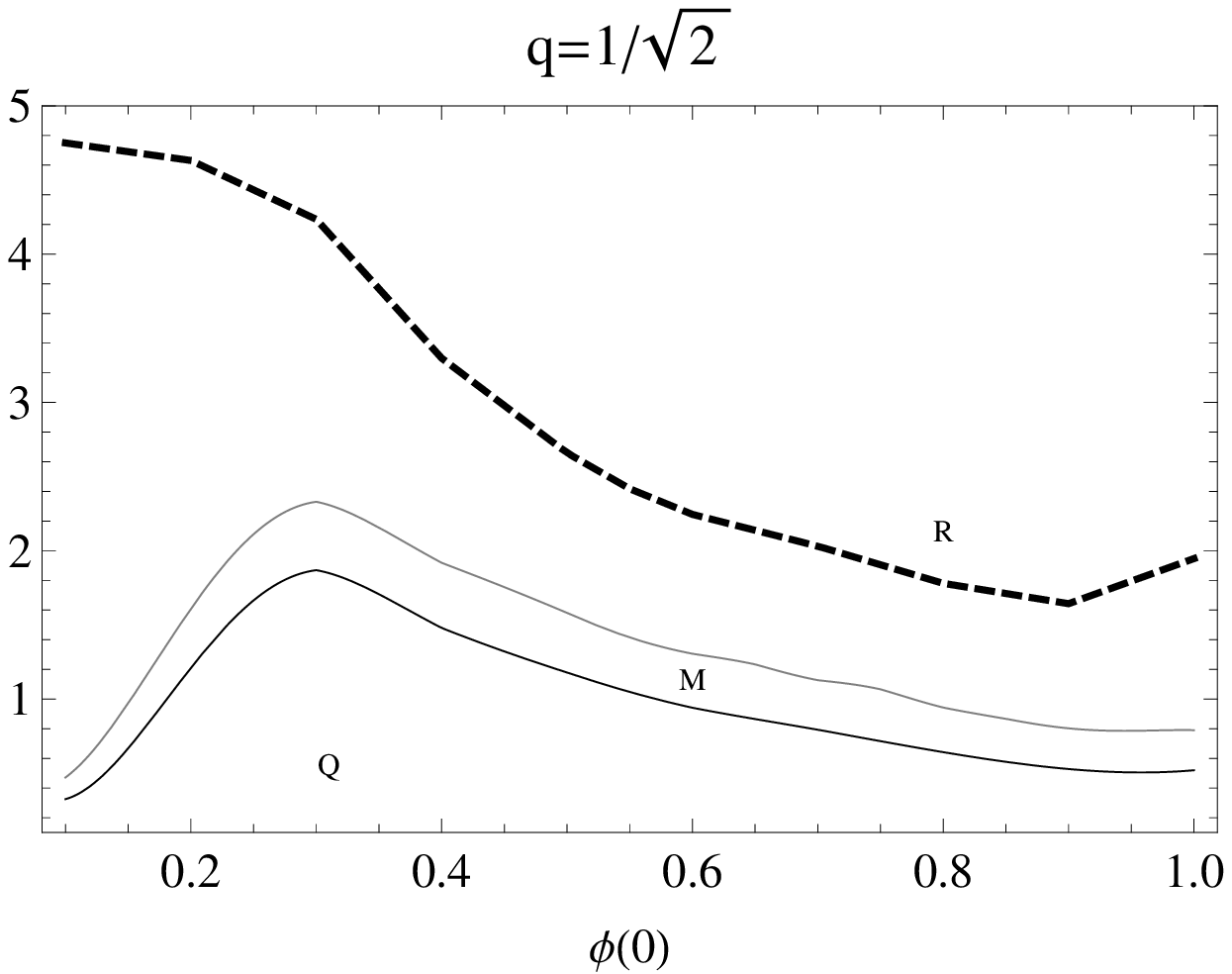}
%\end{minipage}
\end{tabular}
\caption[font={footnotesize,it}]{\footnotesize{The total charge $Q$ (in units of
$\sqrt{8\pi}M_{\mbox{\tiny{Pl}}}/ m$), the radius $R$ (in
units of $1/m$), and the mass $M$ (in units of $\uM$) are
plotted as functions  of the central density $\phi(0)$ and for
different values of the boson charge $q$ (in units of
$\sqrt{8\pi}m/M_{\mbox{\tiny{Pl}}}$).
}}\label{An-g-lo}
\end{figure}

The radius $R$, the total charge $Q$, and the mass $M$ are plotted (in
units of $1/m$, $\sqrt{8\pi} M_{\mbox{\tiny{Pl}}}/m$, and $\uM$,
respectively) in Figs.\il\ref{far-ros-o}--\ref{An-g-lo} as
functions of the central density $\phi(0)$, for different values of
the charge $q$.
We see that the radius, for a fixed
central density, increases as the charge increases (see
Fig.\il\ref{far-ros-o} and Table\il\ref{tab:max-and-min}).

In  Table\il\ref{tab:max-and-min} the maximum values of the total charge
$Q_{\mbox{\tiny{Max}}}$,  for $\phi_{\mbox{\tiny{Max}}}(0)\simeq 0.3$  and for
different $q$ are listed.
For  fixed values of the charge $q$, the total
charge increases with an increase of the central density until it
reaches a maximum value for some  density
$\phi(0)_{\mbox{\tiny{Max}}}$. Then, the value of $Q$ decreases monotonically
as $\phi(0)$ increases. In this way, it is possible to introduce the concept of a
maximum charge $Q_{\mbox{\tiny{Max}}} $ for charged boson stars.

In Fig.\il\ref{far-ros-o}, the charge $Q$  is plotted as
a function of the central density $\phi(0)$ for different
values of the charge $q$.
For  fixed values of the charge $q$, the total
charge increases with an increase of the central density until it
reaches a maximum value for some  density
$\phi(0)_{\mbox{\tiny{Max}}}$. Then, the value of $Q$ decreases monotonically
as $\phi(0)$ increases. In this way, it is possible to introduce the concept of a
critical charge $Q_{\mbox{\tiny{Max}}} $ for charged boson stars.
In  Table\il\ref{tab:max-and-min} the maximum values of the total charge
$Q_{\mbox{\tiny{Max}}}$,  for $\phi_{\mbox{\tiny{Max}}}(0)\simeq 0.3$  and for
different values of $q$ are listed. Let us note that for a fixed central density, to an increase of the
boson charge $q$  corresponds an increase of the maximum
$Q_{\mbox{\tiny{Max}}}$ (see Fig.\il\ref{far-ros-o} and Table\il\ref{tab:max-and-min}).

Fig.\il\ref{An-g-lo} depicts  the total charge $Q$
(in units of $\uQ$), the
radius $R$ (in units of $1/m$), and the mass
$M$ (in units of $\uM$),  as functions  of the central
density $\phi(0)$ and for different values of the boson charge $q$
(in $\uq$).

Note that, for a fixed value of the charge $q$, the mass, the
radius and the charge  are always  positive and to an increase
(decrease) of the total charge there always corresponds an increase
(decrease) of the total mass (and total  particle number).
Both quantities increase as the central density  increases and they reach
a maximum value for the same density
$\phi(0)_{\mbox{\tiny{Max}}}\simeq 0.3$ (see also
Table\il\ref{tab:max-and-min}). Once the maximum is reached, both quantities decrease monotonically
as $\phi(0)$ increases.

\begin{figure}[h]
\centering
\begin{tabular}{cc}
% \begin{minipage}[b]{1cm}
\includegraphics[width=0.5\hsize,clip]{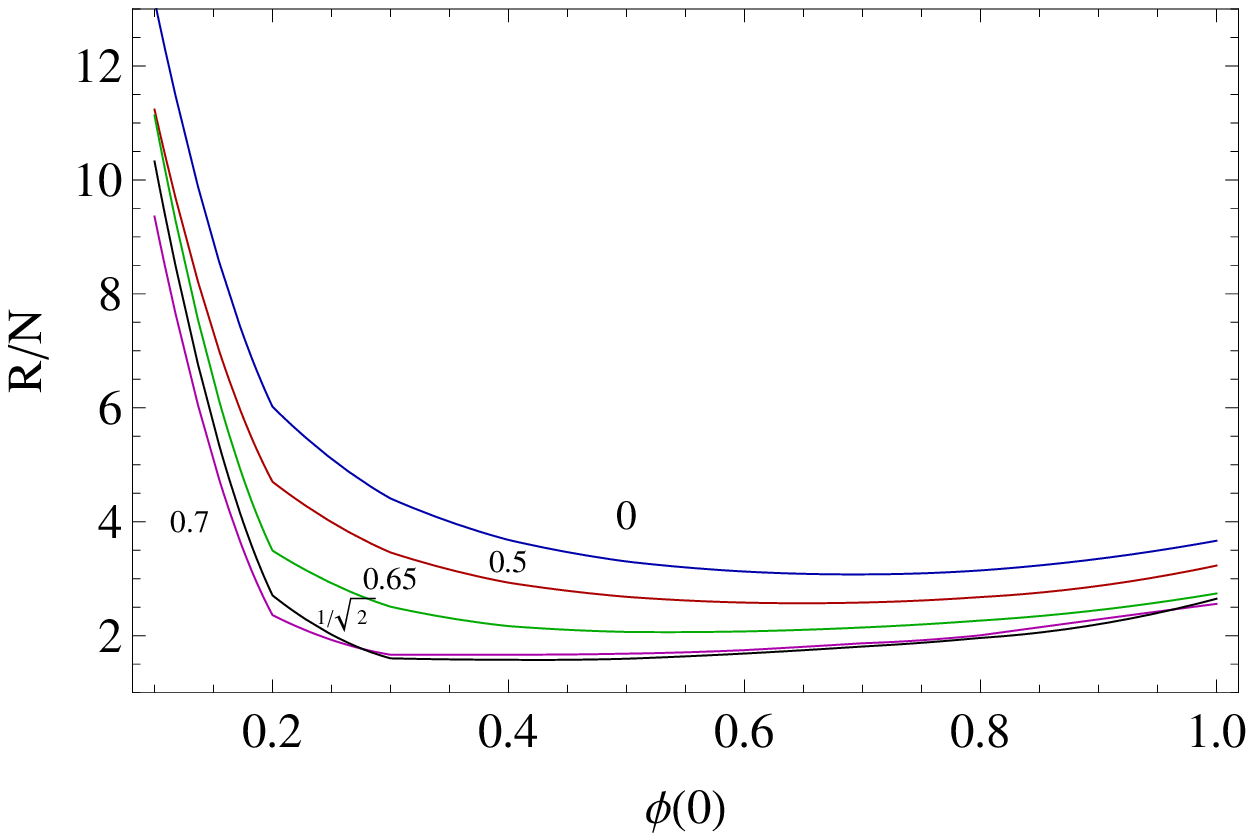}
\includegraphics[width=0.5\hsize,clip]{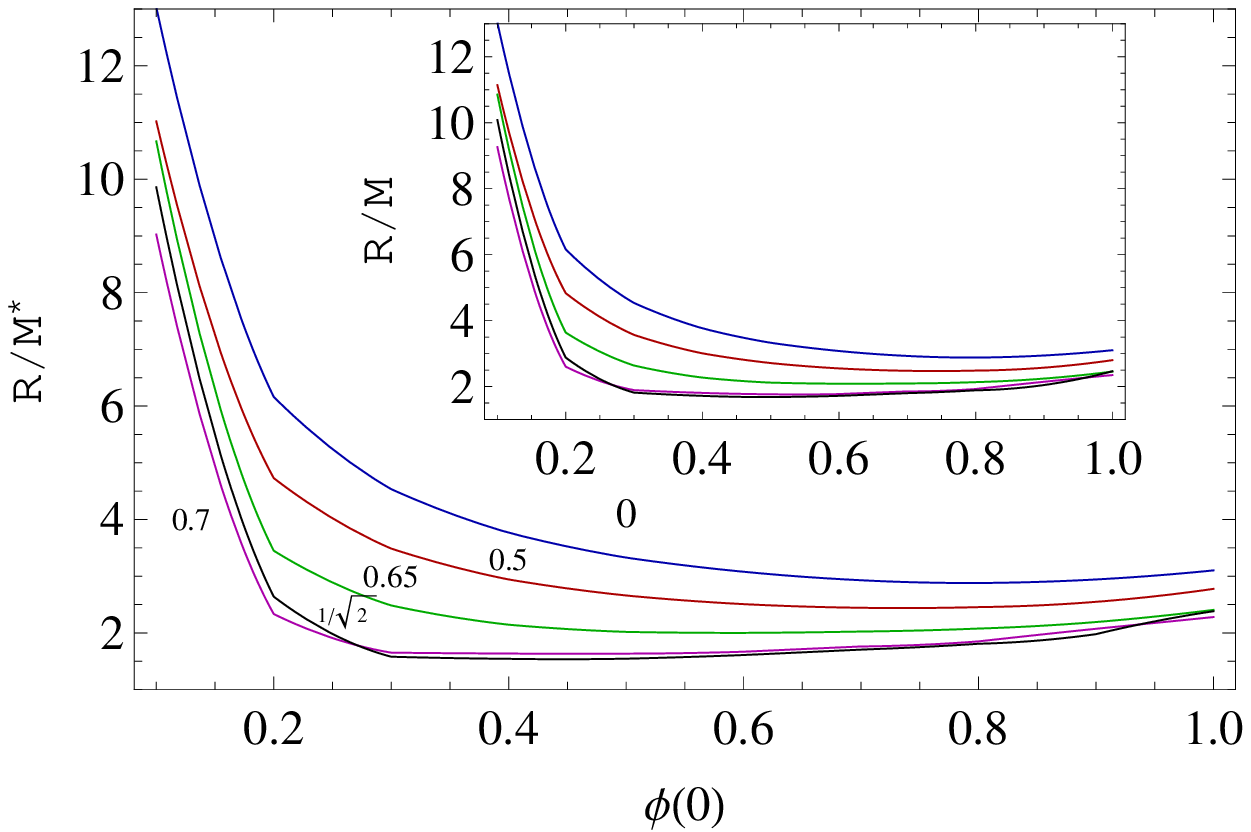}\\
\includegraphics[width=0.5\hsize,clip]{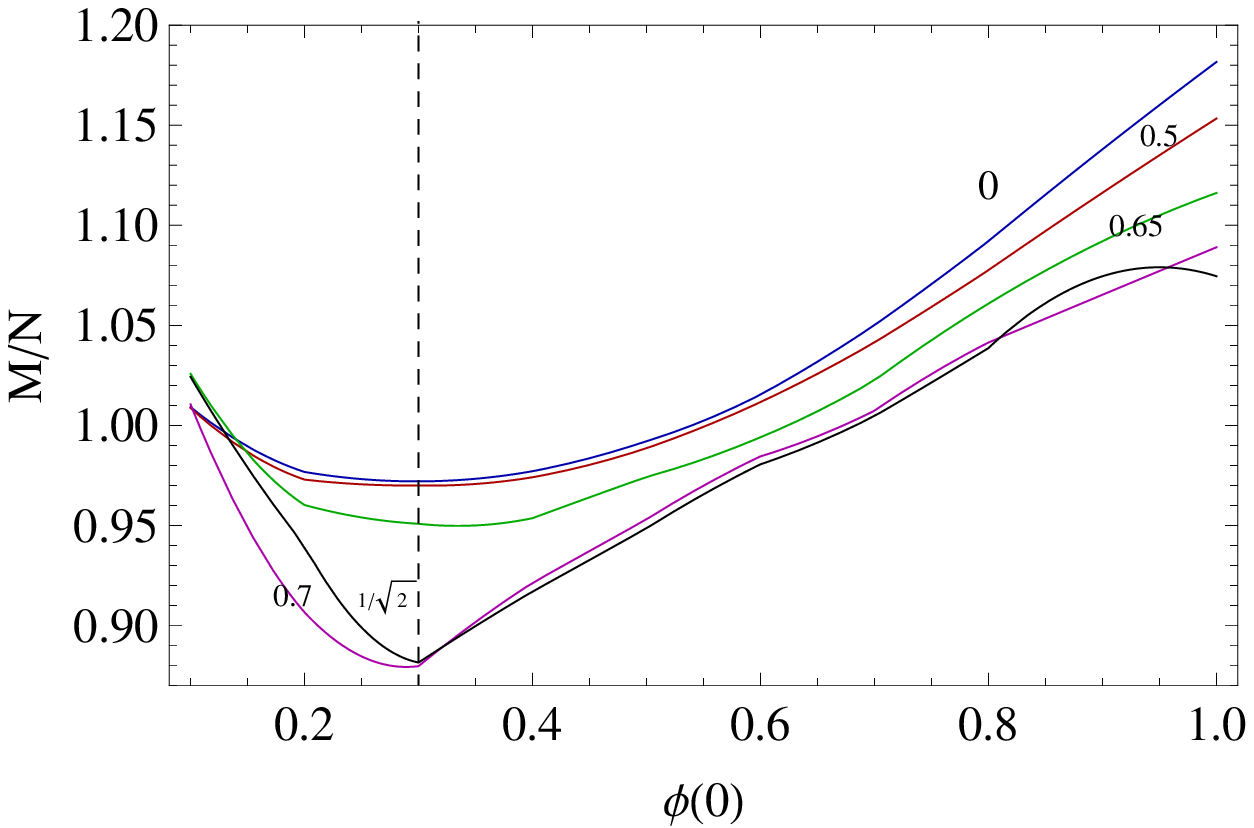}
\includegraphics[width=0.5\hsize,clip]{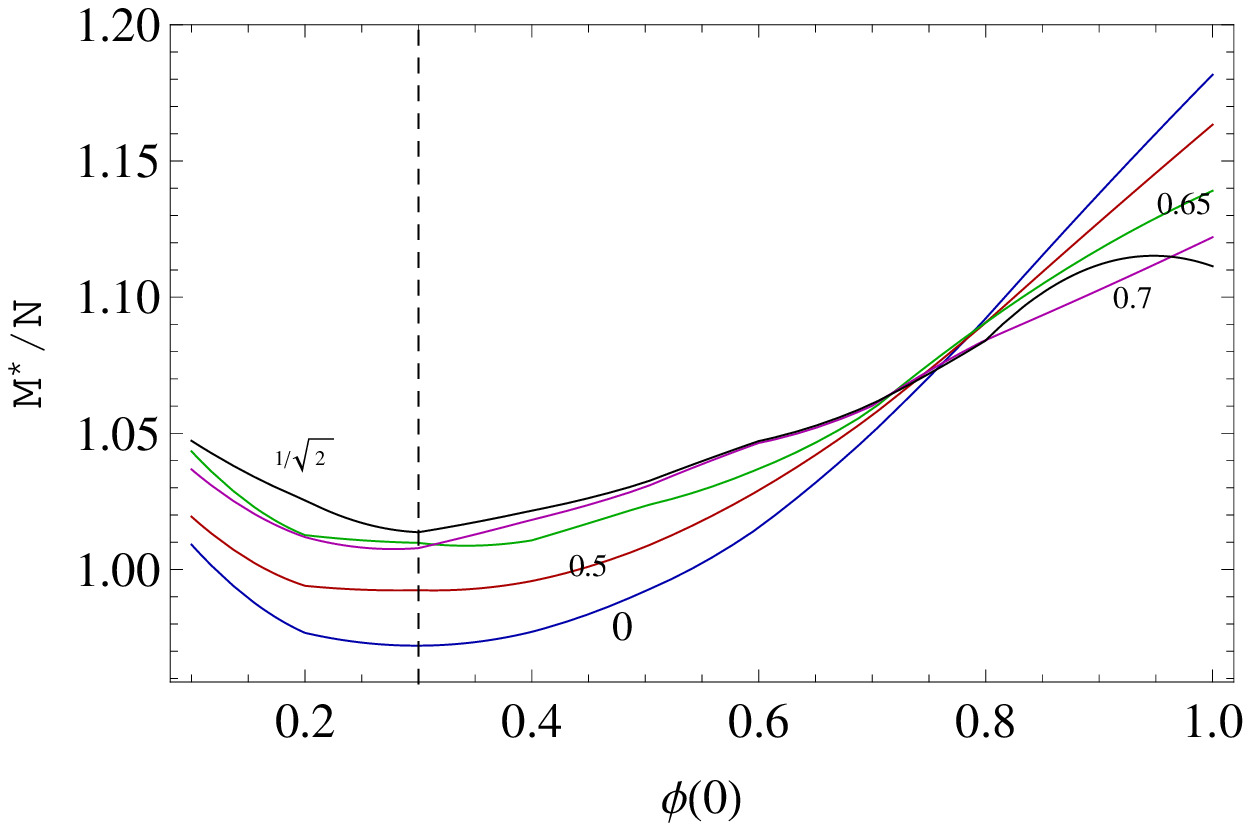}
\end{tabular}
\caption[font={footnotesize,it}]{\footnotesize{(Color online) The ratios $R/N$ (upper left plot),
$R/M^*$ (upper right),  $M/N$ (bottom right) and $M^*/N$ (bottom left) in units of $m/M^2_{\mbox{\tiny{Pl}}}$, $1/M^2_{\mbox{\tiny{Pl}}}$,
and $m$, respectively, are plotted as functions of the central
density, and for different values of the boson charge $q$ (in
$\sqrt{8\pi}m/M_{\mbox{\tiny{Pl}}}$). Dashed line is $\phi(0)=0.3$.
}}.
\label{1-pag-y}
\end{figure}

In Figs.\il\ref{1-pag-y} we show the ratios $R/N$, $R/M$ and $R/M^*$,
$M/N$ and  $M^*/N$ in units of $m/M^2_{\mbox{\tiny{Pl}}}$,
$1/M^2_{\mbox{\tiny{Pl}}}$ and $m$,  respectively, as  functions of
the central density, and for different values of the boson charge
$q$.
For  fixed values of the charge $q$, the
ratio $R/N$, the ratio $R/M$, and $M/N$ decrease as the
central density increases, until they reach a minimum value
$\phi(0)_{\mbox{\tiny{(min,R/N)}}}$, $\phi(0)_{\mbox{\tiny{(min,R/M)}}}$, $\phi(0)_{\mbox{\tiny{(min,M/N)}}}$,  respectively.
After the minimum is reached, all ratios increase
monotonically as the central density increases.

In Table\il\ref{tab:bel-to}, the
minimum values of $R/N$, $M/N$ and $R/M$, $M^*/N$ and $R/M^*$ and  the value of $\phi_{\mbox{\tiny{min}}}$,
are given for different values of $q$.
\begin{table}[h!]
\centering
\begin{tabular}{lcccccccccr}
\toprule
 $q$
&$\frac{R(\phi(0))}{M(\phi(0))}$&$\phi(0)_{\mbox{\tiny{(min,R/M)}}}$ &$\frac{R(\phi(0))}{M^*(\phi(0))}$&$\phi(0)_{\mbox{\tiny{(min,$R/M^*$)}}}$ &
$\frac{M(\phi(0))}{N(\phi(0))}$&$\phi(0)_{\mbox{\tiny{(min,M/N)}}}$&
$\frac{M^*(\phi(0))}{N(\phi(0))}$&$\phi(0)_{\mbox{\tiny{(min,$M^*/N$)}}}$&
$\frac{R(\phi(0))}{N(\phi(0))}$&$\phi(0)_{\mbox{\tiny{(min,R/N)}}}$
\\ \hline \hline
0&2.87985&0.793811&2.87985&0.793811&0.972071& 0.298102&0.972071& 0.298102&3.07529&0.691860
\\
0.50&2.47116&0.742725&2.43821&0.735351&0.969991&  0.291616&0.992362&0.282309&2.56920&0.648890
\\
0.65&2.08637&0.617825&2.00041&0.592344
&0.949828&0.334285&1.00875&0.342184&2.06194&0.535094
\\
0.70&1.76082& 0.541418&1.63117& 0.455206&0.879342&0.289220&1.00752&0.278012&1.66369&0.340757
\\
$1/\sqrt{2}$&1.68120&0.499717& 1.53638&0.446416& 0.881540& 0.300000
& 1.0137& 0.300000& 1.57425& 0.424978
\\
\hline \hline
\end{tabular}
\caption[font={footnotesize,it}]{\footnotesize{The minimum values of the ratios
$R/M$ and $R/M^*$ (in units of $1/M_{\mbox{\tiny{Pl}}}^2$), $ M/N$  and $M^*/N$ (in units of $m$) and $R/N$
(in units of $m/M_{\mbox{\tiny{Pl}}}^2$) as functions of
$\phi(0)$ and for different values of $q$. {We note that to an increase of $q$
corresponds a decrease of the minima of $R/M$, $R/M^*$,
 $M/N$, and $R/N$. Viceversa the $M^*/N$ increases with $q$.}}}\label{tab:bel-to}
\end{table}
Furthermore,  an increase of the boson charge
values corresponds a decrease of the minima of the ratios $R/N$  and $R/M$,
and of the corresponding $\phi_{\mbox{\tiny{min}}}$.
For a fixed value of the central density $\phi(0)$, to a decrease of the
boson charge $q$ corresponds an increase of $R/N$,  $R/M$ and $R/M^*$.
These  ratios decrease as the particle repulsion  increases,
leading to a minimum value for a given central density.
The ratio $M/N$ and $M^*/N$ decrease with an increase of the central density  until it reaches
a minimum value, and then it increases as $\phi(0)$ increases.
The minimum values of $M/N$ decrease as the charge $q$ increases.
On the other side,  from Table\il\ref{tab:bel-to}, we note that the minimum values of $M^*/N$ increases as the charge $q$ increases. This can also be noted in Figs.~\ref{1-pag-y}: $M^*/N$ increases with $q$ until the central density reaches a point $\phi(0)\approx0.75$, at which the lines  $M^*/N$  at different charges match and then $M^*/N$ turns out to be a decreasing function of $q$.  It is clear that the quantity $M/N$ is an indication of the binding energy per particle, $B/N=1-M/N$, in the units we are using. So $M/N>1$ indicates negative binding energies (bound particles) while $M/N<1$ indicates unbound particles, in principle. It can be seen from the lower left panel of Fig.~\ref{1-pag-y} how the misinterpretation of the mass $M$ as the mass of the system would in principle lead to the conclusion that most of the configurations have positive binding energy, since $M/N<1$. Instead, the lower right panel of Fig.~\ref{1-pag-y} shows that indeed most of the configurations have $M^*/N>1$ and have therefore negative binding. However, it can be also seen from this figure that indeed there are configurations for which despite being in the stable branch, $\phi(0)\leq 0.3$, their binding energy is positive for some values of the central density. In contrast, the configurations at the critical point, $\phi(0)\approx 0.3$, and over it, show negative binding energies; this means that objects apparently bound can be unstable against small perturbations, in full analogy with what observed in the mass-radius relation of neutron stars.
For a discussion  on this issue see, for instance, \cite{Kleihaus:2009kr,Kleihaus:2011sx}.

\begin{figure}[h!]
\centering
\begin{tabular}{ccc}
% \begin{minipage}[b]{1cm}
\includegraphics[width=0.33\hsize,clip]{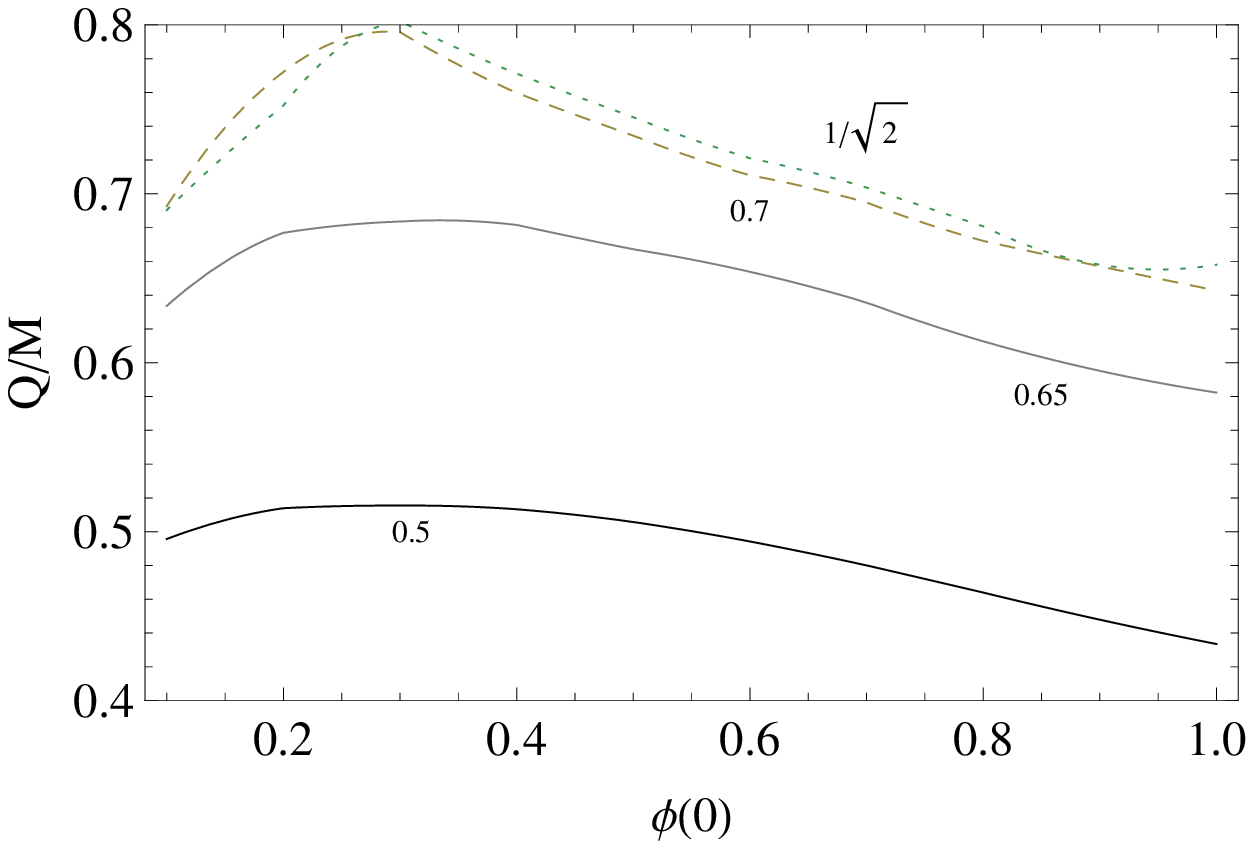}
\includegraphics[width=0.33\hsize,clip]{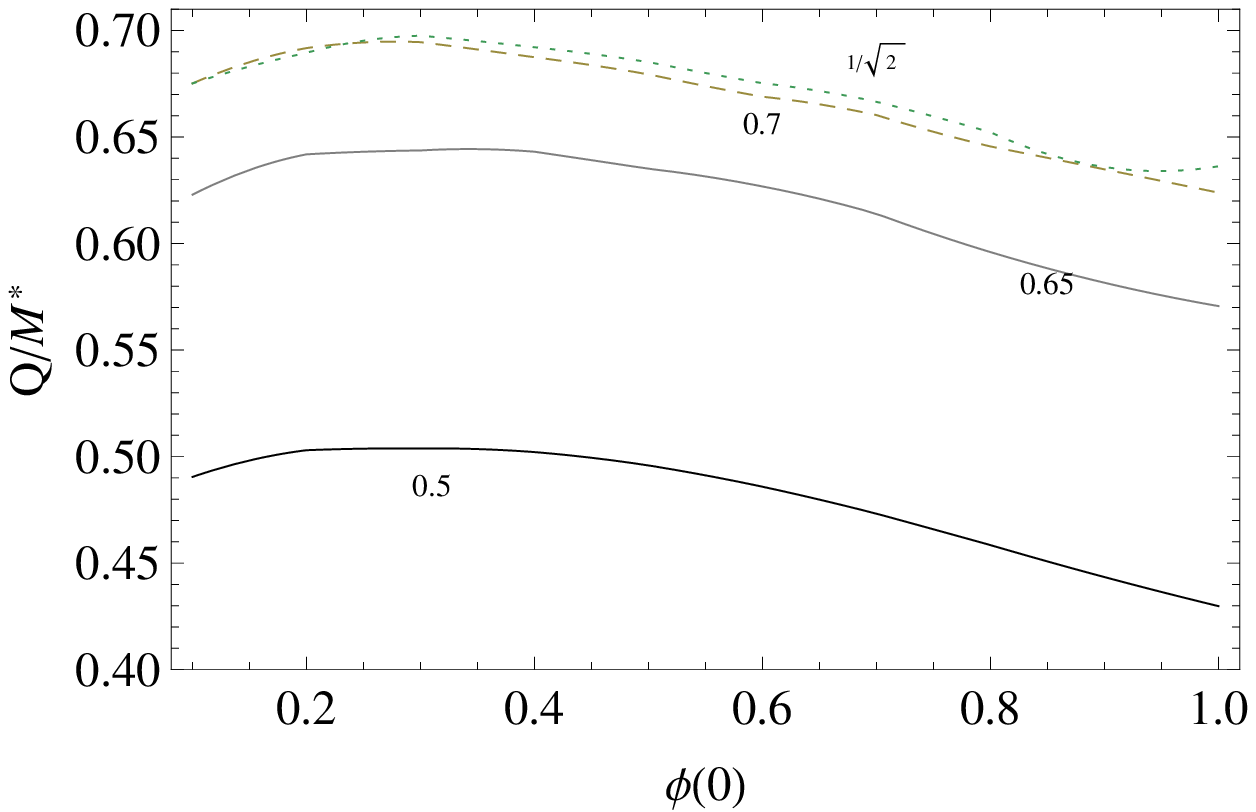}
\includegraphics[width=0.33\hsize,clip]{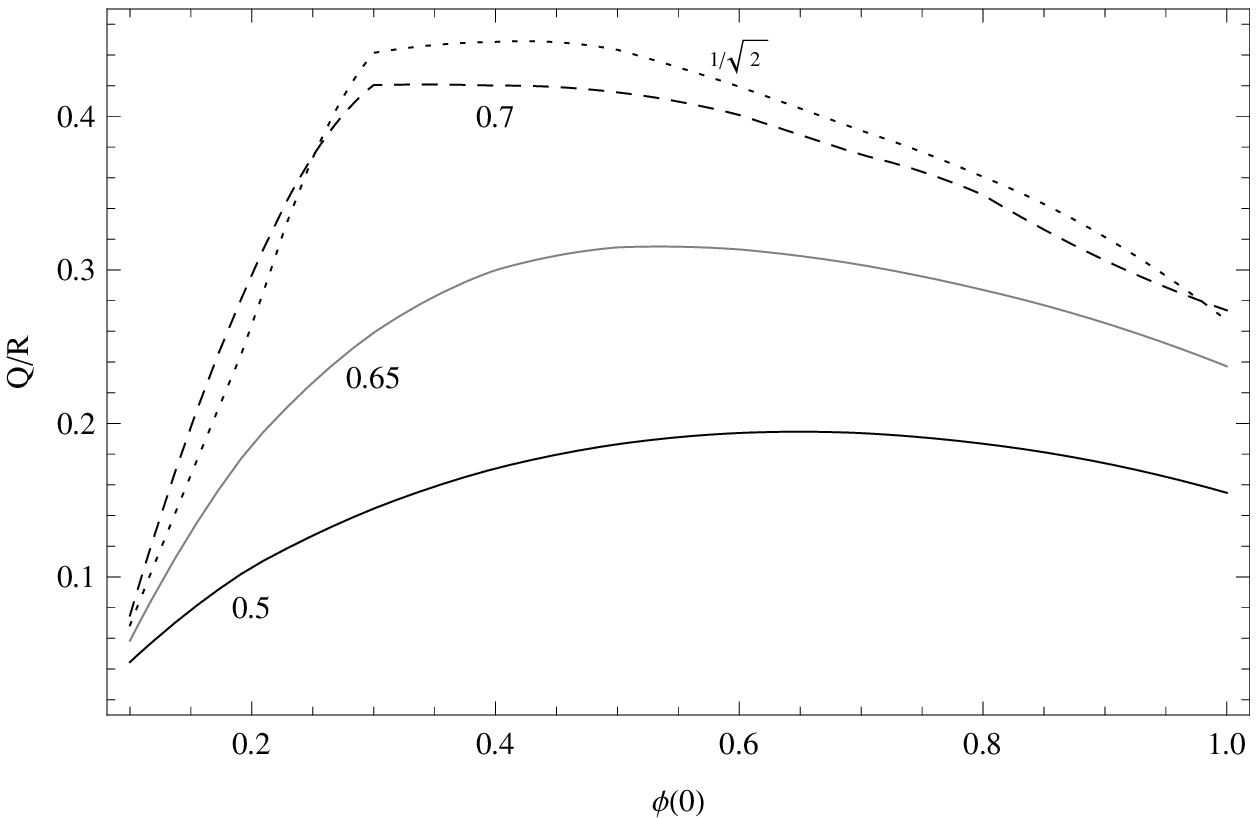}
\end{tabular}
\caption[font={footnotesize,it}]{\footnotesize{The ratios $Q/M$ (left)  and $Q/M^*$ (center)
in units of $\uQM$, and $Q/R$ (right), in units of $\uQR$, as functions
 of the central density, for different values of the boson
charge $q$ (in $\uq$).
}}\label{Ag-08-ven}
\end{figure}

Figs.\il\ref{Ag-08-ven} illustrate the
behavior of the ratios $Q/M$, $Q/M^*$ and $Q/R$
in units of
$\uQM$  and
$\uQR$, respectively,
as functions of
the central density for different values of the boson charge $q$. The maximum values of the charge-to-mass ratio  satisfy  the inequality $Q/M>Q/M^*$ since $M<M^*$ as shown in Fig.~\ref{Ang-9}. We also note that the inequality  $Q/M^*<q/m$ is satisfied for all charges $q$, in particular $ Q/M^*$  never reaches the critical value $q_{crit}/m$; a consequence of the non-zero gravitational binding.

To an increase of the central density corresponds an increase of the
$Q/M$ ($Q/M^*$) ratio, until a maximum value is reached. As the boson
charge $q$ increases, the values of the maximum of $Q/M$ ($Q/M^*$) increase.
Table\il\ref{tab:circ-max}
provides the maximum value of the ratios $Q/M$, $Q/M^*$ and $Q/R$ as
functions of the central density and for different values of the
boson charge $q$.

\begin{table}[h!]
\begin{tabular}{lcccccr}
%\begin{tabularx}{\textheight}{|WWWWW|}
$q$ &$\frac{Q}{M}(\phi(0))$&$\phi(0)_{\mbox{\tiny{(Max,Q/M)}}}$ &$\frac{Q}{M^*}(\phi(0))$&$\phi(0)_{\mbox{\tiny{(Max,$Q/M^*$)}}}$ &
$\frac{Q}{R}(\phi(0)$&$\phi(0)_{\mbox{\tiny{(Max,Q/R)}}}$
\\
\hline \hline
0.50&0.515469&0.291645&0.503848&0.282348&0.194576& 0.648848
\\
0.65&0.684291&0.333835&0.644347&0.34207
&0.315191&0.534444
\\
0.70&0.796011& 0.289606&0.694768& 0.278139&0.420745& 0.340640
\\
$1/\sqrt{2}$&0.802127&0.300000&0.697547&0.300000& 0.449011&0.422762
\\
\hline \hline
\end{tabular}
\centering \caption[font={footnotesize,it}]{\footnotesize{
The maximum value of the ratios $Q/M$,  $Q/M^*$,  and $Q/R$, in units
of $\uQM$ and $\uQR$ respectively, as functions
of the central density and for different values of the boson charge
$q$ (in $\sqrt{8\pi}m/M_{\mbox{\tiny{Pl}}}$).
}}\label{tab:circ-max}
\end{table}

\begin{center}
\begin{figure}[h!]
\centering
\begin{tabular}{cc}
% \begin{minipage}[b]{1cm}
\includegraphics[width=0.5\hsize,clip]{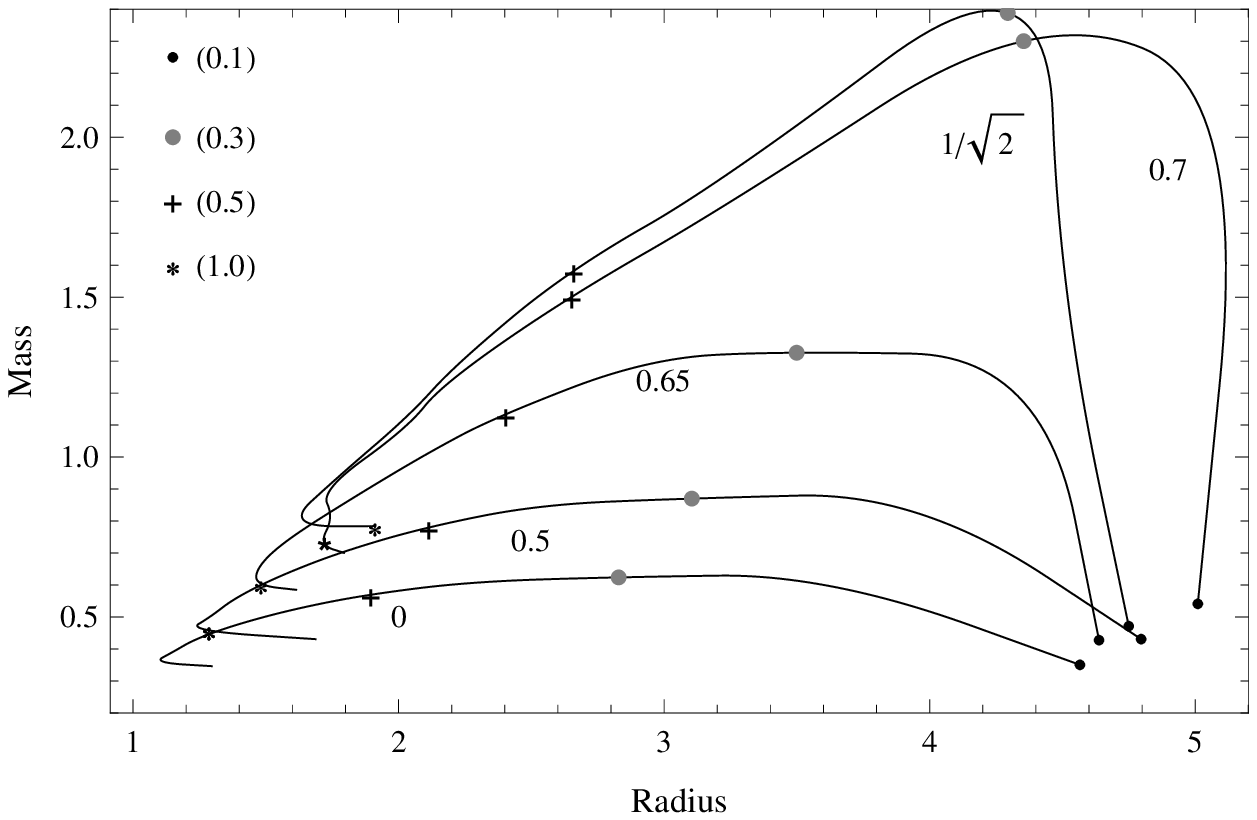}
\includegraphics[width=0.5\hsize,clip]{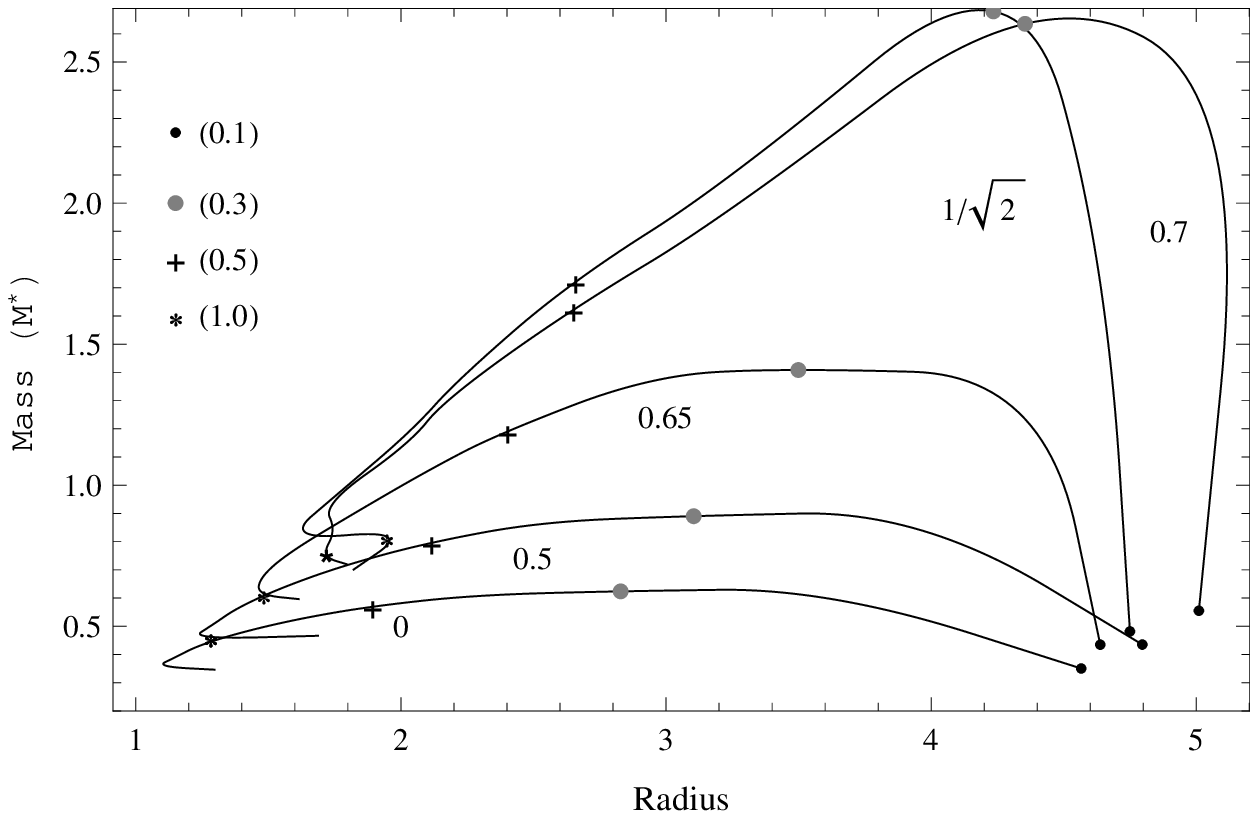}\\
\includegraphics[width=0.5\hsize,clip]{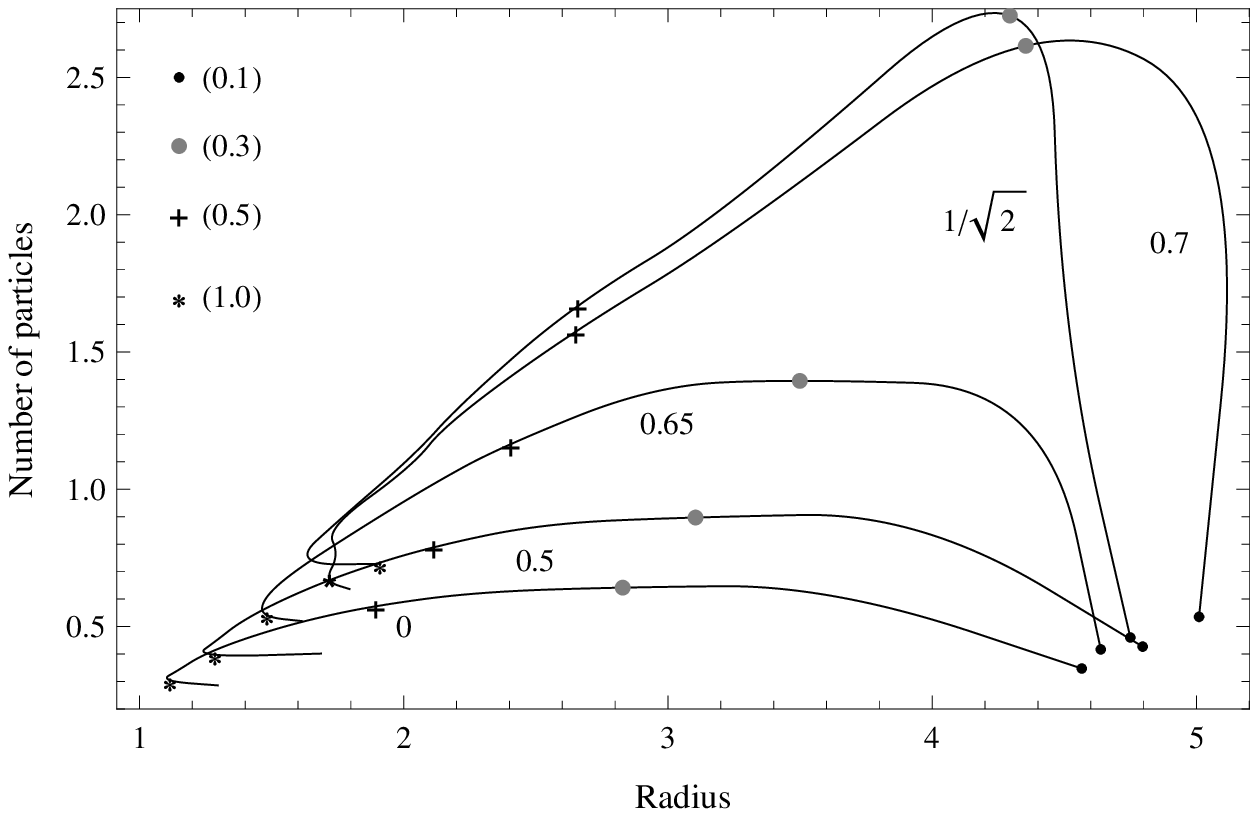}
\includegraphics[width=0.5\hsize,clip]{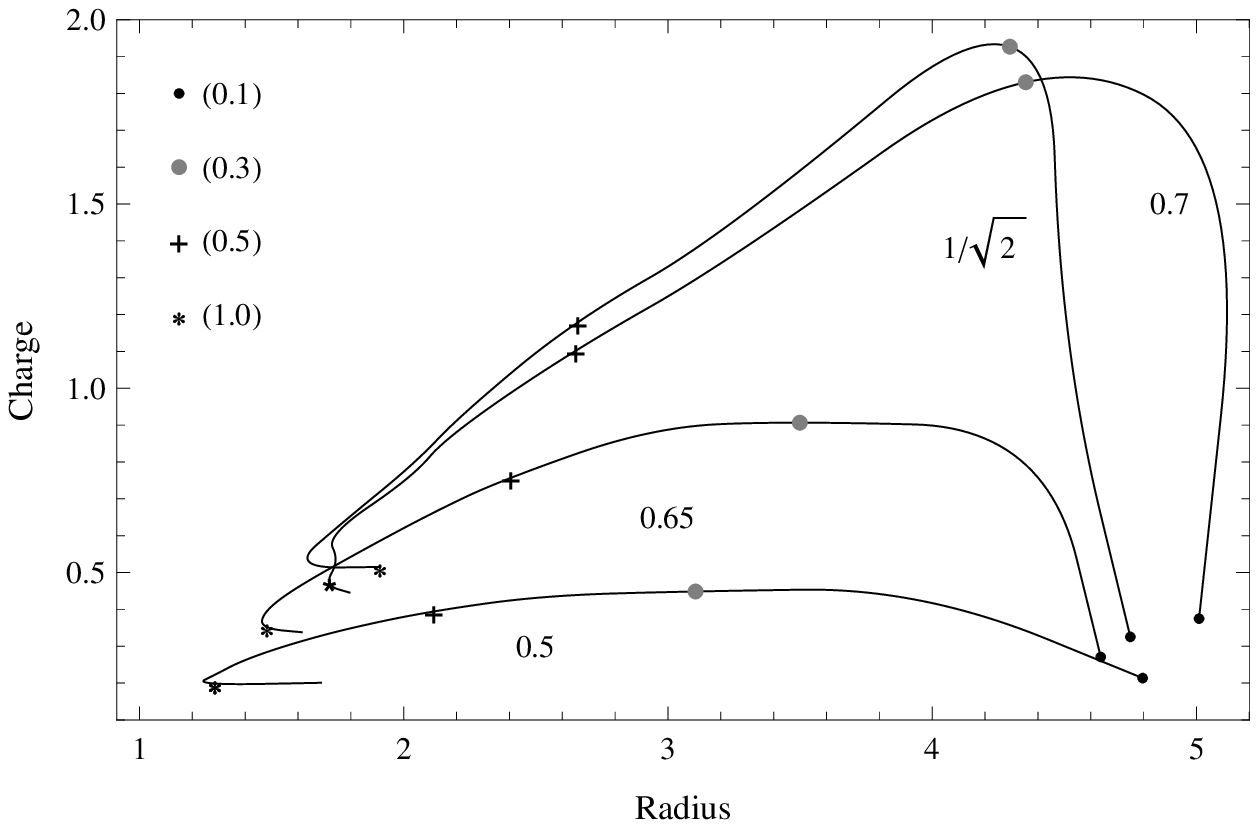}
%\end{minipage}
\end{tabular}
\caption[font={footnotesize,it}]{\footnotesize{The mass $M$ and $M^*$, in units of $\uM$,
the particle number
$N$,  in units of $\uN$, and the charge
$Q$, in units of $\uQ$,   as  functions of
the radius $R$ calculated at a fixed central value $\phi(0)$, in units of
$1/m$, for different values of the charge $q$, in units of
$\sqrt{8\pi}m/M_{\mbox{\tiny{Pl}}}$.
The central density values $\phi(0)$ are represented by markers on the curves and numbers in brackets.
}}\label{J-ane}
\end{figure}
\end{center}

The behavior of the total mass $M$ and $M^*$, particle  number $N$ and charge $Q$ as
functions of the  configuration radius $R$ is also shown in Figs.\il\ref{J-ane}, for different values of the charge $q$.
%%
%\c
We can note  that, for a fixed values of
the charge $q$, the mass, the particle number and the charge,
increase as the radius $R$ increases, until a maximum
value is reached for the same  $R_{\mbox{\tiny{Max}}}$. Then all these quantities
decrease
rapidly as $R$ increases. This means that
the concept of ``critical radius'' $R_{\mbox{\tiny{Max}}}$,
together with a critical mass and a
critical particle number,  for a charged
boson star can be introduced.
The plots also indicate that the presence of a charge $q$ does not change
the qualitative behavior of the quantities.
However, the values of $M_{\mbox{\tiny{Max}}}$ and $N_{\mbox{\tiny{Max}}}$, and of the
corresponding values of $R_{\mbox{\tiny{Max}}}$, are proportional to the value of $q$.
Configurations are allowed only within a finite interval of the radius $R$.  The values of
the minimum and maximum radii are also proportional to the value of the  boson charge $q$.
The critical central density $\phi(0)\simeq 0.3$ represents a  critical point of the curves.
Configurations for $\phi(0)>0.3$, are expected to be unstable, see \cite{Jetzer:1989av,Jetzer:1990wr}.
It is interesting to notice that for small values of the radius, there is a particular range at
which for a specific radius value there exist two possible configurations with different masses
and particle numbers. This behavior has also been found in the case of neutral configurations
and is associated with the stability properties of the system.

\begin{center}
\begin{figure}[b]
\centering
\begin{tabular}{cc}
% \begin{minipage}[b]{1cm}
\includegraphics[width=0.5\hsize,clip]{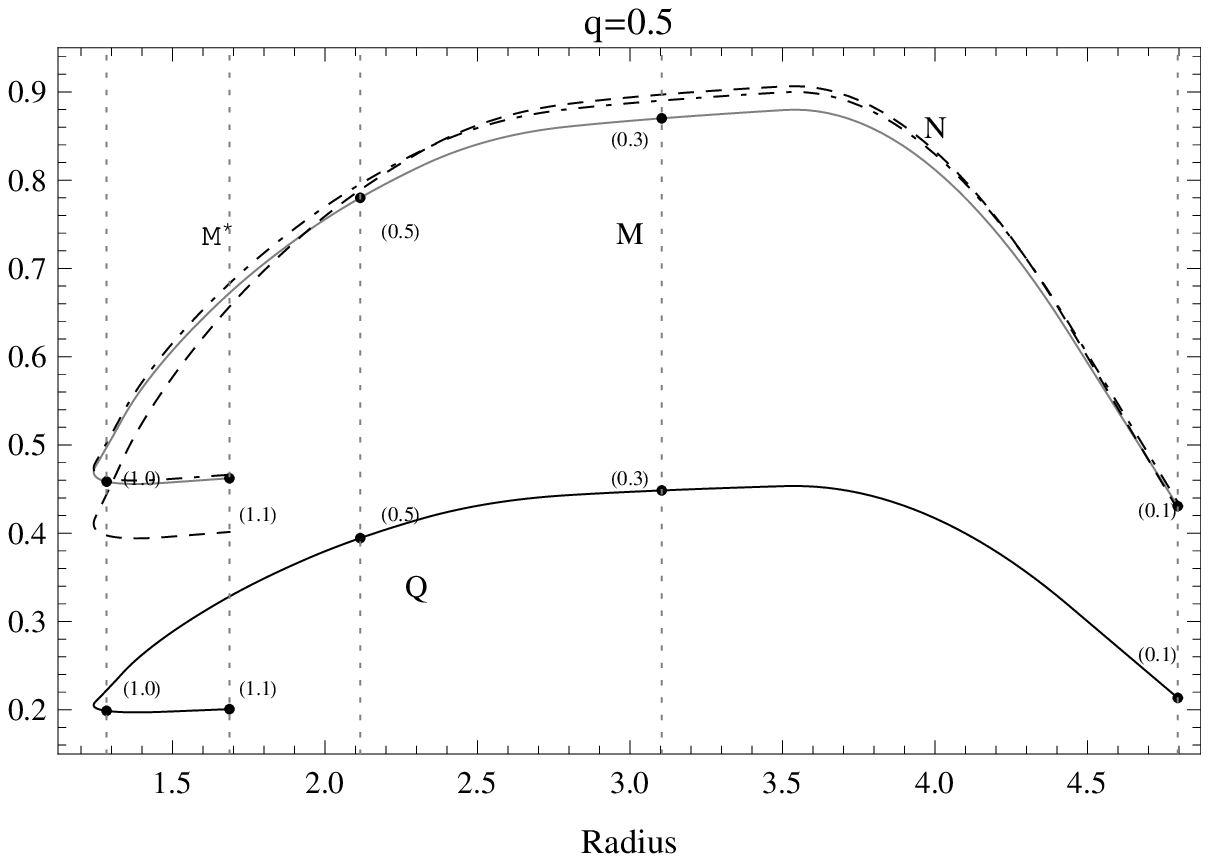}
\includegraphics[width=0.5\hsize,clip]{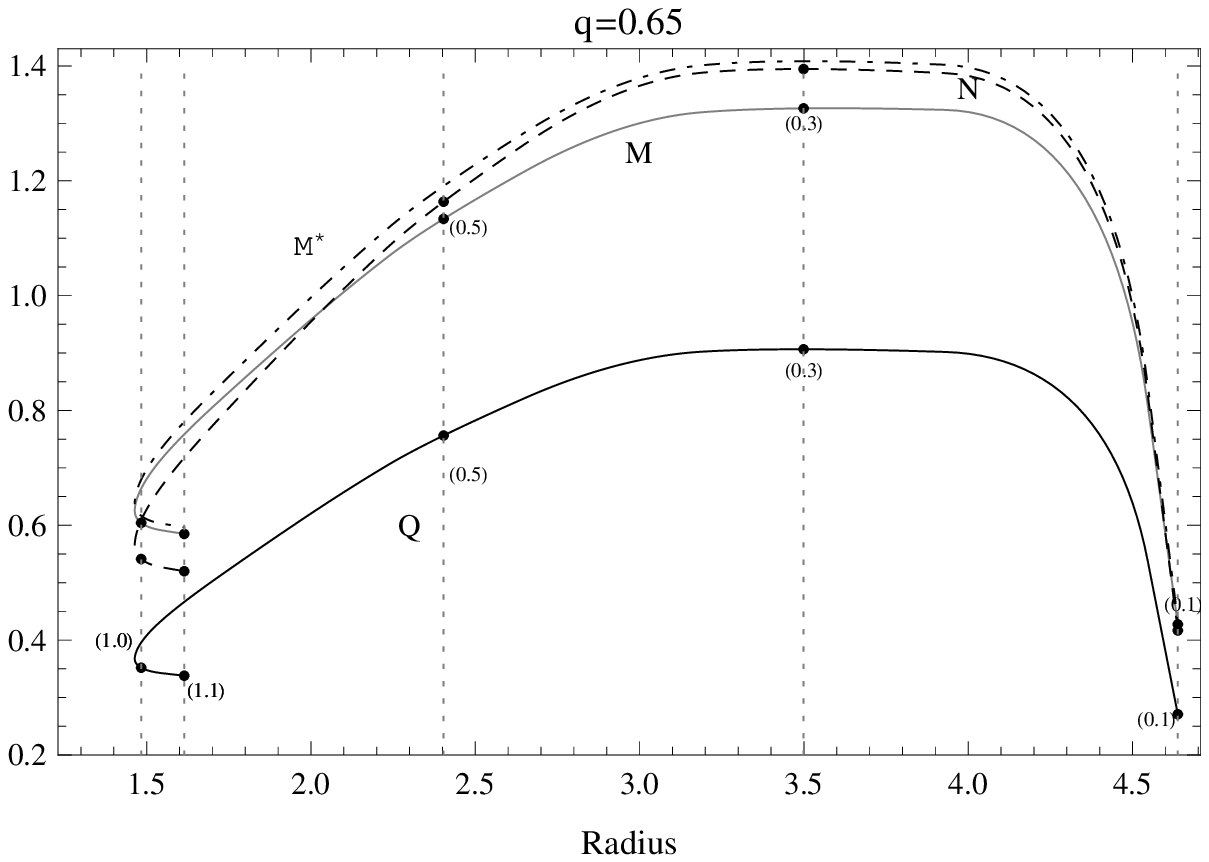}\\
\includegraphics[width=0.5\hsize,clip]{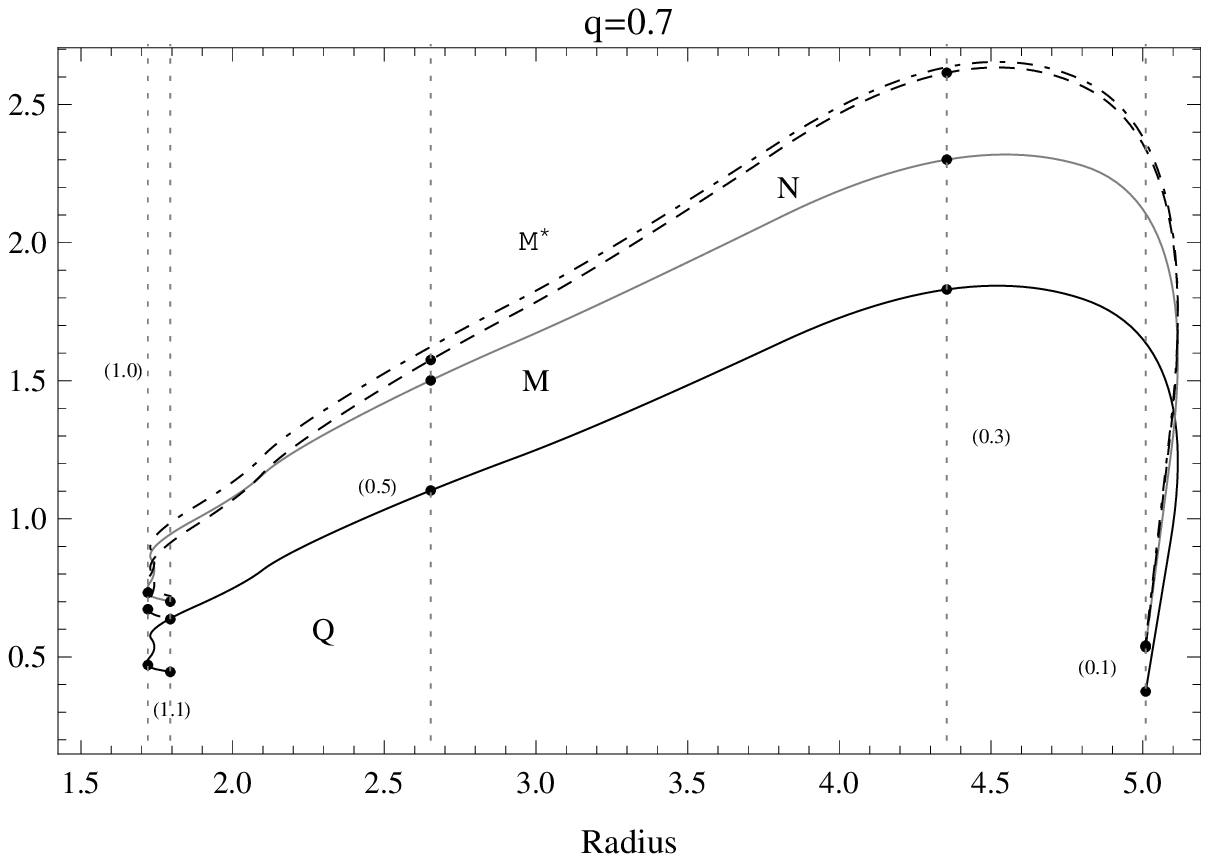}
\includegraphics[width=0.5\hsize,clip]{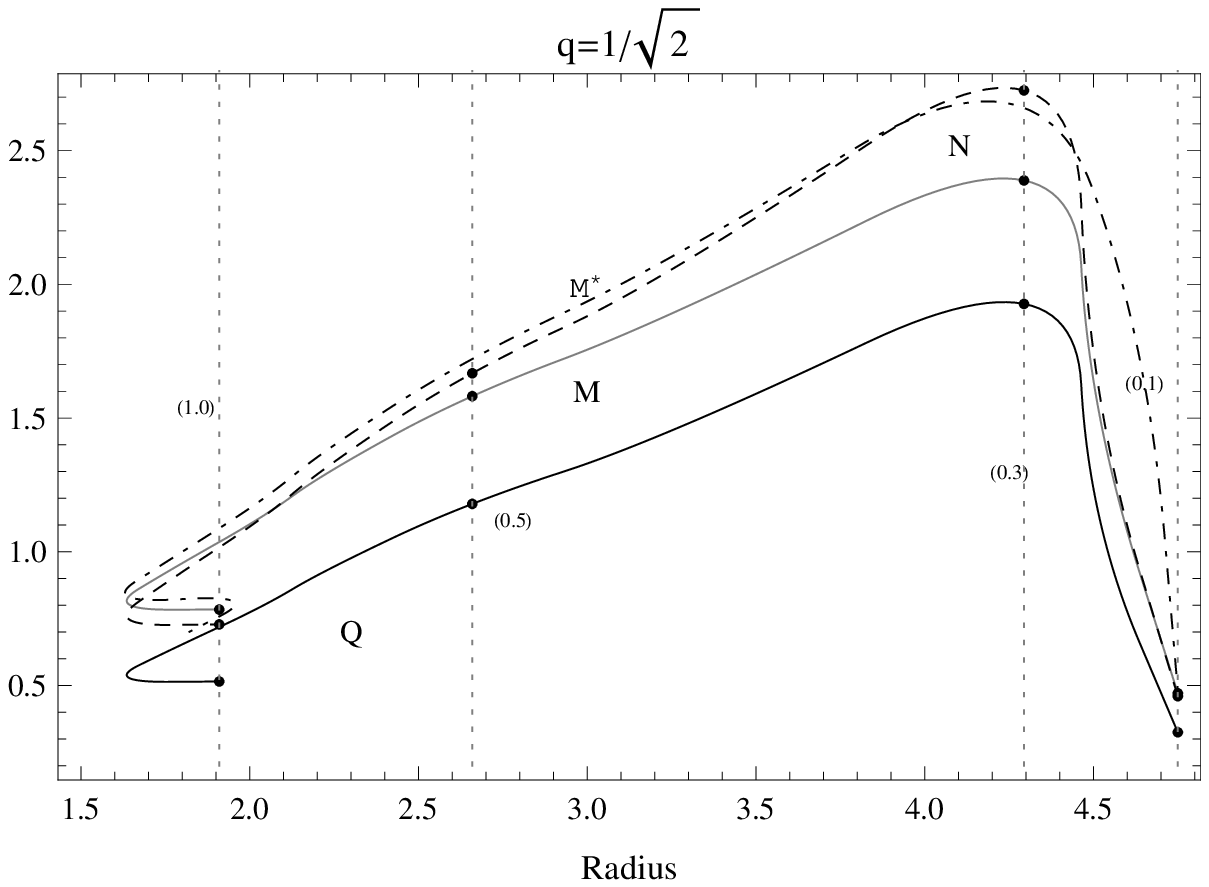}
\\
%\end{minipage}
\end{tabular}
\caption[font={footnotesize,it}]{\footnotesize{The total charge $Q$
in $\uQ$ (black curve), the
total mass  $M$ (gray curve)  and $M^*$ (dotted--dashed curve) in $\uM$, and  the  particle number $N$ (dashed curve) in units of $\uN$ are plotted as  functions of the
radius $R$ (in units of $\uR$) for different values of the charge $q$ (in units of
$\sqrt{8\pi}m/M_{\mbox{\tiny{Pl}}}$). Dotted lines represent the curves  $\phi(0)=const.$; the central density values are designed by points of the curves and numbers in brackets. }}\label{int-te}
\end{figure}
\end{center}

\begin{figure}[b]
\centering
\begin{tabular}{cc}
% \begin{minipage}[b]{1cm}
\includegraphics[width=0.5\hsize,clip]{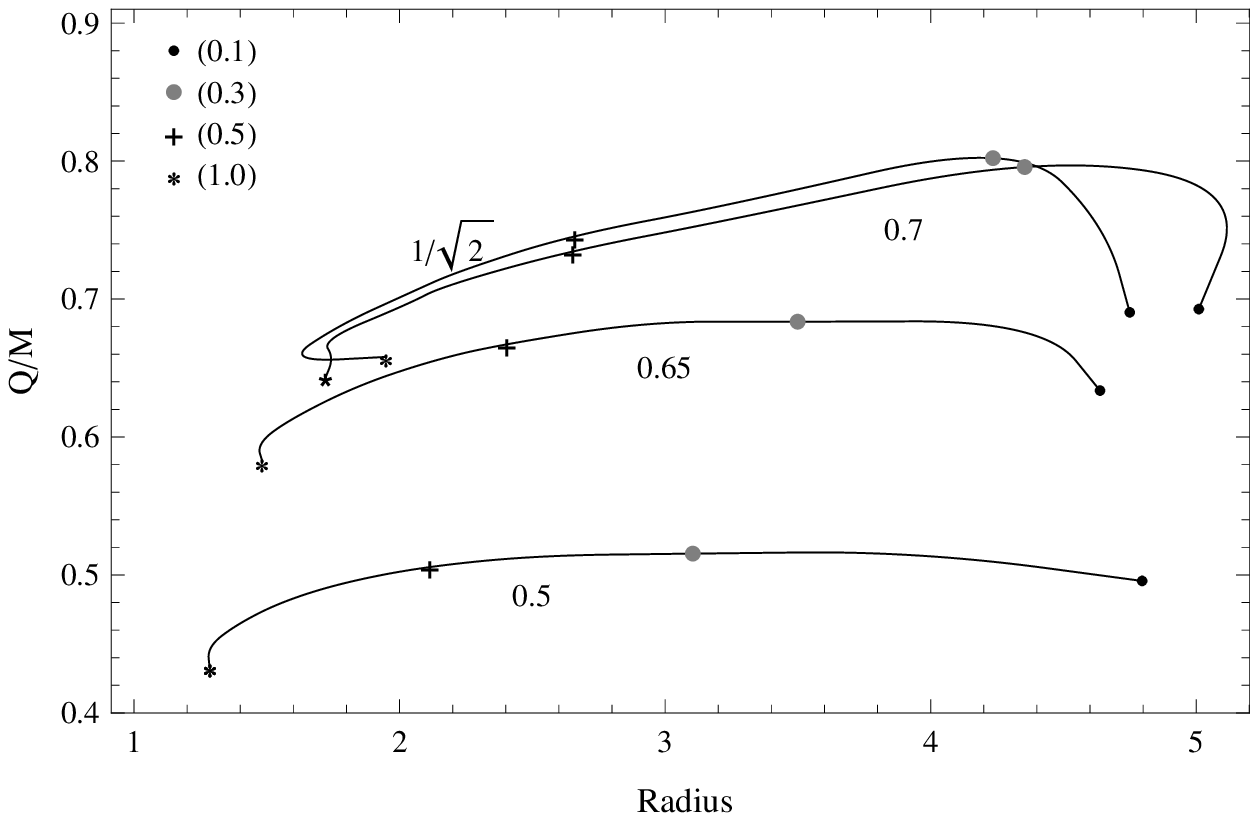}
\includegraphics[width=0.5\hsize,clip]{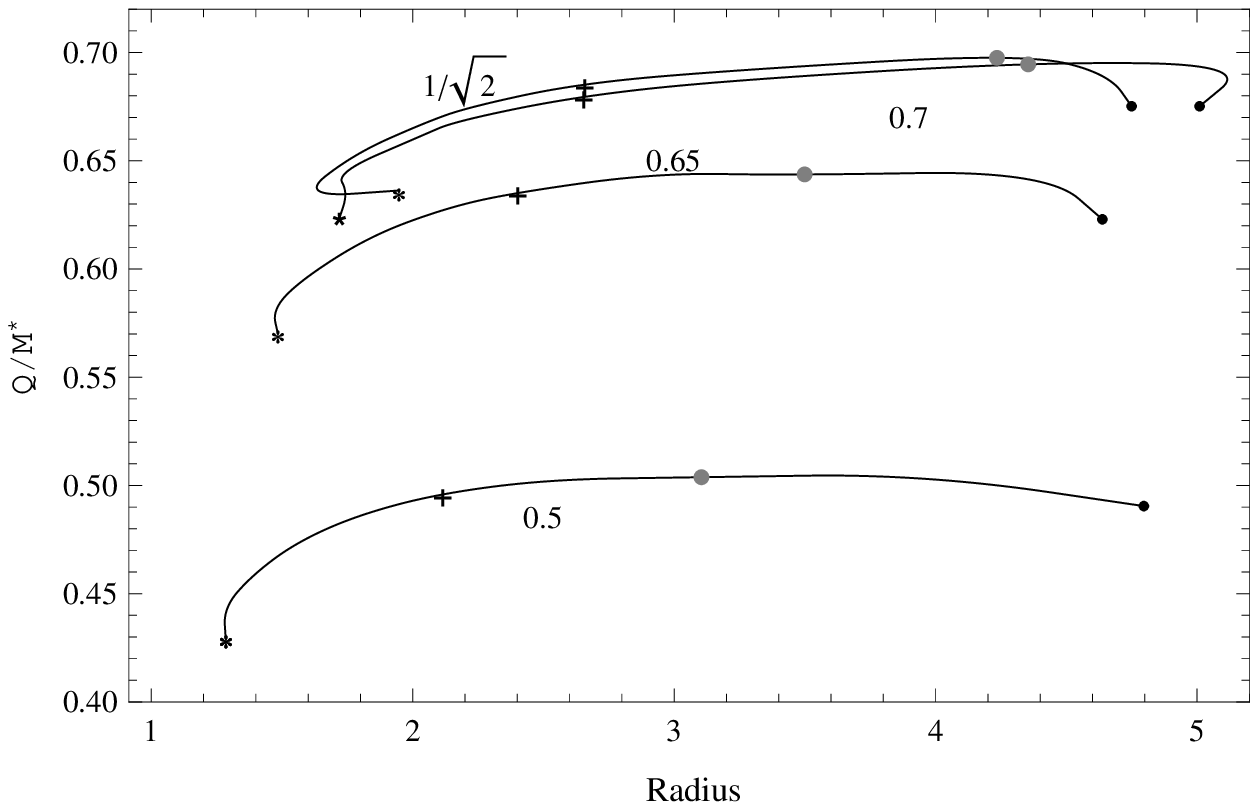}\\
\includegraphics[width=0.5\hsize,clip]{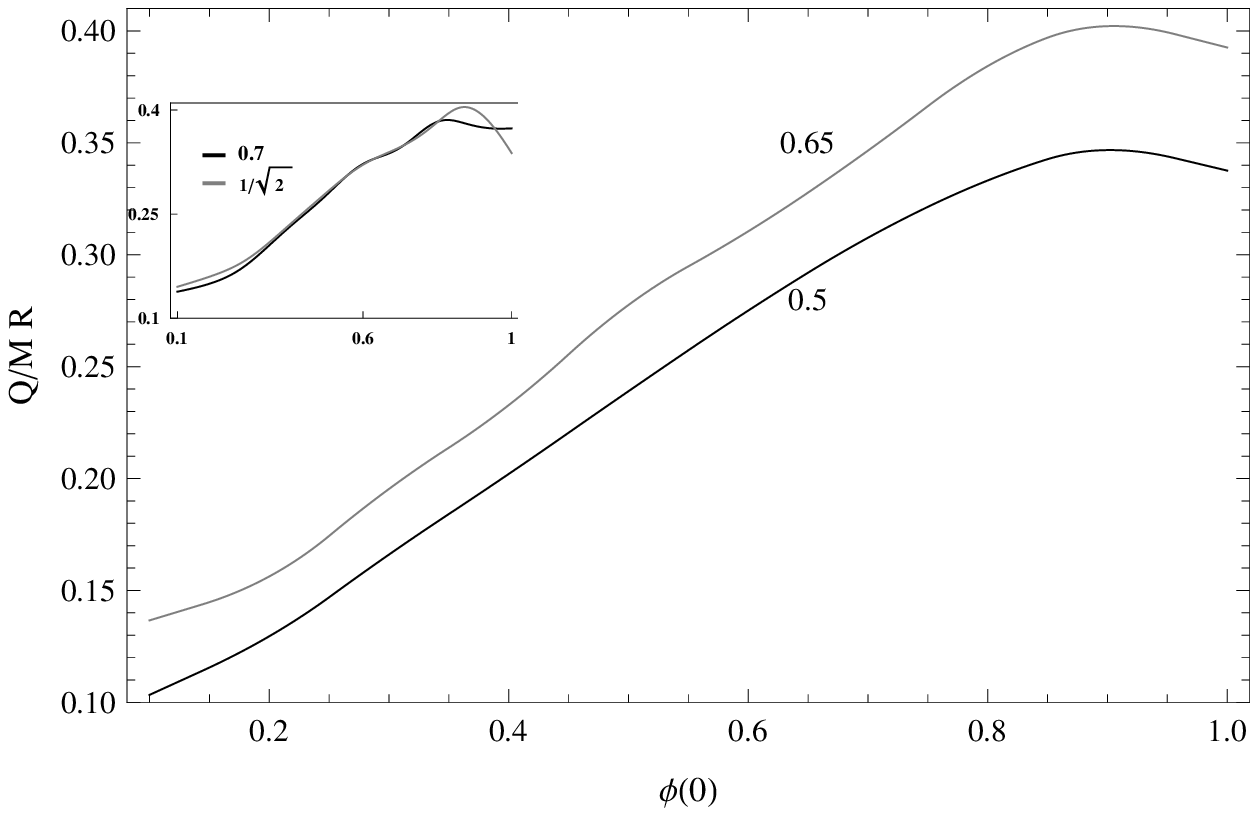}
\includegraphics[width=0.5\hsize,clip]{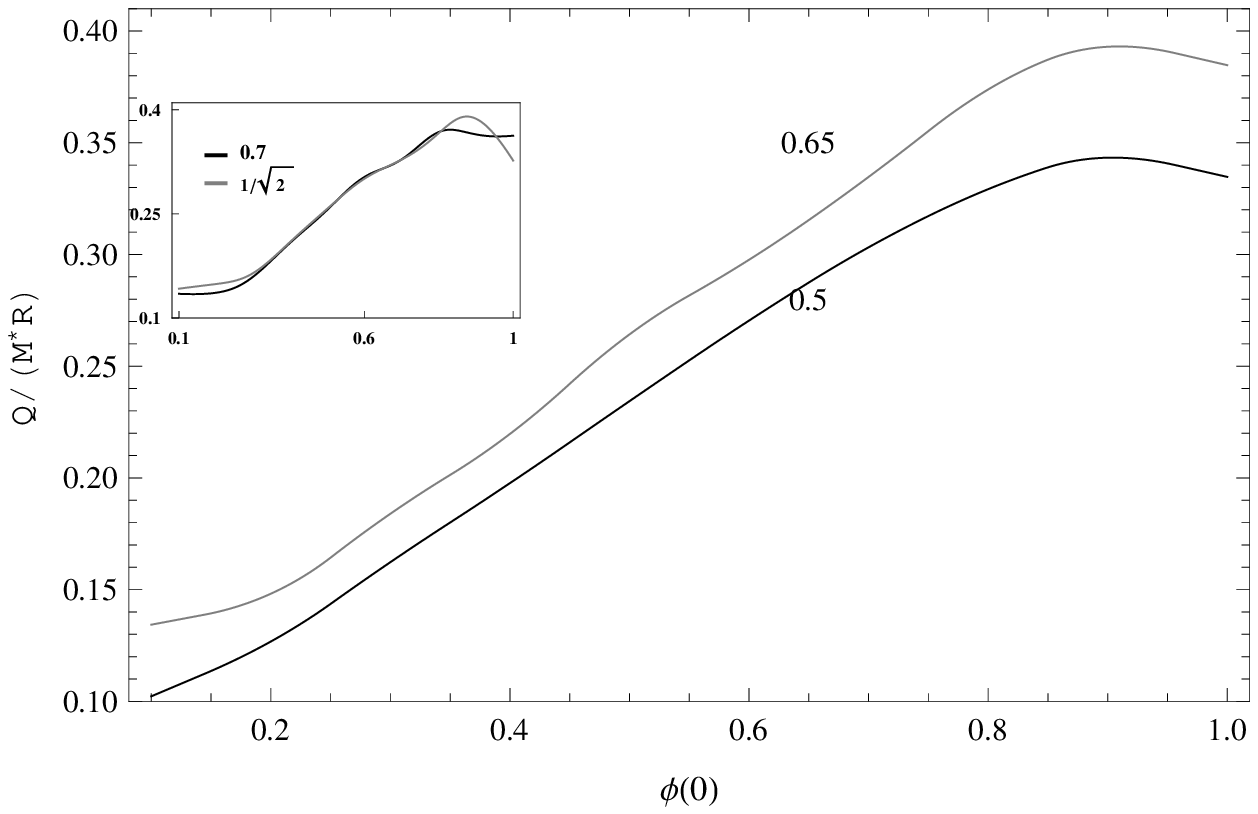}
\end{tabular}
\caption[font={footnotesize,it}]{\footnotesize{Upper plot: The charge-to-mass ratio $Q/M$ and $Q/M^*$
in units of $\uQM$ is plotted in terms of the
radius $R$ (in units of $\uR$) for different values of the charge $q$ (in units of
$\sqrt{8\pi}m/M_{\mbox{\tiny{Pl}}}$).
The central density values $\phi(0)$ are represented by markers on the curves and numbers in brackets.
Bottom plot: The  ratio $Q/(MR)$ and $Q/(M^*R)$
in units of $\uQMR$ is plotted as  a function of the central density
$\phi(0)$ for different values of the charge (in units of
$\sqrt{8\pi}m/M_{\mbox{\tiny{Pl}}}$): $q=0.5$ (black curve) and $q=0.65$ (gray curve).
The inset plot shows the curves $q=0.7$ (black curve) and $q=1/\sqrt{2}$ (gray curve).}}\label{iM}
\end{figure}

Finally, we illustrate the behavior of the physical quantities  for a fixed value of the charge $q$ in
Fig.\il\ref{int-te}.
Figures\il\ref{iM} show the charge-to-mass ratio  as a function of the radius of the configuration
evaluated at different central densities. At the central density $\phi(0)\simeq 0.3$ there exists a critical point on the curve.
To lower central densities correspond configurations with larger radius.
The ratio $Q/(RM)$ and  $Q/(RM^*)$ and increases as $\phi(0)$ increases.
%{B

%%%%%%%%%%%%%%%%%%%%%%%%%%%%%%%%%%%%%%%%%%%%%%%%%%%%%%%%%%%%%%%%%%%%%%%%%%%%%%%%%%%%%%%%%
%%%%%%%%%%%%%%%%%%%%%%%%%%%%%%%%%%%%%%%%%%%%%%%%%%%%%%%%%%%%%%%%%%%%%%%%%%%%%%%%%%%%%%%%%%

\section{Conclusions}
\label{Sec:CBSCONCL}
In this work we studied spherically symmetric charged boson stars.
We have solved numerically  the
Einstein-Maxwell system  of equations coupled to the  general relativistic Klein-Gordon equations of a
complex scalar field with a local $U(1)$ symmetry.

As in the case of neutral boson stars and previous works on charged configurations, we found that
it is possible to introduce the concepts of
critical mass $M_{\mbox{\tiny{Max}}} $ and  critical number
$N_{\mbox{\tiny{Max}}}$. It turns out that the explicit value of these quantities increases
as the value of the boson charge $q$ increases. In previous works \cite{Jetzer:1990wr,Jetzer:1989av},
it was shown that charged configurations are possible for $q<q_{\ti{crit}}\equiv\sqrt{1/2}$ (in units of
$\sqrt{8\pi}m/M_{\mbox{\tiny{Pl}}}$). We performed a more detailed analysis and determined that
bounded charged configurations of self-gravitating  bosons are
possible with a particle charge $q=q_{\ti{crit}}$, and even for higher values localized solutions can exist.

We compared and contrasted both from the qualitative and quantitative point of view the function $M$ given by Eq.~(\ref{F-ro}), often misinterpreted as the mass of a charged system, with the actual mass $M^*$, related to $M$ by Eq.~(\ref{Mstar}), which allows a correct matching of the interior solution at the surface with the exterior Reissner-Nordstr\"om spacetime.

\begin{figure}[h!]
\centering
% \begin{minipage}[b]{1cm}
\includegraphics[width=0.5\hsize,clip]{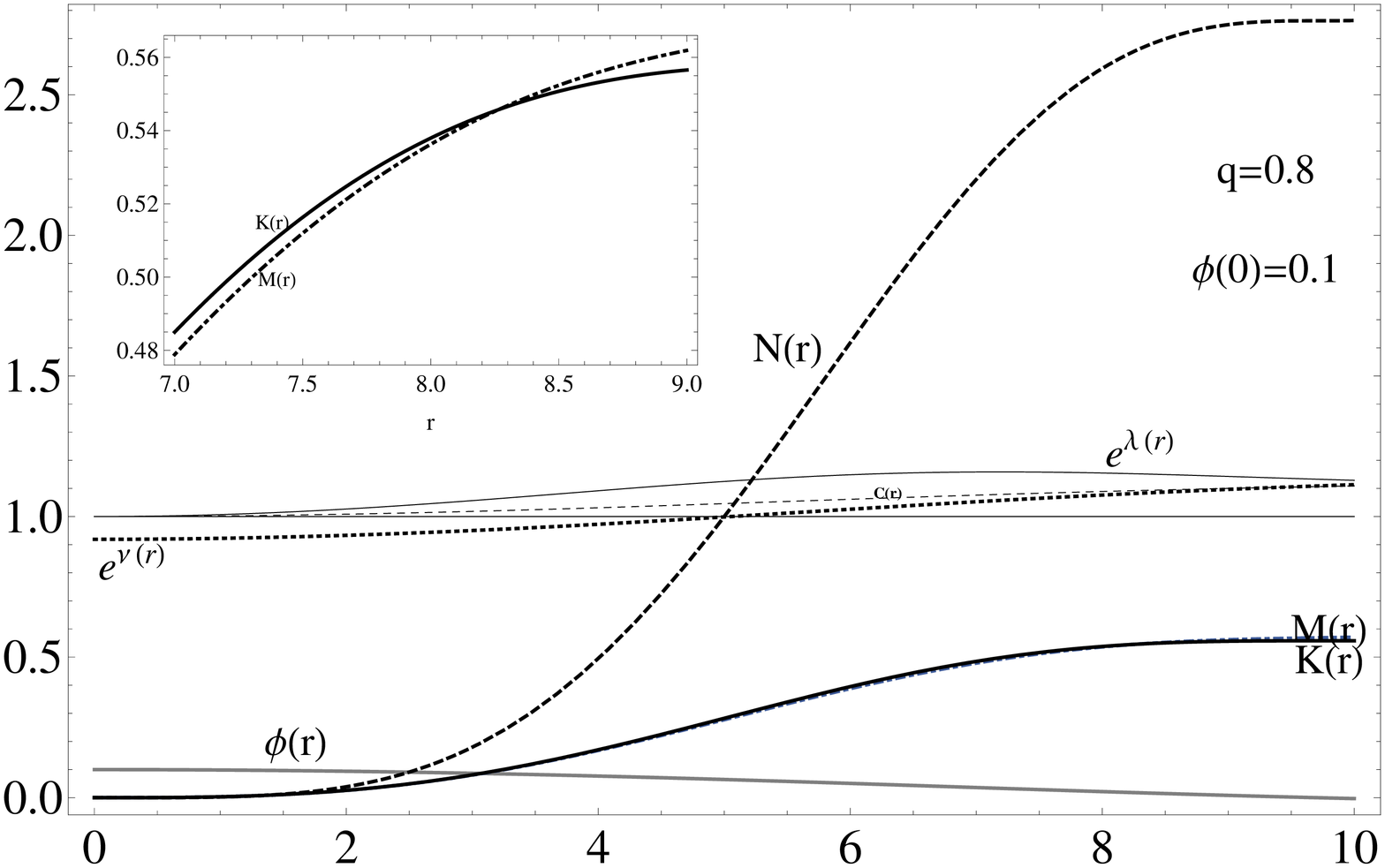}
%\end{minipage}
\caption[font={footnotesize,it}]{\footnotesize{The radial function of the scalar field $\phi(r)$ (gray curve) for
the charge $q=0.8$ in units of $\sqrt{8\pi}m/M_{\mbox{\tiny{Pl}}}$,  $\omega=1.10893$  and for central density $\phi(0)=0.1$, function of the radial coordinate $r$.   Dotted curve is $e^{\nu(r)}$, black curve is $e^{\lambda(r)}$, dashed curve is $C(r)$. Dashed thick curve is $N(r)$. Inside plot is a enlarged view of the curve $M(r)$ (dotted-dashed curve), and $K(r)$ (black thick curve) where $R=\int^{\infty}_0 K'(r) dr$. It is $e^{\lambda}_{\ti{Max}} =1.15874$ in  $r_{\ti{Max}}= 7.18885$.   Moreover the configuration mass $M =0.570768$ and  $M^*=0.590706$ measured in units of $M_{\mbox{\tiny{Pl}}}^{2}/m$.
The particle number $N=0.558146$ , in units of $M_{\mbox{\tiny{Pl}}}^{2}/m^2$, and  the radius
$R=4.95246$
and  the total charge $Q=q N=0.446517$ in units of $1/m$  and $\sqrt{8\pi}M_{\mbox{\tiny{Pl}}}/
m$, respectively.
%that is the integral of $K'(r)$/N.
 Moreover, it is
 $Q/R=0.0901607$,
 $R/N=8.87305$,
$R/M=8.67684$,
$R/M^*=8.38397$,
$M/N=1.02261$,
$M^*/N=1.05833$,
$Q/M=0.78231$,
 $Q/M^*=0.755905$.
%charge $q$ in units of $\sqrt{8\pi}m/M_{\mbox{\tiny{Pl}}}$
Here $R/M$ and $R/M^*$  are in units of $1/M_{\mbox{\tiny{Pl}}}^2$, $M/N$  and $M^*/N$ in units of $m$ and $R/N$
in units of $m/M_{\mbox{\tiny{Pl}}}^2$, $Q/M$,  $Q/M^*$,  and $Q/R$, in units
of $\uQM$ and $\uQR$, respectively}} \label{Fig:q=0.8}
\end{figure}

{By means of numerical integrations
it is possible
 to show that for  $q>q_{\ti{crit}}$
 solutions satisfying the given initial conditions,  without nodes are possible only for small
values of the central density smaller than the critical value $\phi(0)\approx 0.3$} (see e.g.~Fig.\il\ref{Fig:q=0.8}).
%\begin{center}
%\end{center}
 On the other hand, for $q>q_{\ti{crit}}$ and higher central densities the boundary conditions  for zero-node solutions at
the origin are not satisfied and only bounded configurations
with one or more nodes could be possible.

We established
that the critical
central density value corresponding to $M_{\mbox{\tiny{Max}}} $ ($M^*_{\mbox{\tiny{Max}}} $) and
$N _{\mbox{\tiny{Max}}} $ is $\phi(0)_{\mbox{\tiny{Max}}}\simeq 0.3$, independently of the boson charge  $q$.
The critical total mass and number of
particles increase as the electromagnetic
repulsion increases \cite{Jetzer:1990wr} (see \cite{Hod:2010zk,Madsen:2008vq}, and also \cite{Eilers:2013lla,Hartmann:2012gw,Lieb-Pale12},  for a recent discussion on the  charge-radius relation for compact objects).

The total charge of the star increases with an increase of the value of the
central density until it reaches a maximum value at $\phi(0)= \phi(0)_{\mbox{\tiny{Max}}}\simeq0.3$.
As $\phi(0)$ continues to increase, the charge $Q$ decreases
monotonically. In this manner, the concept of a critical charge $Q_{\mbox{\tiny{Max}}} $ for
charged boson stars can be introduced in close analogy to the concept of $N_{\mbox{\tiny{Max}}}$.
In this respect, the value $\phi(0)\simeq 0.3$ plays the role of  a point of
maximum  of the electromagnetic repulsion (as a function of the central density).

In order to have a better
understanding  of these systems for $\phi
\simeq\phi(0)_{\mbox{\tiny{Max}}}$, we studied the behavior of $\phi$ and $g_{11}$ as
functions of $\phi(0)$ and the radial coordinate $r$.
The density $\phi$ increases with larger values of $q$, at fixed  $r$ and fixed central density.
For a fixed value of the boson charge, $g_{11}$ reaches a maximum value
corresponding to a value  $r_{\mbox\tiny{Max}}$ of the radial
coordinate. After this maximum is reached, it decreases monotonically with an
arbitrary increase of $r$. The maximum value of $g_{11}$ depends on the value of
the central density and of the coupling constant $q$. However, this maximum is bound
and reaches its highest value for $\phi_{\mbox{\tiny{Max}}}(0)\simeq 0.3$.

The radius $R$ and the ratios $Q/M$, $M/N$, $R/M$, $Q/M^*$, $M^*/N$, $R/M^*$, $R/N$, $Q/R$ were also studied as
functions of the central density. To the central density value $\phi(0)\simeq 0.3$ corresponds the maxima of the charge $Q$,
the mass $M$ $(M^*)$, the particle number $N$, and of the ratio $Q/M$ ($Q/M^*$). On the other hand, $\phi(0)\simeq 0.3$ corresponds
to the minima of $M/N$, $R/N$ and $R/M$ as well as $M^*/N$ and $R/M^*$.

The effects of the introduction of the mass definition $M^*$ are evident in the analysis of the behavior of $Q/M^*$ and $M^*/N$ with respect to $Q/M$ and $M/N$: we note that the minimum value of $M^*/N$ increases as the charge $q$ increases while  $M/N$  decreases always with $q$. In particular $M^*/N$ increases with $q$ until the central density reaches a point $\phi(0)\approx0.75$, at which the lines  $M^*/N$  at different charges match and then $M^*/N$ turns out to be a decreasing function of $q$.

The maximum values of the charge-to-mass ratio satisfy the inequality $Q/M^*<q/m$ for all charges $q$, in particular $Q/M^*$ never reaches the critical value $q_{crit}/m$. The contrary conclusion would be reached if the misinterpreted mass $M$ were used since the inequality $Q/M>Q/M^*$ is satisfied, i.e. the charge-to-mass ratio $Q/M$ indeed attain values larger than $q_{crit}/m$ (see e.g.~Fig.~\ref{Ag-08-ven}).
To summarize, we found that all the relevant quantities that characterize charged boson stars behave
in accordance with the physical expectations. Bounded configurations are possible only
within an interval of specific values for the bosonic charge and the central density.

%%%%%%%%%%%%%%%%%%%%%%%%%%%%%%%%%%%%%%
%%%%%%%%%%%%%%%%%%%%%%%%%%%%%%%%%%%%%%
\section*{Acknowledgments}
We would like to thank Andrea Geralico for helpful comments and discussions.
One of us (DP) gratefully acknowledges financial support from the A. Della Riccia Foundation and  Blanceflor Boncompagni-Ludovisi, n\'ee Bildt.
This work was supported in part by
CONACyT-Mexico, Grant No. 166391,  DGAPA-UNAM and by CNPq-Brazil.
%\clearpage
%%%%%%%%%%%%%%%%%%%%%%%%%%%%%%%%%%%%%%%%%%%%%%%%%%%%%%%%%%%%%%%%%%%%
%%%%%%%%%%%%%%%%%%%%%%%%%%%%%%%%%%%%%%%%%%%%%%%%%%%%%%%%%%%%%%%%%%%%%

\appendix
\section{Neutral boson stars}
\label{Sec:NBS}
In this Appendix we shall focus on electrically neutral configurations exploring in particular their  global proprieties:
mass, radius and total particle numbers.
Neutral boson stars are gravitationally bound, spherically symmetric,
equilibrium configurations of complex scalar fields $\phi$ \cite{RR}.
It is possible to analyze
their interaction by considering the field equations describing a
system of free particles in a curved space--time with a metric
determined by the particles themselves.

The Lagrangian density of the gravitationally coupled complex scalar field
$\phi$ reads
\begin{equation}\label{R-oma-5}
\mathcal{L}=-\left(g^{\mu\nu}\partial_{\mu}\
\phi\partial_{\nu}\phi^*-m^{2} \phi\phi^*\right)
\end{equation}
where $m$ is the boson mass, $\phi^*$ is the complex conjugate field (see, for example, \cite{RR,Fulling,BD}).
This Lagrangian is invariant under global phase transformation $\phi\rightarrow \exp\left
(i \theta\right)\phi
$ where $\theta$ is a real constant that implies the conservation of  the total
particle number $N$.

Using the variational principle with the Lagrangian (\ref{R-oma-5}), we find the
 following Einstein coupled equations
\begin{equation}\label{rtor-ate}
G_{\mu\nu}\equiv
R_{\mu\nu}-\frac{1}{2}g_{\mu\nu}R=8\pi G_{\ti{N}}T_{\mu\nu}(\phi),
\end{equation}
with the following Klein-Gordon equations
\begin{eqnarray}\label{tot-all}
g^{\mu\nu}\nabla_{\mu}\nabla_{\nu}\phi+m^2\phi&=&0,\\
\label{cor-fe-to}
g^{\mu\nu}\nabla_{\mu}\nabla_{\nu}\phi^*+m^2\phi^*&=&0.
\end{eqnarray}
for the field $\phi$ and its complex conjugate $\phi^*$.

The symmetric energy-momentum tensor is
\begin{equation}\label{ahahhhhah}
T_{\mu\nu}=2\left(|g|\right)^{-1/2}\left(\frac{\partial}{\partial
x^{\alpha}}
\frac{\partial\left(\sqrt{-g}\right)\mathcal{L}}{\partial\left(g^{\mu\nu}/\partial
x^{\alpha}\right)}-\frac{\partial\left(\sqrt{-g}\right)\mathcal{L}}{\partial
g^{\mu\nu}}\right),
\end{equation}
and the current vector is
\begin{equation}\label{ttspo-os}
J^{\mu}=\imath \left\{
\left[\frac{\partial\mathcal{L}}{\partial\left(\partial_{\mu}\phi^{*}\right)}\phi^{*}\right]
-\left[\frac{\partial\mathcal{L}}{\partial\left(\partial_{\mu}\phi\right)}\phi\right]\right\}\ .
\end{equation}

The explicit form of Eq.\il(\ref{tot-all})
\begin{equation}\label{st-ra:mi}
\frac{1}{\sqrt{|g|}}\partial_{i}\left[g^{ij}\sqrt{|g|}\partial_{k}\phi\right]+g^{00}\partial_{0}^{2}\phi+m^{2}\phi=0 \ ,
\end{equation}
can be solved by using  separation of variables
\begin{equation}\label{C-ne-la}
\phi\left(r,\theta,\varphi,t\right)=R(r)Y_{l}^{m}\left(\theta,\varphi\right)e^{-\imath\omega t},
\end{equation}
where $Y_{l}^{m}\left(\theta,\varphi\right)$ is the  spherical harmonic.
%and $\omega\equiv E/\hbar$.
Equation (\ref{C-ne-la}) and
its complex conjugate describe a spherically symmetric bound state
of scalar fields with positive or negative  frequency
$\omega$, respectively\footnote{In the distribution we have considered all the particles are in the
same ground state $(n=1, l=0)$.
}. It ensures that the boson star
space--time remains static\footnote{In the case of a
real scalar field can readily be obtained in this formalism by
requiring $\omega = 0$ due to the condition $\phi =\phi*$.}.

In the case of spherical symmetry, we use as before the general line element
\begin{equation}\label{W-ter}
ds^2=e^\nu dt^2-e^\lambda
dr^2-r^2\left(d\vartheta^2+\sin\vartheta^2d\varphi^2\right)\ ,
\end{equation}
where $\lambda =
\lambda(r)$ and $\nu =\nu(r)$.

Thus, there are only three unknown functions of the radial coordinate $r$ to be determined, the metric function  $\nu, \lambda$, and the radial component
$R$ of the Klein--Gordon field.
From Eq.\il(\ref{st-ra:mi}) we infer the  radial Klein--Gordon
equation
\begin{equation}\label{J-yce}
R''(r)+\left(\frac{2}{r} -\frac{\lambda '(r)}{2}+\frac{\nu
'(r)}{2}\right) R'(r)+ e^{\lambda (r)} \left(-m^2+e^{-\nu (r)}
\omega^2\right) R(r)=0,
\end{equation}
where the prime $(')$ denotes the differentiation with respect to $r$.

The  energy momentum tensor components are (see \cite{RR})

\bea
\label{o-bell-al0}
T^{\phantom\ 0}_{0}&=& \frac{1}{2}\left\{\left[e^{-\nu}\omega^2+m^2\right]R_{01}^{2}+e^{-\lambda}R_{01}^{'2}\right\},
\\
\nonumber
\\ \label{o-bell-al1}
T^{\phantom\ 1}_{1}&=&-\frac{1}{2}\left\{\left[e^{-\nu}\omega^2-m^2\right]R_{01}^{2}+e^{-\lambda}R_{01}^{'2}\right\},
\\\nonumber
\\
\label{o-bell-al2}
T^{\phantom\ 2}_{2}&=& {T}^{\phantom\ 3}_{3}=-\frac{1}{2}\left\{\left[e^{-\nu}\omega^2-m^2\right]R_{01}^{2}-e^{-\lambda}R_{01}^{'2}\right\},
\\
\nonumber
\\
\label{o-bell-al3}
T^{\phantom\ i}_{0}&=&0
\eea

From the expressions (\ref{o-bell-al0},\ref{o-bell-al1}) and
from the Einstein equation (\ref{rtor-ate}) we finally obtain the
following two independent equations
\begin{eqnarray}
\label{spe}
\lambda' &=&\frac{1}{r}\left(1-e^{\lambda}\right)+8\pi G r e^{-\nu
}\left[R^{2}e^{\lambda}\left(
\omega^{2}+m^{2}e^{\nu}\right)+R'^{2}{e^{\nu }}\right],
\\
\nu' &=&-\frac{1}{r}\left(1-e^{\lambda}\right)+8\pi Gre^{-\nu}\left[R^{2}e^{\lambda}\left(
\omega^{2}-m^{2}e^{\nu}\right)+R'^{2}{e^{\nu }}\right],
\label{spe1}
\end{eqnarray}
for the metric fields $\lambda$ and $\nu$, respectively.

To integrate numerically these equations it is convenient to  make the following rescaling of variables:
\begin{equation}\label{long-n}
r\rightarrow \hat{r}/m,\quad\sigma(r)\equiv R(r)/\sqrt{8\pi G_N },\quad\omega\rightarrow\omega m\ .
\end{equation}
Thus,
we finally obtain  from Eqs.\il(\ref{J-yce}) and (\ref{spe}, \ref{spe1}) the following equations
\begin{eqnarray}
% \nonumber to remove numbering (before each equation)
\label{mee-ng}R ''&=&e^{\lambda} R
-\omega^{2}e^{\lambda-\nu} R+R '
\left(-\frac{1}{r}-\frac{e^{\lambda}}{r}+e^{\lambda} r
R^2\right)
\\\label{sol-vo1}
\lambda'&=& \frac{1}{r}\left(1-e^{\lambda}\right)+r e^{-\nu
}\left[R^{2}e^{\lambda}\left(
\omega^{2}+e^{\nu}\right)+R'^{2}{e^{\nu }}\right],\\
\label{sol-vo2}\nu
'&=&-\frac{1}{r}\left(1-e^{\lambda}\right)+re^{-\nu}\left[R^{2}e^{\lambda}\left(
\omega^{2}-e^{\nu}\right)+R'^{2}{e^{\nu }}\right],
\end{eqnarray}
for the radial part of the scalar field $R$ and
the  metric coefficients $\lambda$ and $\nu$ in the dimensionless variable $\hat{r}$.

The initial and boundary conditions we impose are
\bea
R(\infty)=0, \quad R'(\infty)=0,\quad
\mbox{and}\quad R(0)=\mbox{constant},\quad R'(0)=0.\label{Pro-sts}
\eea
in order to have a localized particle distribution  and
\begin{eqnarray}
\label{la-pri:ra}
\lambda(0)&=&0,\\
\label{Cong-na}\nu(\infty)&=&0,
\end{eqnarray}
to get asymptotically the ordinary Minkowski metric (\ref{la-pri:ra}), and to satisfy the regularity condition (\ref{Cong-na}).

We calculate the mass of system as
\begin{equation}\label{Der-ier}
M=4\pi \int_0 ^{\infty } \rho r^{2}dr,
\end{equation}
where the density $\rho$, given by $T_0^0$, is
\begin{equation}\label{Ar-d-re}
\rho =\frac{1}{2}\left[R^2\left(1+\omega^2 e^{-\nu}\right) +e^{-\lambda} R'^2\right].
\end{equation}
The particle number   is determined by the following
normalization condition
\begin{equation}\label{a-per-det}
N=\int_0 ^{\infty }\left\langle J^0\right\rangle
\left(-g\right)^{1/2}d^3x.
\end{equation}
which, using Eq.\il(\ref{ttspo-os}), becomes
\begin{equation}\label{a-per-det1}
N=\int_0^{\infty }r^{2} \omega e^{(\lambda-\nu
)/2}R^{2} dr.
\end{equation}
The mass  $M$ is measured in units of $M_{\mbox{\tiny{Pl}}}^{2}/m$, the
particle number $N$ in units of $M_{\mbox{\tiny{Pl}}}^{2}/m^2$, $\omega$ in units of $m c^2$ and the radius of the configuration is
in units of $\hbar/(m c)$.

\begin{center}
\begin{figure}[h!]
\begin{tabular}{lll}
% \begin{minipage}[b]{1cm}
\includegraphics[width=0.33\hsize,clip]{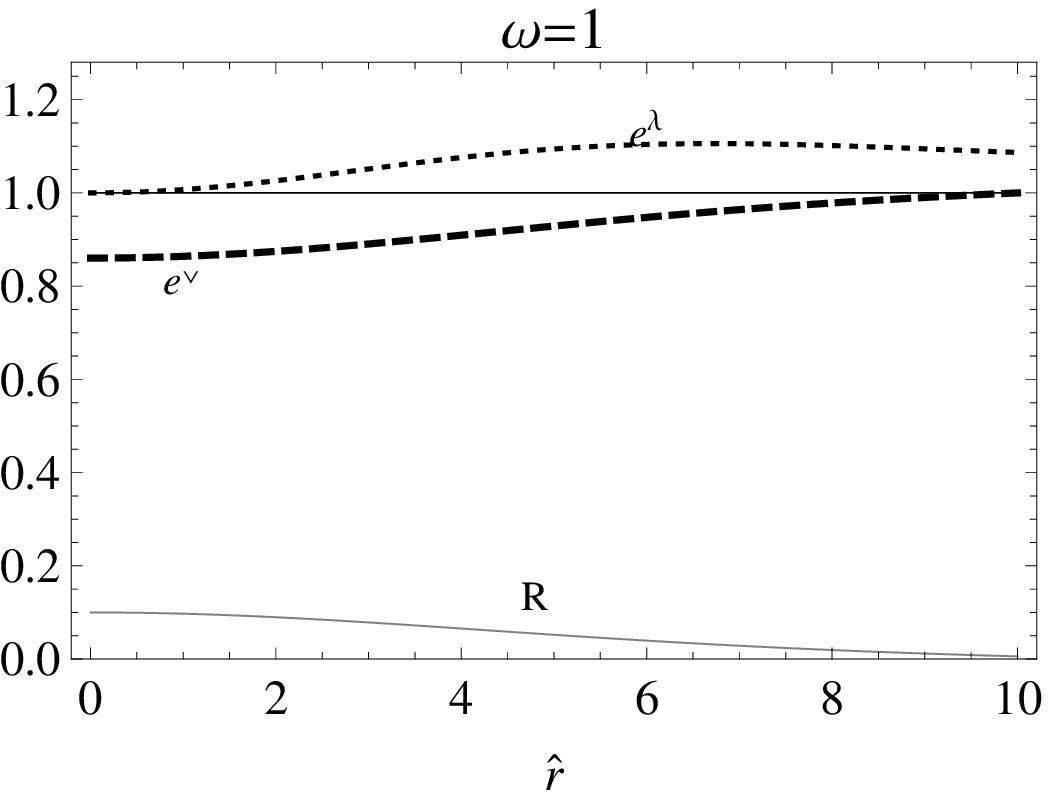}% "%" necessario
&\includegraphics[width=0.33\hsize,clip]{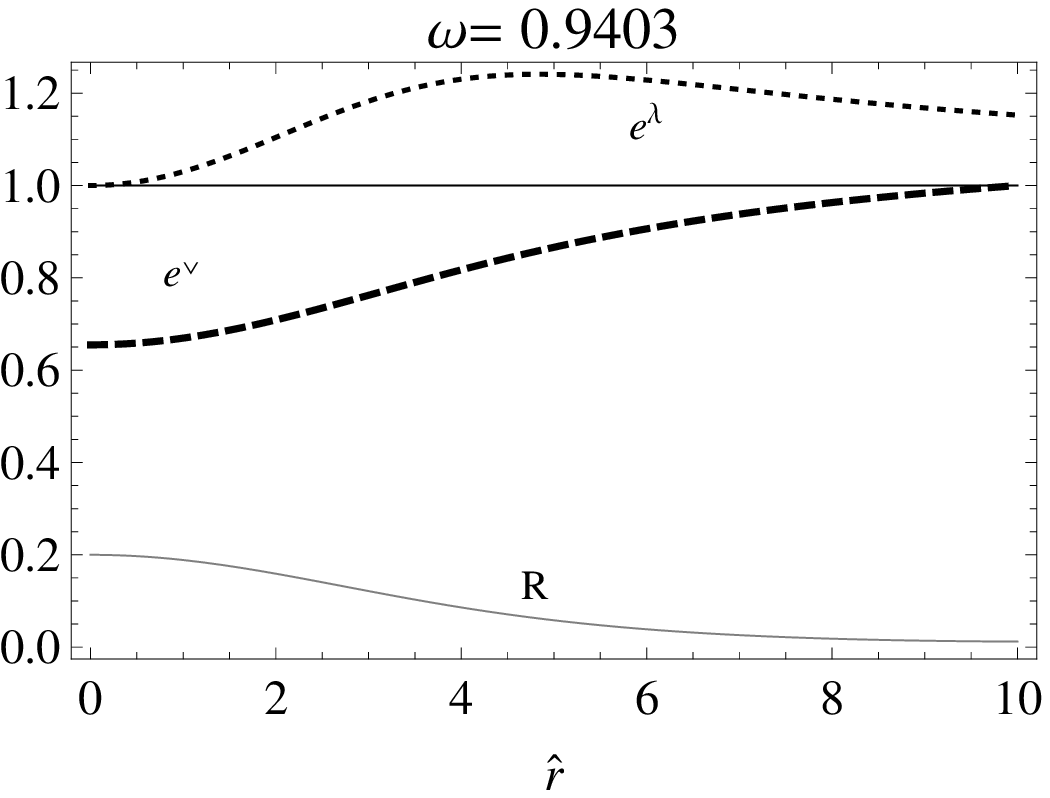}
&\includegraphics[width=0.33\hsize,clip]{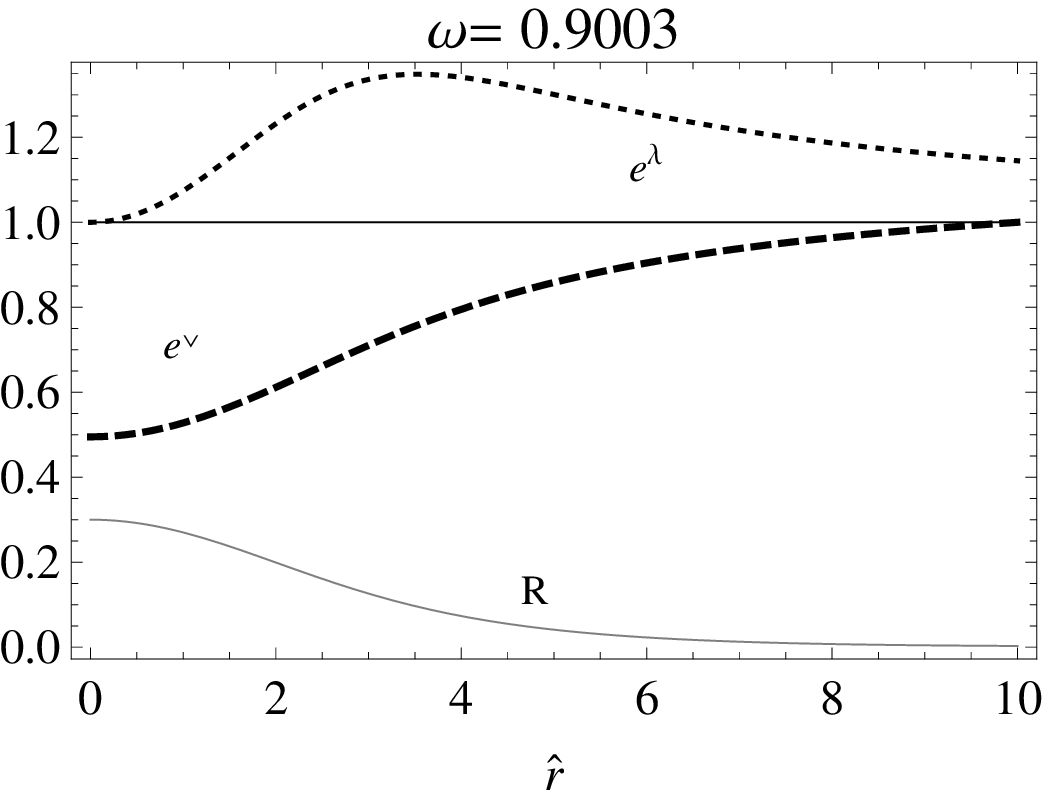}\\%(a)&(b)&(c)\\
\includegraphics[width=0.33\hsize,clip]{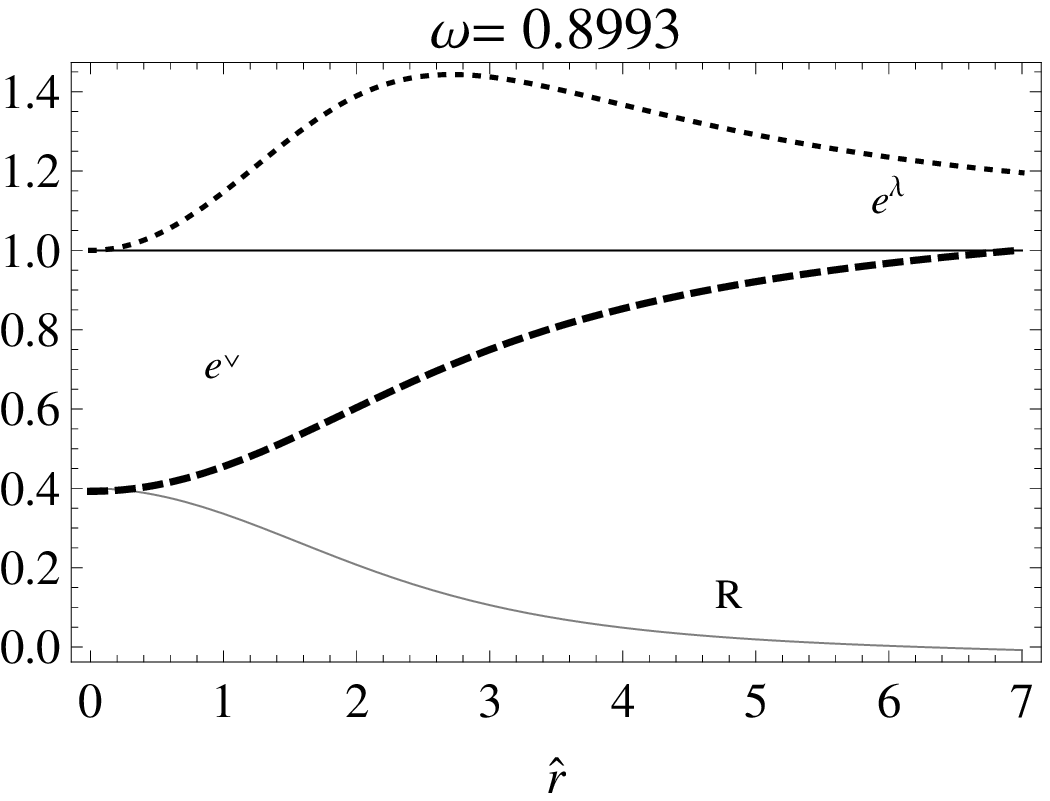}&
\includegraphics[width=0.33\hsize,clip]{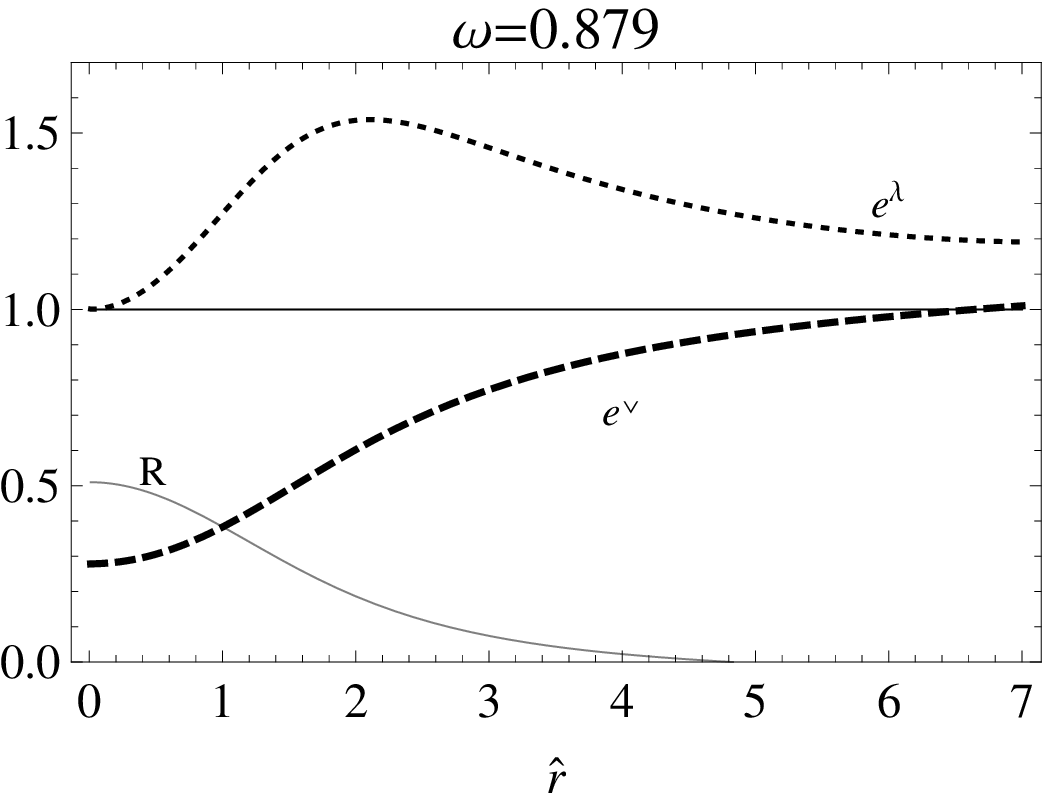}&
\includegraphics[width=0.33\hsize,clip]{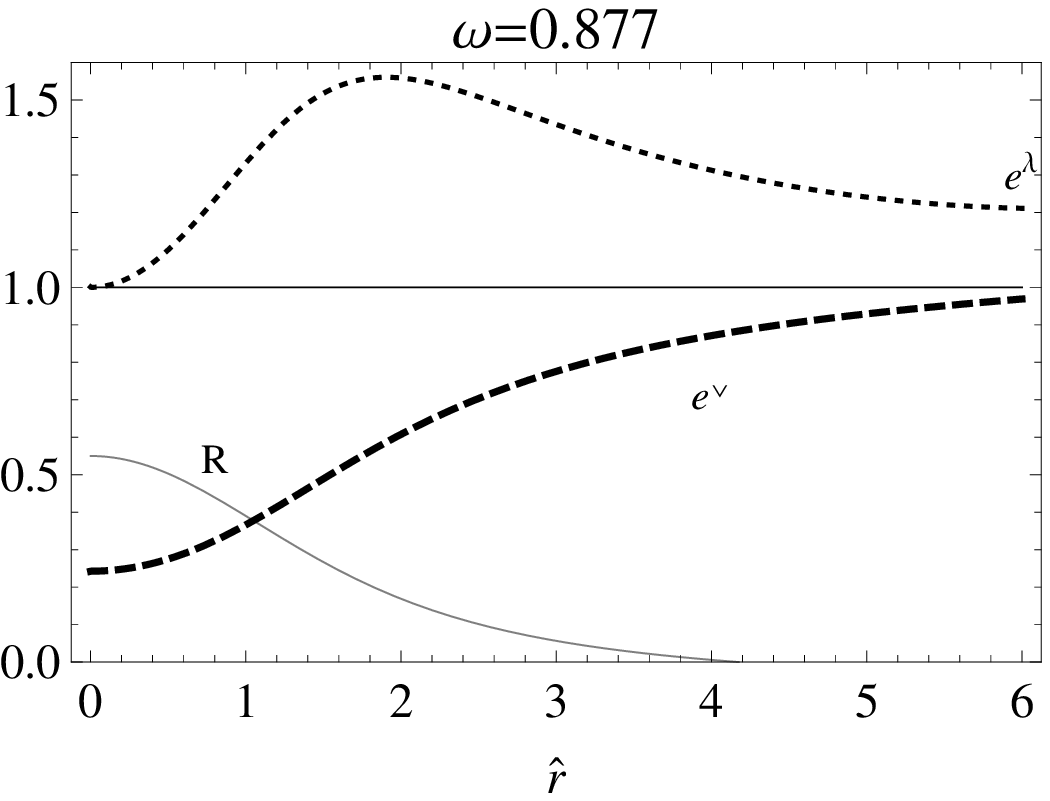}\\%(d)&(e)&(f)
%\end{minipage}
\end{tabular}
\caption{The radial function $R$  of the Klein--Gordon field (gray line),  the  metric coefficient $g_{11}=-e^{\lambda}$ (dotted line) and the function
$e^{\nu}=g_{00}$ (dashed line)  are
plotted as functions of $\hat{r}$ (dimensionless) $\hat{r}=r/m$ for selected values of
the redial function $R(r)$ at the origin and different values of the
eigenvalue $\omega$ in units of $m c^2$. }\label{Mak-oo}
\end{figure}
\end{center}

\begin{center}
\begin{figure}[h!]
\begin{tabular}{cc}
% Requires \usepackage{graphicx}
\includegraphics [scale=0.7]{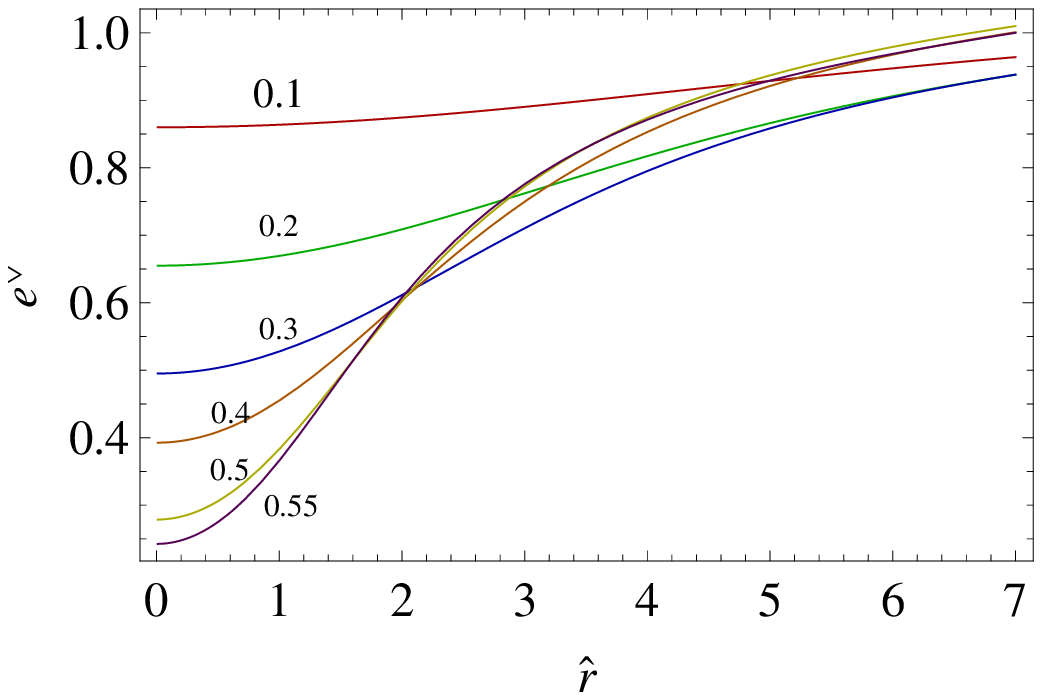}
\includegraphics [scale=0.7]{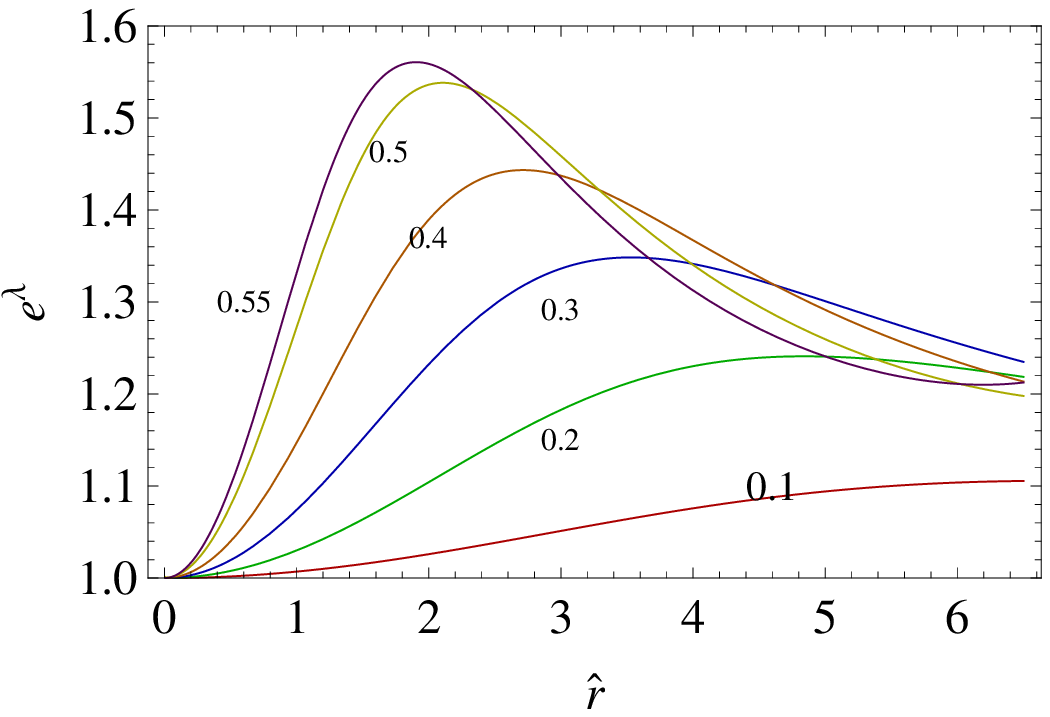}
\end{tabular}
\caption{(Color online) The metric coefficient $g_{11}=-e^{\lambda}$  (right plot) and the function
$e^{\nu}=g_{00}$ (left plot)  are
plotted  as functions of $\hat{r}$ (dimensionless) $\hat{r}=r/m$  for selected values
of the radial function $R(r)$ at the origin.
} \label{fig:carcta-lot}
\end{figure}
\end{center}

\begin{figure}[h]
\begin{center}
% Requires \usepackage{graphicx}
\includegraphics [scale=.61]{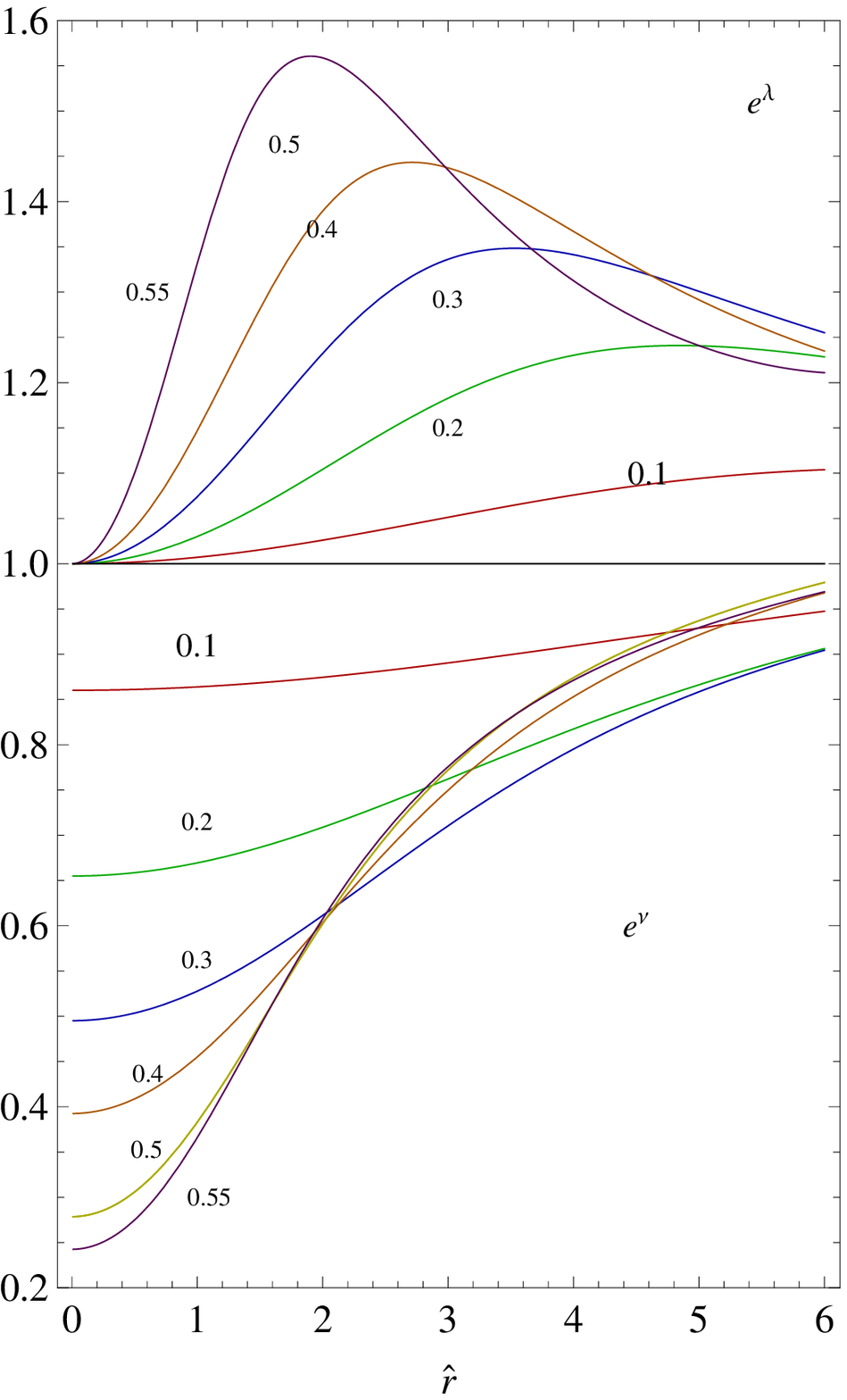}
\caption{(Color online) The metric coefficient $e^{\lambda}=-g_{11}$ and the function
$e^{\nu}=g_{00}$   are
plotted  as functions of $\hat{r}$ (dimensionless) $\hat{r}=r/m$  for selected values of
the redial function $R(r)$ at the origin.
See also Table\il\ref{tab:grosse}.}
\label{fig:Con-so}
\end{center}
\end{figure}

\begin{figure}[h]
\begin{center}
\includegraphics [scale=1]{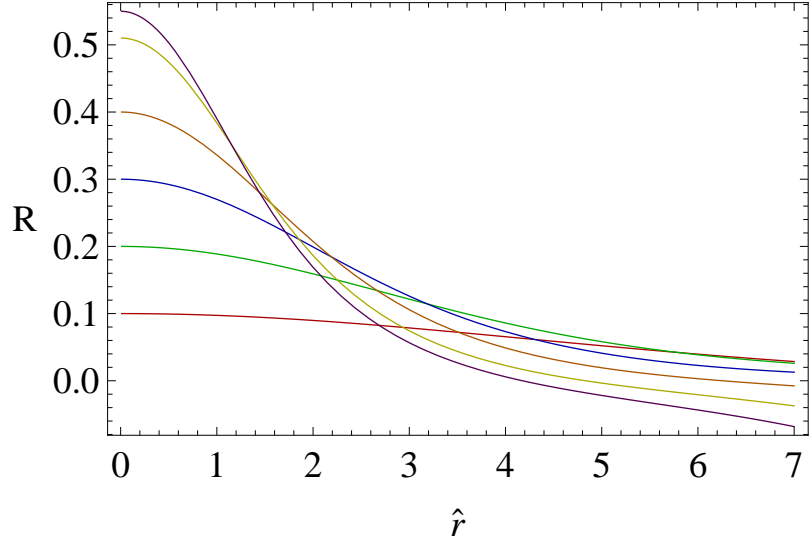}
\end{center}
\caption{(Color online) The radial function $R$ is plotted as a function of $\hat{r}$ (dimensionless) $\hat{r}=r/m$  for selected values of $R$ at the origin. See also Table\il\ref{tab:grosse}.}
\label{fig:Radiale}
\end{figure}

In the numerical analysis we obtain a maximum value of $g_{11}$ from which we determine the  ``effective
radius'' $R_{\ti{eff}}$ of the distribution as the  radius, $r_{g_{11}^{\ti{Max}}}$, corresponding to the maximum of
$g_{11}$ \cite{RR, Schunck:2003kk}. We carried out a numerical integration for
different values of the radial function $R$ at the origin. We give
some numerical values  in Table\il\ref{tab:grosse}.
\begin{table}[h!]
%\begin{center}
\resizebox{0.7\textwidth}{!}{%
\begin{tabular}{lcccrrrrrr}
%\begin{tabularx}
%\textheight{WWWWWWWWW}
\hline\hline
 $R\left(0\right)$ &$\omega$ $\left(mc^2\right)$&$g^{\ti{Max}}_{11}$&
$r_{g_{11}^{\ti{Max}}}$$(\hbar/mc)$
&  $g^{\ti{min}}_{00} $& $g_{00}\left(r_{\ti{Max}}
\right) $ &N
$(m^{-2} M_{Pl}^{2})$&M $(M_{Pl}^{-2}/m)$
\\ \hline \hline
0.10&1.0000&1.10572&6.73590&0.860104&0.959895&0.338031&0.334027
\\
0.20&0.9403&1.24099&4.84330&0.654933&0.859126&0.625526&0.602570
\\
0.30&0.9003&1.34844&3.52866&0.495123&0.757922&0.644172&0.623620
\\
0.40&0.8993&1.44335&2.71423&0.392571&0.712016&0.575796&0.573405
\\
0.51&0.8790&1.53397&2.09463&0.276904&0.619017&$\checkmark$&$\checkmark$
\\
0.55&0.8770&1.56106&1.90578&0.242491&0.587189&$\checkmark$&$\checkmark$
\\
\hline \hline
\end{tabular}
}
\caption{Numerical results for   neutral boson stars. $R\left(0\right)$ is the
value of the radial part of the wave function at the origin.
The mass at infinity has been computed by Eq.\il(\ref{Der-ier}). The value is given
in units  $M_{Pl}^2/m=\left( \hbar^{2}cG^{-1}m^{-1}\right)$. The eigenvalue
$\omega$, measured in units of $mc^2$, where $m$ is the  boson mass,
has been determined by requiring that the redial part $R$ goes to
zero at infinity. The radius of the distribution (units $\hbar
c^{-1}m^{-1} $) has been defined to be the value
$r_{g_{11}^{\ti{Max}}}$ of the radial coordinate corresponding to
the maximum of $g_{11}$. The minimum $g_{00}$ is attained at the
origin.}\label{tab:grosse}
\end{table}
We have fixed some values  for $R$ at the origin and a random value
for the eigenvalue $\omega$. We solved all the three equations
simultaneously, looking for the value of $\omega$ for which the
radial function decreases exponentially, reaching the value zero at
infinity.
We have plotted some results in Fig.\il\ref{Mak-oo}
and in Figs.\il\ref{fig:carcta-lot},\ref{fig:Con-so} and \ref{fig:Radiale} where the profiles are shown in terms of the radial variable.
%
%\end{figure}

\begin{figure}[h!]
\begin{tabular}{c}
% Requires \usepackage{graphicx}
%\includegraphics [scale=0.9]{PlotnMmezzodoppio.eps}(a)\\
\includegraphics [scale=1]{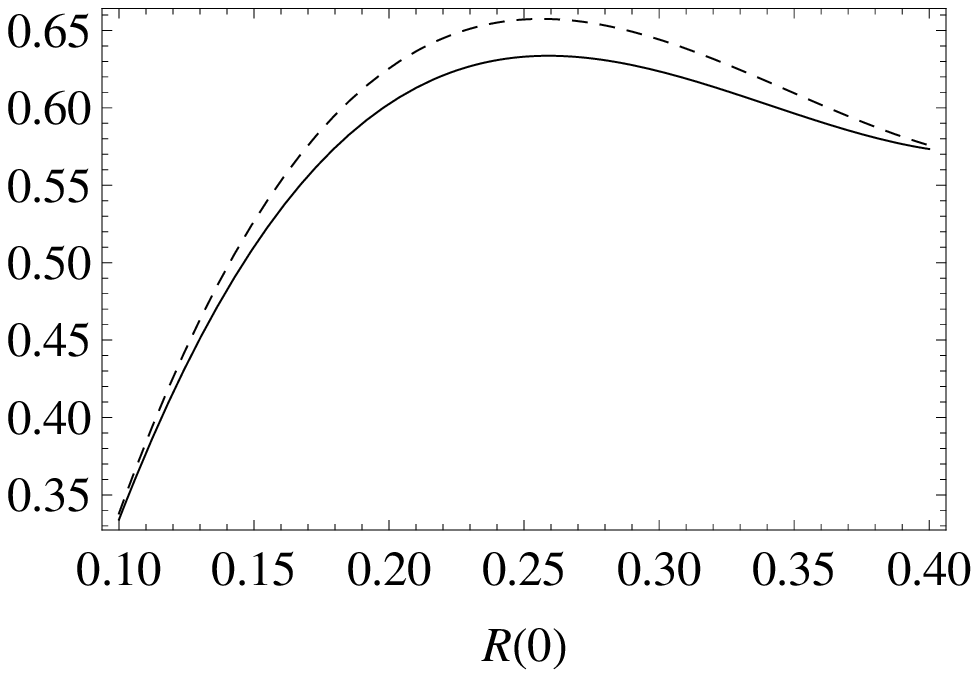}
\end{tabular}
\caption{The mass at infinity $M$ in units of $M_{Pl}^{2}/m$
(solid line), and  the particle number $N$ in units of $M_{Pl}^{2}/m^2$
(dashed line) are plotted as functions of the central density
$R(0)$.  Note that there
exists a maximum value of the mass
$M\simeq0.635626 M_{Pl}^{2}/m$ for the central density
$R(0)_{\mbox{\tiny{Cri}}}=0.277619$ above which there are no static
solutions.  There exists
a maximum value of the particle number
$N_{\mbox{\tiny{Cri}}}=0.658438 M_{Pl}^{2}/m^2$ for the central density
$R(0)_{\mbox{\tiny{Cri}}} = 0.278289$.}
\label{nat-vi-y}
\end{figure}

The mass at infinity and the total number of particle always stays
positive. To an
increase (decrease) of the number of particles always corresponds an
increase (decrease) of the mass at infinity  (see Fig.\il\ref{nat-vi-y}).
 The concept of
critical mass is introduced since the total  particle number and the mass at infinity
 (as a function of the central density, see Fig.\il\ref{nat-vi-y}) reaches a
maximum value $N_{\mbox{\tiny{Cri}}}=0.658438 M_{Pl}^{2}/m^2$ and
$M_{\mbox{\tiny{Cri}}}=0.635626 M_{Pl}^{2}/m$, respectively,
for a specific central density $R(0)_{\mbox{\tiny{Cri}}}$:
\bea
\label{a}
N_{\mbox{\tiny{Cri}}}=0.658438 M_{Pl}^{2}/m^2,\qquad& R(0)_{\mbox{\tiny{Cri}}} = 0.278289,\\
\label{paz-a} M_{\mbox{\tiny{Cri}}}=0.635626 M_{Pl}^{2}/m,\qquad&
R(0)_{\mbox{\tiny{Cri}}}=0.277619.
\eea
For further details, see also \cite{Jetzer:1988af}.
%

%\begin{center}
%\end{center}

%\section{The case $q=0.8$}
%

\end{document}